\documentclass[11pt]{article}
\usepackage{graphicx}
\usepackage{graphics}
\usepackage{dsfont}
\usepackage{epsfig}
\usepackage{bbm}
\usepackage{amsmath,amssymb,amsthm,amscd}
\usepackage[sort&compress,square,numbers]{natbib}
\usepackage{float}
\usepackage{braket}
\usepackage{tikz}

\numberwithin{equation}{section}

\restylefloat{table}

\usepackage{hyperref}

\begin{document}

\baselineskip=15pt
\begin{titlepage}
\begin{flushright}
CERN-TH-2020-081\\
UUITP-14/20
\end{flushright}

\medskip

\begin{center}
{\LARGE {\bf Non-Simply-Connected Symmetries in 6D SCFTs}}\\[12pt]
\bigskip

Markus Dierigl$^{\,\text{a}}$, Paul-Konstantin~Oehlmann$^{\,\text{b}}$, Fabian Ruehle$^{\,\text{c,d}}$
\bigskip

\vspace{0.cm}
{
	{\it${}^{\text{a}}$ Department of Physics and Astronomy, University of Pennsylvania, \\ Philadelphia, PA 19104, USA } \\ markusd@sas.upenn.edu\\[.5em]
	{\it ${}^{\text{b}}$ Department of Physics and Astronomy,~Uppsala University,\\~Regementsv\"agen 1, 77120 Uppsala,~Sweden}\\paul-konstantin.oehlmann@physics.uu.se\\[.5em]
	{\it ${}^{\text{c}}$ CERN Theory Department, 1 Esplanade des Particules,\\ CH-1211 Geneva, Switzerland} \\ 
	{\it ${}^{\text{d}}$ Rudolf Peierls Centre for Theoretical Physics, University of Oxford\\Department of Physics, Parks Road, Oxford OX1 3PU, United Kingdom}
	fabian.ruehle@cern.ch
}
\end{center}

\bigskip\bigskip

\begin{abstract}\noindent
Six-dimensional $\mathcal{N}=(1,0)$ superconformal field theories can be engineered geometrically via F-theory on elliptically-fibered Calabi-Yau 3-folds. We include torsional sections in the geometry, which lead to a finite Mordell-Weil group. This allows us to identify the full non-Abelian group structure rather than just the algebra. The presence of torsion also modifies the center of the symmetry groups and the matter representations that can appear. This in turn affects the tensor branch of these theories. We analyze this change for a large class of superconformal theories with torsion and explicitly construct their tensor branches. Finally, we elaborate on the connection to the dual heterotic and M-theory description, in which our configurations are interpreted as generalizations of discrete holonomy instantons.
\end{abstract}

\end{titlepage}
\clearpage
\setcounter{footnote}{0}
\setcounter{tocdepth}{2}
\tableofcontents
\clearpage

\section{Introduction}

Six-dimensional superconformal field theories (SCFTs) have played a prominent role in high energy physics in recent years. Since there are no SCFTs in dimensions beyond six \cite{Nahm:1977tg}, and those in dimension six are necessarily strongly coupled, these theories are highly interesting already in their own rights, see e.g.\ \cite{Witten:1995gx, Witten:1995ex, Strominger:1995ac, Seiberg:1996vs, Ganor:1996mu}. Moreover, 6d SCFTs can be used to derive a vast network of lower-dimensional SCFTs via compactification, see \cite{Ohmori:2015pua, Ohmori:2015pia, DelZotto:2015rca, Morrison:2016nrt, Razamat:2016dpl, Apruzzi:2016nfr, Bah:2017gph, DelZotto:2017pti, Apruzzi:2018oge, Bhardwaj:2018yhy, Bhardwaj:2018vuu,Bhardwaj:2019fzv} for a partial list of compactifications of 6d $\mathcal{N} = (1,0)$ theories. In this sense, the 6d SCFTs can be understood as master theories whose investigation is paramount in order to understand the generation and connection of SCFTs in general.

Of great importance in the investigation of 6d SCFTs is their construction within string theory frameworks and especially F-theory \cite{Vafa:1996xn, Morrison:1996na, Morrison:1996pp}. In fact, F-theory has lead to a vast number of 6d SCFTs using a classification of base geometries describing the tensor branch deformation of the singular theories and taking into account the possibility of singular fibers \cite{Heckman:2013pva, Heckman:2015bfa}, see also \cite{Bhardwaj:2015oru, Bertolini:2015bwa, Bhardwaj:2015xxa, Bhardwaj:2018jgp} and the review \cite{Heckman:2018jxk}. In most of these models, the construction of the 6d SCFTs was achieved by tuning a Weierstrass model: The appearance of non-Higgsable clusters on rational curves with negative self-intersection~\cite{Morrison:2012np}, as well as the collisions of singular divisors, lead to superconformal field theories \cite{DelZotto:2014hpa}. While the Weierstrass model, together with monodromy data \cite{Grassi:2011hq}, is sufficient to extract the non-Abelian part of the gauge and flavor \textit{algebras},\footnote{Note that F-theory in general only leads to a subgroup of the maximal flavor algebra, cf.~\cite{Bertolini:2015bwa}.} it is usually not enough to determine the full gauge and flavor \textit{groups} on the tensor branch. For works discussing also Abelian flavor symmetries see \cite{Lee:2018ihr, Apruzzi:2020eqi}.

In order to determine the full non-Abelian group structure, one needs to access additional information encoded in the presence of torsional sections, see e.g.\ \cite{Aspinwall:1998xj, Mayrhofer:2014opa, Cvetic:2017epq, Hajouji:2019vxs} and also \cite{Monnier:2017oqd}. These impose certain factorization constraints on the parameters in the Weierstrass model. Equivalently, in these models the full $\text{SL}(2,\mathbb{Z})$ monodromy is reduced to an orbit of an appropriate congruence subgroup, see \cite{Hajouji:2019vxs}. With the torsional sections enforcing a modding out of a finite discrete group embedded into the center of the non-Abelian group factors, this leads to a restriction on the gauge and flavor representations that appear on the tensor branch of the corresponding SCFTs. Moreover, the resulting non-Abelian groups are no longer simply-connected. Therefore, SCFTs respecting the presence of a non-trivial Mordell-Weil torsion will have restrictions on their tensor branch and other deformations, see e.g.\ \cite{Heckman:2015ola, Mekareeya:2016yal, Heckman:2016ssk, Heckman:2018pqx}.

In this paper, we will extend the classification of 6d SCFTs to models with non-trivial Mordell-Weil torsion. In this way, we can classify the gauge and flavor symmetry groups appearing on the tensor branch of 6d SCFTs rather than their algebras. As in the unrestricted case, the major building blocks are given by non-Higgsable clusters and the collision of two singular divisors, now, in the presence of torsional sections. Scanning different theories and their compatible singularity enhancements, we find a class of allowed deformations which respect the Mordell-Weil torsion. This can be regarded as a starting point for the development of constructive approaches as described for example in \cite{Heckman:2018pqx} and has the potential for a full classification of 6d SCFTs with respect to their global group structure, harnessing the full power of F-theory. 

The global structure also modifies the possibility of compactifications to lower dimensions, since reducing the group structure (see e.g. \cite{Anderson:2019kmx} for compact examples) naturally leads to a large possibility  for background fluxes \cite{tHooft:1979rtg, Aharony:2013hda}, see also \cite{Tachikawa:2013hya, Garcia-Etxebarria:2019cnb} for considerations with more supersymmetry and the connection of the global group in 4d to flux data in the 6d theory. In fact, some of the theories we find here have already been anticipated in \cite{Ohmori:2018ona} and used to compactify from six to four dimensions with fluxes of non-trivial Stiefel-Whitney class. While in \cite{Ohmori:2018ona} the authors use a field-theoretic intuition based on the operator spectrum, our construction sheds light on the UV-complete origin of the global data in terms of the string theory construction within F-theory.

The rest of this work is organized as follows. In Section \ref{sec:rev} we review the construction of F-theory models with non-trivial and finite Mordell-Weil group related to the presence of torsional sections or the restriction of the monodromy groups. The approach to the construction of general 6d SCFTs in F-theory is briefly described in Section \ref{sec:SCFT}. While not essential for our later analysis, we further point out alternative approaches in string constructions to access the global realization of gauge and flavor groups, which should be understood as a staring point for further investigations in the M-theory and heterotic realization of 6d SCFTs. In Section \ref{sec:NHC} we begin our systematic approach for the construction of globally constrained 6d SCFTs by investigating non-Higgsable clusters in the presence of torsional sections. These become important building blocks for the analysis of non-simply-connected 6d SCFTs, which is carried out in Section \ref{sec:NSCFTs} and \ref{sec:zoo}. We conclude in Section \ref{sec:concl} and present details of some of the results, as well as exotic models, in the Appendices.

\section{Mordell-Weil Torsion and Restricted Monodromies}
\label{sec:rev}

In this section we briefly review F-theory models with non-trivial Mordell-Weil group. In F-theory on Calabi-Yau 3-folds, singularities over non-compact curves in the base lead to flavor symmetries, whereas singularities over compact curves lead to gauge symmetries of the 6d effective theory. In the presence of non-trivial Mordell-Weil torsion, the structure of these groups is modified. More precisely, the presence of torsional sections in the geometry of the elliptically-fibered 3-folds leads to a restriction of the allowed fiber monodromies\footnote{Moreover, a conjecture was made in \cite{Klevers:2014bqa,Oehlmann:2016wsb}, that fibrations with torsion are related to genus-one fibrations via mirror duality in the fiber.}. These restrictions impose constraints on the available matter representations and lead in general to a non-simply-connected non-Abelian structure of the symmetries in the model. For a discussion of Mordell-Weil torsion in F-theory see also \cite{Aspinwall:1998xj, Mayrhofer:2014opa, Hajouji:2019vxs} as well as the review \cite{Weigand:2018rez}.

\subsection{F-Theory and Weierstrass Models}
F-theory compactifications can be thought of as a non-perturbative generalization of type IIB string constructions, where the axio-dilaton is identified with the complex structure parameter $\tau$ of an auxiliary torus $\mathcal{E}$, \cite{Vafa:1996xn, Morrison:1996na, Morrison:1996pp}. The generalization proceeds by allowing the axio-dilaton to vary over the physical compactification space $B$, sourced by the back-reaction of generalized 7-branes, so-called $(p,q)$-branes. The whole geometry can be described as an elliptic fibration $X$
\begin{align}
\begin{array}{rcl}
\mathcal{E} & \rightarrow & X \\
&&\,\downarrow \pi \\
&& B \, ,
\end{array} 
\end{align}
which, in order to preserve the minimal amount of supersymmetry, needs to be Calabi-Yau. In the following, we are interested in the description of 6d $\mathcal{N} = (1,0)$ supersymmetric field theories decoupled from gravity. For this purpose, we choose $B$ to be a non-compact, complex two-dimensional K\"ahler manifold.

The main tool for the description of the elliptically-fibered Calabi-Yau 3-fold $X$ is going to be the Weierstrass model, which defines $X$ by the hypersurface equation
\begin{align}
y^2 = x^3 + f x z^4+ g z^6 \,.
\end{align}
Here, $x$, $y$, and $z$ are projective coordinates in $\mathbb{P}_{2,3,1}$, and the coefficients $f,g$ are sections of multiples of the anti-canonical class $-K$ of the base,
\begin{align}
f \sim - 4 K \qquad g \sim - 6 K \, .
\end{align}
They parametrize the local complex structure $\tau$ up to SL$(2,\mathbbm{Z})$ transformations. By construction, the Weierstrass model has a smooth section, called the zero-section, given by $s_0:~[1:1:0]$. Along the discriminant locus $\{\Delta=0\}$ with
\begin{align}
\Delta = 4f^3 + 27 g^2 \sim -12 K \,,
\end{align}
the elliptic fiber degenerates. The type of singularity is determined by the vanishing order of $f$, $g$, and $\Delta$, as well as possible monodromy actions \cite{Grassi:2011hq}. This classification is summarized in Table~\ref{tab:KodTate}. For generalization see also \cite{Katz:2011qp}. 
\begin{table}
\begin{center}
{\footnotesize
\begin{tabular}{|c|c|c|c|c|c|}\hline
Fiber Type & ord (f) & ord(g)&$ord(\Delta) $& monodromy cover & algebra \\ \hline
$I_0$ & $\geq 0$ &$\geq 0 $&$0$ & - & - \\ \hline 
$I_1$ & $ 0$ &$  0 $&$1$ & - & - \\ \hline  
$I_m$ & $0$ &$0 $&$m$ & $\psi^2+(2 g/2f)|_{z=0}$ &$ \mathfrak{sp}([m/2])$ or $\mathfrak{su}(m)$ \\ \hline  
$II$ & $\geq 1$ &$1 $&$2$ & $-$ & - \\ \hline  
$III$ & $  1$ &$\geq 2 $&$3$ & $-$ & $\mathfrak{su}(2) $  \\ \hline  
$IV$ & $  \geq 2$ &$2 $&$4$ & $ \psi^2 - (g/z^2)|_{z=0}$ & $\mathfrak{sp}(1) $ or $\mathfrak{su}(3) $   \\ \hline  
$I_0^*$ & $  \geq 2$ &$ \geq 3 $&$6$ & $ \psi^3 + \psi (f/z^2)|_{z=0} \,\psi  +(g/z^3)|_{z=0}$ & $\mathfrak{g}(2) $ or $\mathfrak{so}(7) $ or $\mathfrak{so}(8) $    \\ \hline  
$I_{2n-5}^*$, $n>2$ & $ 2$ &$ 3 $&$2n+1$ & $ \psi^2 +  \frac{1}{4} (\Delta/z^{2n+1})(2 z f/ 9 g)^3|_{z=0}$ & $\mathfrak{so}(4n-3) $ or $\mathfrak{so}(4n-2) $    \\ \hline  
$I_{2n-4}^*$, $n>2$ & $ 2$ &$ 3 $&$2n+2$ & $ \psi^2 +  (\Delta/z^{2n+2})(2 z f/9 g)^2|_{z=0} $ & $\mathfrak{so}(4n-1) $ or $\mathfrak{so}(4n ) $    \\ \hline  
$IV^*$ & $  \geq 3$  &$4 $&$8$ & $ \psi^2 - (g/z^4)|_{z=0}$ & $\mathfrak{f}_4 $ or $\mathfrak{e}_6 $   \\ \hline  
$III^*$ & $  3$ &$\geq 5 $&$9$ & $-$ & $\mathfrak{e}_7 $  \\ \hline  
$II^*$ & $ \geq 4$ &$ 5 $&$10$ & $-$ & $\mathfrak{e}_8 $  \\ \hline  
non-min & $ \geq 4$ &$ \geq 6 $&$ \geq 12$ & $-$ & -   \\ \hline  
\end{tabular}
}
\caption{\label{tab:KodTate} Kodaira-Tate classification of singular fibers and local gauge algebras. }
\end{center}
\end{table}

In codimension one, these degenerations of the fiber precisely correspond to the loci of $(p,q)$ 7-brane stacks in $B$, whose world-volume gauge algebras correspond to their Tate fiber type. Since the gauge-kinetic function of a 7-brane is proportional to the volume of the divisor it wraps, we distinguish the cases where they have finite or infinite volume:
\begin{itemize}
\item Branes that wrap compact divisors admit dynamical gauge fields and encode gauge symmetries.
\item Branes that wrap non-compact divisors encode non-dynamical background gauge connections and host flavor symmetries.
\end{itemize}
To engineer a consistent 6d supersymmetric field theory with the described ingredients, cancellation of various anomalies must be taken care of by the 6d version of the Green-Schwarz mechanism \cite{Green:1984bx, Sagnotti:1992qw}. Since we work in a non-gravitational setup, anomaly cancellation only needs to be imposed for the gauge part of the symmetries, while the flavor sector (or would-be gravitational sector prior to decoupling) can be anomalous. For a general account of anomaly cancellation in F-theory see \cite{Park:2011ji}. 

In the type IIB picture, matter multiplets naturally arise from open strings stretching between two stacks of 7-branes. These states become massless where the branes intersect. In the F-theory picture, the 7-branes wrap codimension-one loci in the base, over which the fiber degenerate. Brane intersections hence correspond to codimension-two loci in the base. Since two loci over which the fiber degenerates intersect, the fiber singularity becomes worse at the intersections. 

From the type IIB picture, matter multiplets transform in the bi-fundamental representation of the two 7-brane stacks. However, since in F-theory more general $(p,q)$ branes and strings are possible, more general representations can occur. To deduce the precise representations $\mathbf{R}$ of the matter multiplets, one can employ two different techniques: The first way is to fully resolve the geometry and use the duality to M-theory on the Coulomb branch to geometrically extract the weights from M2-branes wrapping holomorphic curves (e.g. see \cite{Marsano:2011hv}). The second approach is due to Katz and Vafa \cite{Katz:1996xe} and is closer to the IIB picture by interpreting the two intersecting 7-brane stacks as a deformation of the configuration where they were parallel and on top of each other. Even though Kodaira's (or Tate's) classification is strictly speaking only valid in codimension one, one can usually start from the enhanced algebra $\mathfrak{g}_{IJ}$ dictated by the codimension-two vanishing orders, and break it by tilting one part of the brane stack such that two stacks with gauge algebra $\mathfrak{g}_I$ and $\mathfrak{g}_J$ arise. In field theory, such a tilting corresponds to a VEV in the adjoint representation and therefore the representation at codimension two can be read off from the decomposition 
\begin{align}
\text{adj}(\mathfrak{g}_{IJ})\to\bigoplus_{I,J} (\mathbf{R}_{I},\mathbf{R}_{J})\oplus (\text{adj}(\mathfrak{g}_{I}),\mathbf{1})\oplus (\mathbf{1},\text{adj}(\mathfrak{g}_{J})) \,.
\end{align}
In most cases, this is enough to deduce matter representations. More care has to be taken when considering non-simply laced groups. Here the additional source of monodromy has to be taken into account when deducing the resulting representations \cite{Aspinwall:2000kf,Grassi:2011hq}.

\subsection{Non-Simply-Connected Groups from Mordell-Weil Torsion}
In this work we focus on the global structure of non-Abelian gauge and flavor symmetries in the six-dimensional F-theory models. In general, these groups will be non-simply-connected which means that
\begin{align}
\pi_1 (\mathcal{G}) = T \,,
\label{eq:globalT}
\end{align}
where $\mathcal{G}$ describes the flavor and gauge symmetries of the model. In the following, we focus on models without Abelian gauge group factors, thus $T$ is a finite group. By starting with the simply-connected group $\mathcal{G}^*$ related to the algebras determined by the codimension-one singularities, \eqref{eq:globalT} can be understood as modding out part of the center, see e.g.\ \cite{Aharony:2013hda}, of the flavor and gauge symmetries,
\begin{align}
\mathcal{G} = \frac{\mathcal{G}^*}{T}\,.
\end{align}
Note that if $T$ is only a subgroup of the centers of the individual factors in $\mathcal{G}^*$, the group action has to be further specified. One way to deduce it is to study the matter fields of the theory, which have to transform trivially under $T$.

In F-theory, the appearance of a non-simply-connected gauge group is related to the presence of extra sections in the elliptic fibration. In general, additional sections are related to the Mordell-Weil group of the elliptic fiber. Extra rational sections lead to Abelian $\text{U}(1)$ symmetries \cite{Cvetic:2018bni, Weigand:2018rez} and would correspond to the free part of the Mordell-Weil group, which we do not discuss here. Extra torsional sections are related to the torsion part of the Mordell-Weil group, which is encoded in the discrete group $T$, see e.g.\ \cite{Aspinwall:1998xj, Mayrhofer:2014opa}. Adding a $k$-torsional section $k$ times to itself (addition is meant as an element of the Mordell-Weil group) will map it back to the zero-section of the elliptic fibration. Therefore, there are no extra degrees of freedom associated to torsional sections. 

The presence of these torsional sections imposes restrictions on the coefficients in the Weierstrass model \cite{Aspinwall:1998xj}. For compact Calabi-Yau 3-folds\footnote{This is also true for higher-dimensional Calabi-Yau manifolds, but elliptically fibered K3's admit more freedom \cite{Hajouji:2019vxs}.} with no extra rational sections the possibilities for the Mordell-Weil group are quite constrained. The only possibilities are:
\begin{align}
\begin{split}
T&= \mathbb{Z}_N \, , \qquad\qquad\!\! N \in \{2,3,4,5,6 \}\,, \\
T&= \mathbb{Z}_2 \times  \mathbb{Z}_{2 N}\,, \quad  N \in \{1,2 \} \, ,\\
T&= \mathbb{Z}_3 \times  \mathbb{Z}_{3} \, .
\end{split}
\end{align}
The most general Weierstrass models for these torsions have been constructed in \cite{Aspinwall:1998xj} and proven to be exhaustive for compact 3-folds in \cite{Hajouji:2019vxs}. For convenience we reproduce the restricted Weierstrass coefficients for all possibilities in \eqref{eq:torsionfrak} in Appendix~\ref{sec:EnhancedWSFs}. In local models, also more exotic torsion models can be constructed, as illustrated in an example in Appendix~\ref{sec:ExoticTorsion}.

Mordell-Weil torsion can also be understood in terms of the allowed monodromies of the axio-dilaton field, i.e.\ the complex structure of the elliptic fiber. While this can be any $\text{SL}(2,\mathbb{Z})$ element in a generic F-theory model, it is restricted to a proper subgroup thereof (or an $\text{SL}(2,\mathbb{Z})$ orbit of the subgroup) in the presence of torsion. Since the monodromies directly correspond to stacks of 7-branes in the base, one can already anticipate that this will restrict the allowed symmetry groups as well as matter representations. Indeed, one finds that the simply-connected group $\mathcal{G}^*$ must have a center which is compatible\footnote{Note that the action of $T$ on the center of some of the factors in $\mathcal{G}^*$ might be trivial, see e.g.\ \cite{Baume:2017hxm}.} with the discrete Mordell-Weil torsion $T$. In addition, the matter representations have to respect the structure as well and thus have to be uncharged under $T$.

As a simple example consider the symmetry algebra $\mathfrak{su}_n$. If $T$ is trivial, all matter representation are allowed and there is no restriction. However, if the group is $\text{SU}(n) / \mathbb{Z}_k$ where $k$ divides $n$, representations that transform under the $\mathbb{Z}_k$ subgroup of the center are forbidden. This is not only relevant for the presence of dynamical charged states, but also on the level of line operators \cite{Aharony:2013hda, Gaiotto:2014kfa}. For product groups, the structure can be even more versatile. For example, if the global group structure is given by $[\text{SU}(n) \times \text{SU}(n)] / \mathbb{Z}_k$, fields in the bi-fundamental representation are allowed (among others), while fundamentals of one group which are singlets of the other are forbidden.

Beyond restricting the allowed symmetry groups in the six-dimensional theory, the restricted Weierstrass models in \eqref{eq:torsionfrak} also lead to singularities of the fiber that appear generically. For example in the $\mathbb{Z}_2$ case, $a_4$ appears quadratically in $\Delta$ and the zeros of this section will directly lead to $I_2$ fibers and hence to $\mathfrak{su}_2$ algebras. These generic singularities can also be seen by studying the modular curve of the restricted monodromy groups~\cite{Hajouji:2019vxs}.

\section{6d SCFTs}
\label{sec:SCFT}

In this section we review different ways of generating 6d SCFTs form string theory. We start with a discussion of the construction via F-theory, which we will then extend to include Mordell-Weil torsion in later sections. We further describe some of the elements in M-theory as well as heterotic string theories. While we do not claim to have a full understanding of the global characterization of the symmetry group in M-theory, we hint at some interesting possibilities for future investigations in that direction.

\subsection{6d SCFTs and F-Theory}
6d SCFTs are non-gravitational theories that are strongly coupled and contain tensionless strings. In the F-theory framework, decoupling gravity is associated to decompactifying the base manifold $B$, as already mentioned in Section \ref{sec:rev}. The anti-selfdual strings originate from D3 branes that wrap curves in $B$. Since the tension of the 6d strings is associated to the volume of the curves these have to collapse in the SCFT limit. For that to be possible at finite distance in moduli space they have to be contractible, imposing strong constraints on the overall geometry. 

Using this reasoning, it is possible to classify all configurations that lead to 6d SCFTs in terms of a collection of (possibly intersecting but still contractible) curves in a complex two-dimensional, non-compact K\"ahler manifolds, \cite{Heckman:2013pva}. For all these models the SCFT point can be reached by a continuous deformation of the curve volumes, which are given in terms of the VEVs of scalar fields in tensor multiplets of the 6d theory. This is where the name tensor branch of the theory comes from. On generic points of the tensor branch the base manifold $B$ is smooth. 

Relevant information about the base is contained in the intersection matrix of the compact curves $C_a$, given by
\begin{align}
\label{eq:ChargeMatrix}
\Omega_{a b} = C_a \cdot C_b \,,
\end{align}
which has to be negative definite in order for the model to lead to a 6d SCFT, see \cite{Mumford1961}. Simultaneously, $\Omega$ describes the string charge lattice of anti-selfdual strings in the 6d effective theory, which couple to the 2-form fields in the tensor multiplets. The allowed charge lattices for such configurations is quite restricted and plays a central role in the classification of 6d SCFTs \cite{Heckman:2015bfa}.

The intersection pattern~\eqref{eq:ChargeMatrix} can enforce certain singularities in the elliptic fiber. These singularities can be further enhanced by tuning complex structure parameters, cf.\ also~\cite{Heckman:2015bfa}. We focus on theories without frozen singularities, see~\cite{Tachikawa:2015wka,Bhardwaj:2018jgp} for a discussion of the latter. Of course many models of this type are connected by deformations with subsequent RG flow, see e.g.\ \cite{Heckman:2015ola, Heckman:2016ssk, Mekareeya:2016yal, Heckman:2018pqx}.

The essential building blocks of the F-theory construction are non-Higgsable clusters (NHCs) \cite{Morrison:2012np} and superconformal matter (SCM) theories \cite{DelZotto:2014hpa, Heckman:2014qba}. The former can be separated into single-curve NHCs and multi-curve NHCs. Both involve curves with self-intersection smaller than $(-2)$, as will be discussed in section~\ref{sec:NHC}. Superconformal matter theories can be understood as the collision of two irreducible components of the discriminant locus $\Delta$. Usually this would lead to hypermultiplets in certain matter representations\footnote{Hence the name superconformal matter as a non-perturbative generalization of bi-fundamental matter.} \cite{Katz:1996xe}, but in the case of SCM collisions, the vanishing order at the intersection point is
\begin{align}
(4,6,12) \leq \text{ord}(f,g,\Delta) < (8,12,24) \,.
\label{eq:vanord}
\end{align}
Therefore, a crepant resolution of these models require resolving not only in the fiber, but also blowing up the base manifold\footnote{Another alternative is a non-flat resolution of the fiber which encodes part of the tensor branch structure \cite{Buchmuller:2017wpe,Dierigl:2018nlv,Paul-KonstantinOehlmann:2019jgr} and can be used to explore 5D SCFTs via the M-theory duality \cite{Apruzzi:2018nre,Apruzzi:2019vpe,Apruzzi:2019opn,Apruzzi:2019enx}.}, which corresponds to a deformation that moves into the tensor branch of the theory as stated above. The upper limit in \eqref{eq:vanord} is imposed by the fact that codimension-two singularities with $\text{ord}(f,g,\Delta) \geq (8,12,24)$ do not allow for a crepant resolution: after blowing up such a codimension-two locus in the base, which reduces the vanishing order in codimension one by $(4,6,12)$, one still finds a codimension-one locus (i.e.\ the blow-up divisor) of vanishing order $(4,6,12)$ or worse. Often a single blow-up is not enough and one uncovers a full set of new tensor multiplets forming the SCM sector.

For illustrational purposes, we discuss a simple example of an SCM theory given by the collision of a $II^*$ fiber with $\mathfrak{e}_8$ algebra over the non-compact divisor defined by $\{ u = 0 \}$ and an $I_1$ fiber on $\{ v = 0 \}$, which intersect at the origin
\begin{align}
y^2 = x^3 + x u^4(v-3) +   u^5 (2 u-v) \, .
\end{align}
The collision leads to a non-minimal $(4,6,12)$ singularity that can be resolved with a single blow-up, which introduces a curve of self-intersection $(-1)$. More explicitly, the blow-up along $\{ e_1 = 0 \}$ reads
\begin{align}
\begin{split}
u& \rightarrow \tilde{u}\, e_1 \, , \quad v\rightarrow \tilde{v}\, e_1\,, \quad y \rightarrow \tilde{y}\, e_1^3 \, , \quad x \rightarrow \tilde{x}\, e_1^2 \,.
\end{split}
\end{align}
After taking the proper transform, the new Weierstrass equation is 
\begin{align}
\widetilde{y}^2 = \widetilde{x}^3 + \widetilde{x}\, \widetilde{u}^4( \widetilde{v}\, e_1-3) + \widetilde{u}^5 (2 \widetilde{u}-\widetilde{v}) \,.
\end{align}
The resulting geometry is smooth since now $\tilde{u} \tilde{v}$ is in the Stanley-Reisner ideal, indicating the fact that both coordinates cannot vanish at the same time. The resulting theory is the tensor branch of the SCM with just a single tensor. This is the minimal example of a 6d SCFT and it is usually referred to as the E-string theory due to its flavor group being $\text{E}_8$.

The full set of 6d SCFTs can be composed of several of the above components connected by $(-1)$ curves, leading to a vast set of 6d SCFTs. The goal of this paper is to use the geometric F-theory description of these 6d SCFTs and their fundamental building blocks and include torsional sections, which leads to a global specification of non-Abelian gauge and flavor symmetries on the tensor branch. The global constraints can already be anticipated by studying close relatives of the E-string theory above, namely discrete holonomy instanton theories (which also feature prominently in \cite{Aspinwall:1998xj}).

The E-string theory admits deformations which are described by finite Abelian subgroups $T$ of $\text{E}_8$. The $\text{E}_8$ flavor symmetry in turn is broken to the centralizer of $T$. The full list of such possibilities is given in Table \ref{tab:E8Broke}. The discrete holonomy instanton theories are understood as collisions of two or multiple flavor branes, constituents of the original brane configuration. These lead to flavor groups forming a subgroup of $\text{E}_8$ with a $(4,6,12)$ singularity at the collision point. For example, the $\mathbb{Z}_5$ discrete holonomy instanton theory corresponds to a breaking\footnote{To obtain the breaking pattern one can delete the root with Dynkin index $5$ from the affine Dynkin diagram of $\text{E}_8$. This extends to other breaking patterns with other discrete groups.} of $\text{E}_8$ to the subgroup $[\text{SU}(5) \times \text{SU}(5)]/\mathbb{Z}_5$. This exact setup will appear in the discussion of SCFTs respecting $\mathbb{Z}_5$ Mordell-Weil torsion.

Let us mention that flavor symmetries might get modified at the SCFT point. Moreover, occasionally only part of the full flavor symmetry expected from field theory arguments is realized geometrically, see e.g.\ \cite{Bertolini:2015bwa} for the rank one cases. Note that in rare cases the flavor symmetry can also be reduced, as was e.g.\ discussed in \cite{Ohmori:2015pia}. The same can happen in the case of 6d SCFTs with non-trivial Mordell-Weil torsion. In cases where the enhancement is realized geometrically, we can track the global group structure throughout the RG flow. Often these situations can be engineered as the collision of two stacks of flavor branes which can explicitly be performed within the Mordell-Weil torsion models we discuss. As a consequence, also the flavor-enhanced models are affected by the action of the torsion $T$. Sometimes, however, these enhancements lead to non-minimal singularities in codimension one that violate the Calabi-Yau condition and the corresponding flavor group enhancement is impossible in the presence of torsional sections.

In cases where the enhancements are not realizable by tuning the geometry, one needs to employ field theoretic arguments. One possibility is to investigate the allowed flavor symmetry fluxes on non-trivial spacetime manifolds, as done e.g.\ in \cite{Ohmori:2018ona}. Since a full understanding of the different possible flavor enhancements requires an inclusion of Abelian flavor symmetries, which we do not treat here, we will stick to a discussion of the geometric realization of the group structure and leave a discussion of the field theoretic arguments to future work. Nevertheless, we often find differences of the tensor branches in theories which only differ by their flavor groups (but their algebra), which confirms that global data is preserved under the RG flow.

\begin{table}
\centering
\renewcommand{\arraystretch}{1.3}
\begin{tabular}{|c|c|}\hline
$T$ & Flavor group \\ \hline \hline
$\mathbb{Z}_2$ & $\text{Spin}(16)/\mathbb{Z}_2$\,, $\quad[\text{E}_7 \times \text{SU}(2) ]/\mathbb{Z}_2$ \\ \hline 
$\mathbb{Z}_3$ & $\text{SU}(9)/\mathbb{Z}_3$\,, $\quad[\text{E}_6 \times \text{SU}(3)]/\mathbb{Z}_3$ \\  \hline
$\mathbb{Z}_4$ & $[\text{SU}(8)\times \text{SU}(2)] /\mathbb{Z}_4$\,, $\quad[\text{Spin}(10) \times \text{Spin}(6)] /\mathbb{Z}_4$ \\ \hline
$\mathbb{Z}_5$ & $[\text{SU}(5) \times \text{SU}(5)]/\mathbb{Z}_5$ \\ \hline
$\mathbb{Z}_6$ & $[\text{SU}(6) \times \text{SU}(3) \times \text{SU}(2)] /\mathbb{Z}_6$  \\ \hline \hline
$\mathbb{Z}_2 \times \mathbb{Z}_2 $ & $[\text{Spin}(12) \times \text{Spin}(4)]/[\mathbb{Z}_2 \times \mathbb{Z}_2]$\,, $\quad[\text{Spin}(8) \times \text{Spin}(8)]/[\mathbb{Z}_2 \times \mathbb{Z}_2]$\\ \hline 
$\mathbb{Z}_2 \times \mathbb{Z}_4 $ & $[\text{SU}(2)^2 \times \text{SU}(4)^2]/[\mathbb{Z}_2 \times \mathbb{Z}_4]$ \\ \hline  
$\mathbb{Z}_3 \times \mathbb{Z}_3 $ & $\text{SU}(3)^4 / [\mathbb{Z}_3 \times \mathbb{Z}_3]$ \\ \hline  
\end{tabular}
\caption{\label{tab:E8Broke}Flavor groups of $\text{E}_8$ discrete holonomy instanton. The surviving flavor group is given by the centralizer of the discrete subgroup in $\text{E}_8$ \cite{Aspinwall:1998xj}.}
\end{table}

Before we initiate the investigation of 6d SCFTs with Mordell-Weil torsion in Sections~\ref{sec:NHC},~\ref{sec:NSCFTs},~and \ref{sec:zoo}, we first want to mention alternative descriptions of 6d SCFTs in heterotic string theory as well as M-theory constructions.

\subsection{The Heterotic Perspective}
The heterotic duals of F-theories with discrete holonomy instantons can be understood in the following way. F-theory on a genus-one fibered 3-fold is dual to a heterotic theory on a genus-one fibered K3 surface. Let us describe the case where the base of $X$ is a $\mathbb{P}_F^1$ fibration over\footnote{The subscripts $F$ and $H$ are used to distinguish the two $\mathbb{P}^1$s, which we call~\cite{Braun:2018ovc} F-theory $\mathbb{P}_F^1$ and heterotic~$\mathbb{P}_H^1$.} $\mathbb{P}_H^1$, i.e.\ a Hirzebruch surface $\mathbb{F}_n$. This structure induces a K3 fibration of $X$ in addition to the genus-one fibration. The heterotic limit is recovered when the volume of $\mathbb{P}_F^1$ becomes small compared to that of $\mathbb{P}_H^1$. This leads to a stable degeneration of the 3-fold $X$, where each generic K3 fiber of $X$ degenerates into a union of two rational elliptic surfaces that intersect along an elliptic curve. The heterotic K3 is then the elliptic fibration over $\mathbb{P}_H^1$ whose fiber is isomorphic to the elliptic fiber at the intersection of the ruled surfaces. Thinking of $\mathbb{P}_F^1$ as a circle fibration over an interval, the $\mathbb{P}_F^1$ turns into the Ho\v{r}ava-Witten interval of the M-theory picture with the two $\text{E}_8$ walls sitting at the zeros of the two homogeneous coordinates of $\mathbb{P}_F^1$. 

It is now possible to take an instanton on one of the $\text{E}_8$ walls and pull it off into the bulk as an NS5-brane via a small instanton transition. In terms of the heterotic Bianchi identity for the Kalb-Ramond field $B$ with field strength $H$,
\begin{align}
\text{ch}_2(V) - \text{ch}_2(TX) = n\,,
\end{align}
where $\text{ch}_2(\cdot)$ denotes the second Chern character and $n$ denotes the number of NS5-branes, the process can be thought of as lowering $\text{ch}_2(V)$ by one and balancing it with an NS5-brane on the right hand side. We will restrict to unbroken supersymmetry, which means that we cannot use anti-NS5-branes, i.e.\ $n\geq0$

If the heterotic K3 is singular, one can also consider discrete holonomy instantons on the $\text{E}_8$ walls with fractional instanton numbers. These cases are discussed for K$3/\mathbbm{Z}_n$ and for $T^4/\mathbbm{Z}_n$ in~\cite{Aspinwall:1998xj} and~\cite{Ludeling:2014oba}, respectively. One should note here that the genus-one fiber of the Calabi-Yau 3-fold, together with the $\mathbb{P}_H^1$, correspond to the heterotic K3 compactification geometry.\footnote{Note that this is different from the type II limit, where the elliptic fiber is not part of geometry, but the whole base is.} Hence, the singularities lead to extra components in the discriminant of the the elliptic fibration of the ruled surfaces in the stable degeneration. This means that the discrete holonomy instantons are accompanied by extra (non-perturbative) gauge groups and matter representations. One can also coalesce point-like instantons with these fractional instantons, or coalesce several fractional instantons; if the charges of the coalesced instantons sum up to one, they can become a standard point-like instanton which can be moved away from the orbifold singularity. The occurrence of extra matter or tensor multiplets in such transitions is somewhat subtle, since extremal transitions can be blocked by a $B$ field (which corresponds to RR modes in the dual Type IIA)~\cite{Aspinwall:1998he}.

\subsection{The M-Theory Perspective} 
In the following, we describe the M-theory perspective of 6d SCFTs and their building blocks. The two main ingredients for their discussion are the M-theory realization of (higher-rank) E-string theories as well as the SCM theories.

The higher-rank E-string can be formulated in terms of M-theory on a half-space. The spacetime boundary is associated to an $\text{E}_8$ wall. Instantons in this $\text{E}_8$ background connection can be shrunk to a point, 
a small instanton, as in the heterotic description above. 
\begin{figure}
\centering
\includegraphics[width=0.8\textwidth]{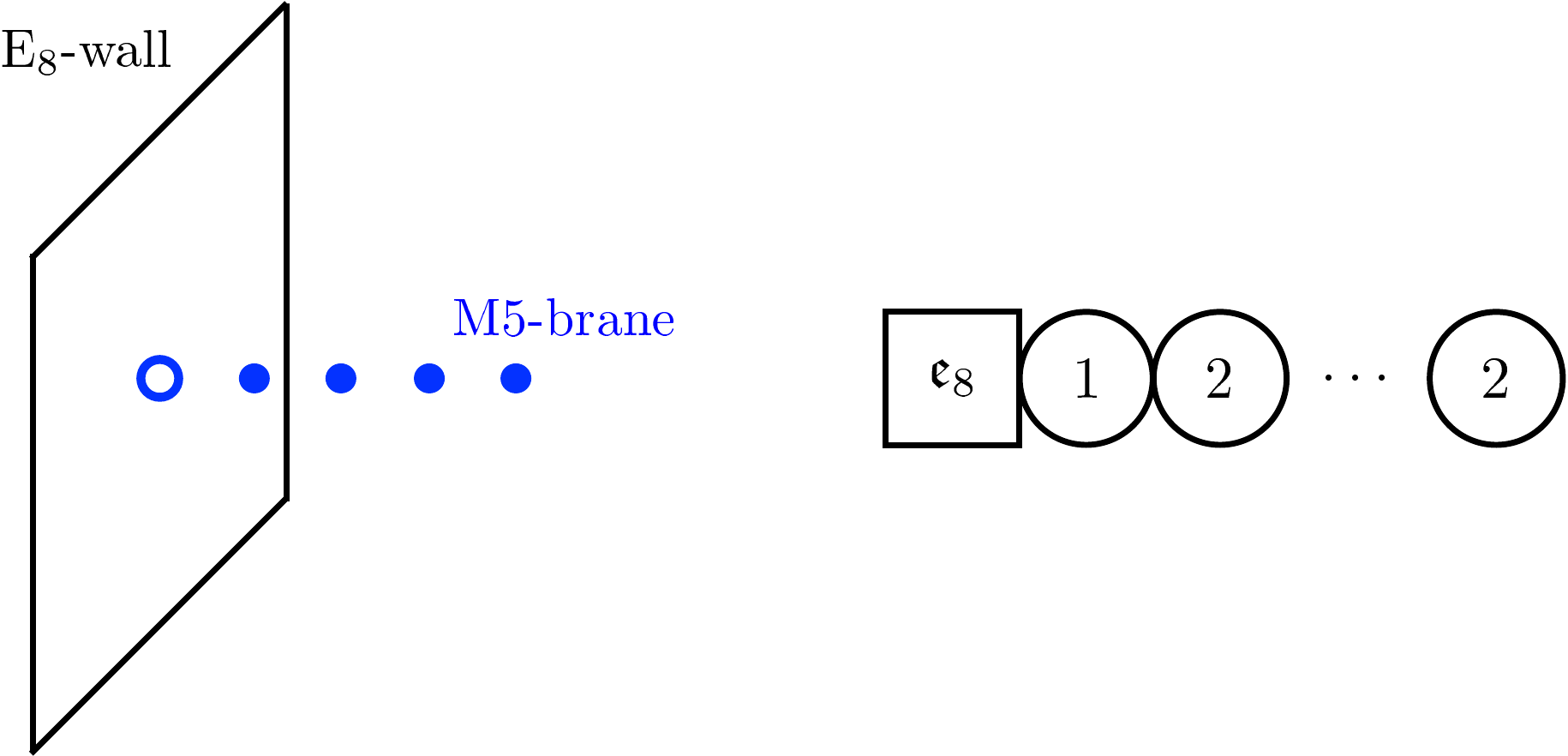}
\caption{M-theory setup for a higher-rank E-string theory and its F-theory geometry.}
\label{fig:highrkE}
\end{figure}
From this singular point in moduli space, which is associated to the origin of the Higgs branch, the small instanton can be dragged into the bulk as an M5-brane. The situation is depicted in Figure \ref{fig:highrkE}. The corresponding graph of the dual F-theory geometry\footnote{The squares indicate flavor symmetries and the the circles with number $m$ encode compact curves with self-intersection $(-m)$ and mutual intersection according to the picture. If the fiber becomes singular over the compact curves, we further give the corresponding gauge algebra above the circle.} is also given in Figure~\ref{fig:highrkE}. It has a curve with self-intersection $(-1)$ connected to a flavor algebra given by $\mathfrak{e}_8$ and a chain of curves with self-intersection $(-2)$.

\begin{figure}
\centering
\includegraphics[width=0.95\textwidth]{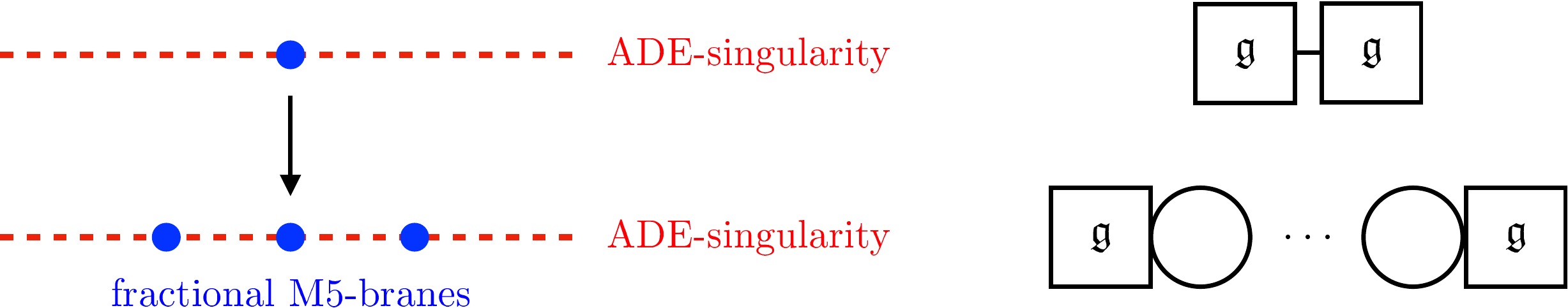}
\caption{M-theory setup for a superconformal matter.}
\label{fig:SCM}
\end{figure}

The second configuration we will need in the following is the $\mathfrak{g}$-type superconformal matter theory. In M-theory, it can be engineered by placing an M5-brane on an ADE-singularity of type $\mathfrak{g}$, see Figure \ref{fig:SCM}. 
Depending on the specific type of the ADE-singularity, the M5-brane can split up into fractions corresponding to a deformation onto the tensor branch of the theory. In the process, one finds compact curves of negative self-intersection and the corresponding gauge algebras in the F-theory picture. The flavor symmetry is described by a non-dynamical seven-dimensional super-Yang-Mills theory, which appears naturally in M-theory on an ADE-singularity.

\begin{figure}
\centering
\includegraphics[width=0.8\textwidth]{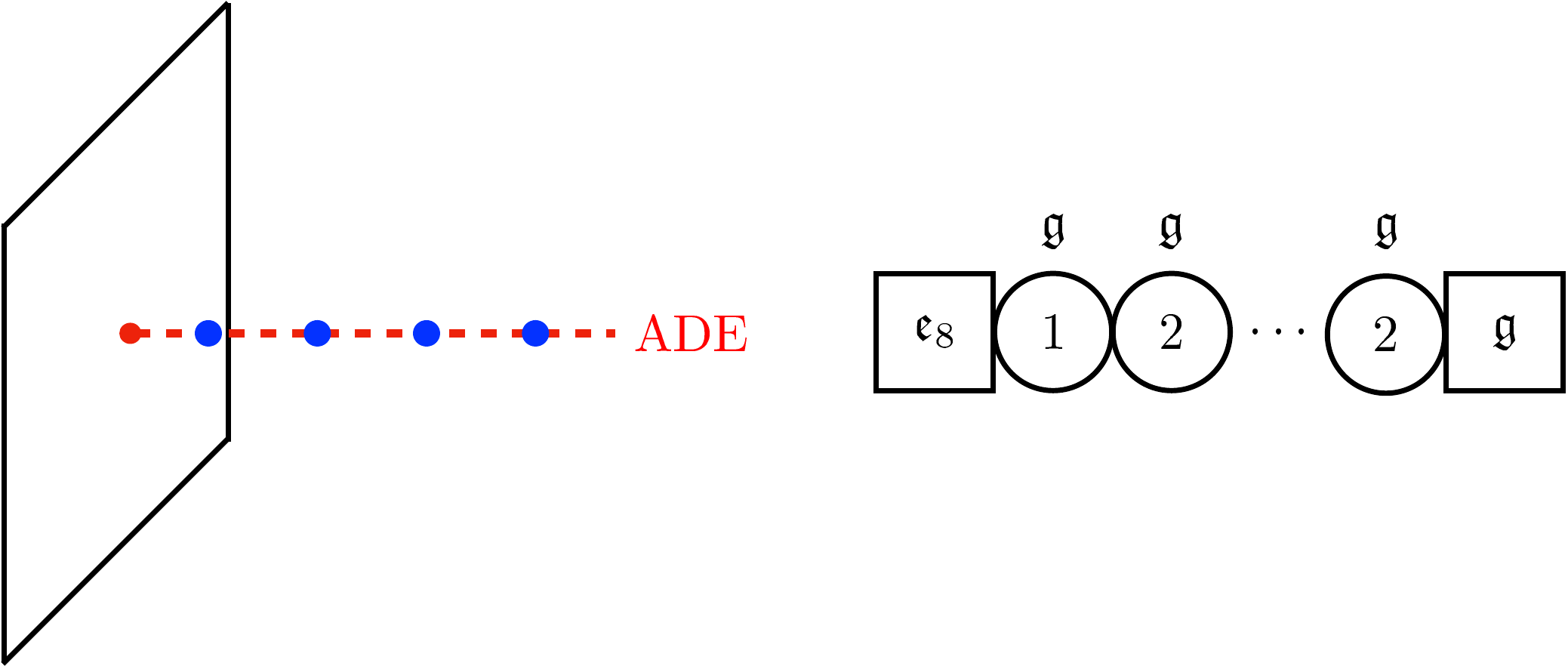}
\caption{Realization of the orbi-instanton theory in M-theory on the partial tensor branch.}
\label{fig:orbiinst}
\end{figure}

One can also combine the two ingredients, which leads to a setup that was called an \textit{orbi-instanton} theory in \cite{Heckman:2018pqx}. It is formulated as the intersection of an ADE-singularity in the bulk that intersects the $\text{E}_8$-wall bounding spacetime. Pulling M5-branes into the bulk along the singularity, one finds the theory depicted in Figure \ref{fig:orbiinst}.
In this way the gauge algebra $\mathfrak{g}$ is realized over each of the compact curves of negative self-intersection in the F-theory geometry and further appears as a flavor symmetry on the far right. In order to obtain a smooth geometry, one might have to further blow up the base manifold. As above, this corresponds to a fractionalization of the M5-branes on the ADE-singularity.

With these theories as building blocks, one can engineer six-dimensional SCFTs by switching on deformations that break the flavor symmetries to subgroups. These deformations can be associated to homomorphisms $\mathfrak{su}_2 \rightarrow \mathfrak{g}$, as well as group homomorphisms $\Gamma_{\text{ADE}} \rightarrow \text{E}_8$. The former objects can be interpreted as semi-simple and nilpotent deformations, and thus contains T-brane data~\cite{Heckman:2015ola, Mekareeya:2016yal, Heckman:2016ssk, Heckman:2018pqx}. The latter are associated to the possibility of having a non-trivial $\text{E}_8$ background connection on the boundary of spacetime, and are therefore directly related to discrete holonomy instantons. Of course, once a global group structure is imposed, these deformations of the theory are also subject to consistency requirements, which will be investigated elsewhere. Here, we just illustrate these restrictions in a simple example corresponding to M5-branes on an $\text{A}_5$ singularity, as e.g.\ in Figure~\ref{fig:SCM}.

In the case of $\mathfrak{su}_n$-type flavor symmetries, the nilpotent deformations are classified by a partition of $n$. For the example $n = 6$, there are 11 possible partitions given by
\begin{align}
\mathcal{P}(6) = \{ [6], [5,1], [4,2], [4,1^2], [3^2], [3,2,1], [3,1^3], [2^3], [2^2,1^2], [2,1^5], [1^6] \} \,.
\end{align}
Denoting a partition by $[ \mu_1^{d_1}, \dots, \mu_k^{d_k}]$, the preserved flavor symmetry is given by
\begin{align}
\mathfrak{g} = \mathfrak{s} \big( \bigoplus_j \mathfrak{u}_{d_j} \big) \,.
\end{align}
If we want to mod out a global $\mathbb{Z}_3$ quotient, we see that only those $d_j$ which are a multiple of three are allowed. This reduces the allowed partitions to
\begin{align}
\mathcal{P}_{\mathbb{Z}_3}(6) = \{ [2^3], [1^6] \} \,.
\end{align}
For the case $[ 1^6 ]$, the flavor group is not broken at all. For the other case, we can work out the effects on the gauge algebras close to the deformed flavor group. The corresponding part of the quiver is given by
\begin{align}
\begin{array}{c}
\includegraphics[scale=0.6]{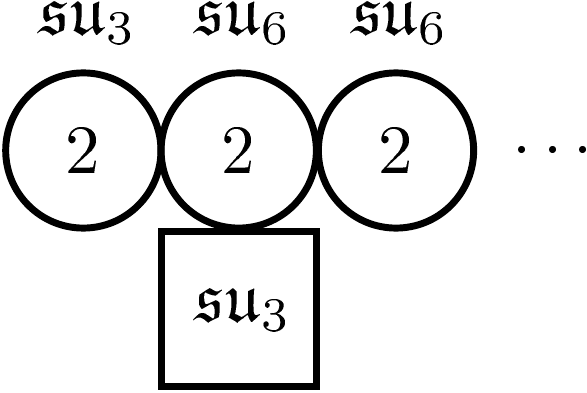}\end{array}
\end{align}
We see that this configuration is indeed consistent with a global $\mathbb{Z}_3$ quotient.

Moreover, we have seen above that the discrete holonomy is only possible if the $\text{E}_8$ wall contains an orbifold singularity, which is associated to the singular geometry in the heterotic dual. This singularity naturally extends into the bulk as an ADE-singularity and introduces further gauge degrees of freedom in the M-theory description. It is therefore suggestive to assume that the induced group structure also extends into the bulk and in F-theory corresponds to the existence of a torsional section. While the global realization of the gauge group is not well understood in general, there are some interesting advances in higher supersymmetric setups \cite{Garcia-Etxebarria:2019cnb}. Before we go to the F-theory analysis, we want to suggest how to access some of the global data in the M-theory formulation. The full analysis of these systems, however, is left for future work.

\begin{figure}
\centering
\includegraphics[width=0.25\textwidth]{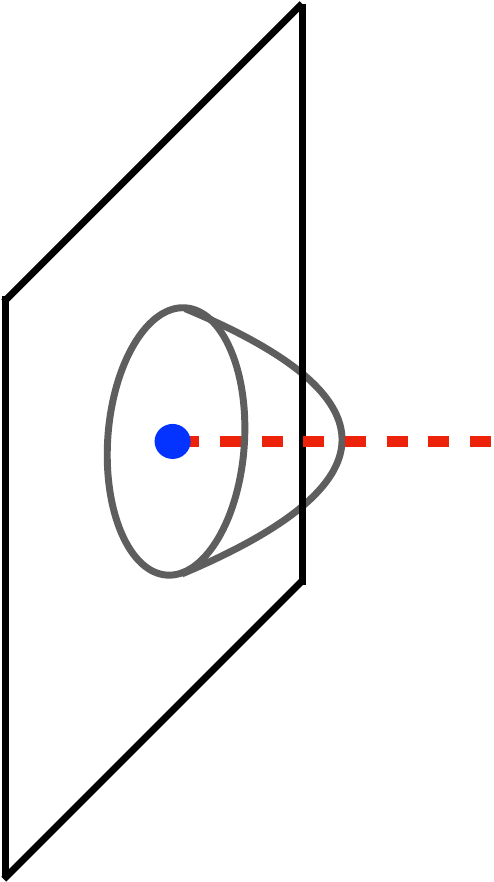}
\caption{4-chain on which the M-theory flux $G_4$ evaluates to the fractional instanton number, i.e.\ the number of M5-branes.}
\label{fig:4chaininst}
\end{figure}

Consider an $\text{E}_8$ wall with a $\mathbb{Z}_3$ singularity, which corresponds to the collision with an $\text{A}_2$ singularity from the bulk. On this background, one can switch on a discrete holonomy instanton that breaks the $\text{E}_8$ flavor symmetry to $[\text{E}_6 \times \text{SU}(3)] / \mathbb{Z}_3$. The corresponding discrete holonomy instanton carries a charge whose fractional part is $\tfrac{2}{3}$, see \cite{Aspinwall:1998xj}. Now consider the 4-chain $\Sigma$ depicted in Figure \ref{fig:4chaininst}, which surrounds the discrete holonomy instanton.
In this way one finds that
\begin{align}
\int_{\Sigma} G_4 = \oint_{S^3/\Gamma_{\text{ADE}}} C_3 \,.
\end{align}
Since we choose $\Sigma$ such that no additional M5-branes in the bulk contribute, one finds fractional values of the integral of the M-theory 3-form $C_3$ over the Lens space at infinity also in the bulk. Usually, one would say that $\text{A}$-type singularities cannot host a fractional holonomy in $C_3$, which is related to the allowed instanton numbers in the 7d super-Yang-Mills theory~\cite{deBoer:2001wca, Tachikawa:2015wka, Ohmori:2015pua}. However, if the gauge group on the $\text{A}_2$ singularity is $\text{SU}(3) / \mathbb{Z}_3$ rather than $\text{SU}(3)$, there are fractional instantons of instanton number $\tfrac{k}{3}$. These can be traced back to holonomies which commute in $\text{SU}(3) / \mathbb{Z}_3$ but not in $\text{SU}(3)$ and thus have a direct connection to the construction of triples for the known cases of fractional holonomies in $C_3$. Therefore, it seems that the fractional instantons of non-simply-connected gauge groups have a direct relation to the fractionalization of branes, their induced fluxes, and holonomies. They are also directly related to frozen singularities, cf.~\cite{Tachikawa:2015wka, Bhardwaj:2018jgp}.

\section{Non-Higgsable Clusters}
\label{sec:NHC}

In this section we begin with our investigation of 6d SCFTs with non-trivial Mordell-Weil torsion in the simple setups of non-Higgsable clusters. The properties of the geometry of NHCs enforces the presence of gauge factors on these compact curves with a minimal matter spectrum preventing further Higgs transitions. In the following, we present a brief recap of the derivation of NHCs and then turn to their realization with Mordell-Weil torsion.

\subsection{Non-Higgsable Clusters Without Torsion}
\label{subsec:NHCnotorsion}
\begin{figure}
\centering
\includegraphics[width=0.9\textwidth]{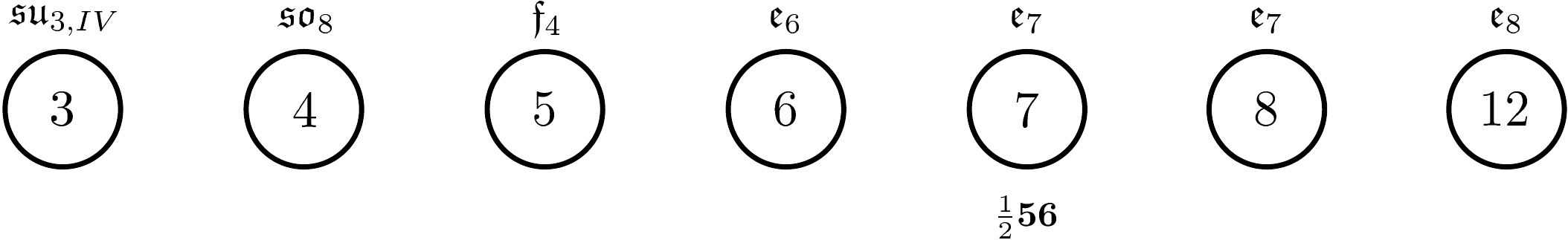}
\caption{Gauge algebra and matter content for single-curve NHCs.}
\label{fig:NHCs}
\end{figure}

Given an irreducible, effective curve $C$ with negative self-intersection $C \cdot C = -m < 0$ in a base $B$ of an elliptically-fibered Calabi-Yau 3-fold, any other effective curve $D$ on $B$ with $D \cdot C < 0$ is necessarily non-reduced and contains $C$ as an irreducible component,
\begin{align}
D = C + D' \,,
\end{align}
with $D'$ effective \cite{Morrison:2012np}. Knowing that the anti-canonical class of the base, $-K$, has to be effective and that for genus $g = 0$ curves one has
\begin{align}
- K \cdot C = 2 + C \cdot C \,,
\label{eq:genus}
\end{align}
we deduce that for $C \cdot C < -2$ the anti-canonical class needs to contain $C$. Since the discriminant locus of an elliptic fibration is a section of $\Delta = - 12 K$, it also has to contain $C$. This indicates that the fiber degenerates over $C$ and in general there are non-trivial gauge degrees of freedom localized on $C$. The same logic holds for $f = - 4 K$ and $g = - 6 K$ in the Weierstrass form of the elliptic fibration, leading to a minimal degeneration of the fiber over $C$ and an associated gauge algebra. In general, one can make the ansatz
\begin{align}
- n K = k C + D \,,
\end{align}
with $C \cdot D \geq 0$. Plugging this into \eqref{eq:genus} one has
\begin{align}
- n K \cdot C = 2 n - n m = - k m + D \cdot C \,,
\end{align}
and can solve for the smallest possible $k$, which in turn determines $D \cdot C$. Defining $C$ to be given by $\{ z = 0 \}$ in the base and parametrized by the coordinates $v_1$ and $v_2$ with
\begin{align}
v_1 \sim v_2 \,, \quad z \sim v_1^{-m} \,,
\end{align}
this can be rephrased as
\begin{align}
- n K \sim z^k P_r (v_1, v_2) \,.
\end{align}
Here, $P_r$ is a general degree $r = 2n - nm + k m$ polynomial in the coordinates $v_1$ and $v_2$. For example for $r=2$ this is $P_2 = P_{2,1} v_1^2 + P_{2,2} v_1 v_2 + P_{2,3} v_2^2$ with $P_{2,i} \in \mathbb{C}$. The minimal gauge algebra and matter content derived in this way is summarized in Figure~\ref{fig:NHCs}.

\begin{figure}
\centering
\includegraphics[width=0.75\textwidth]{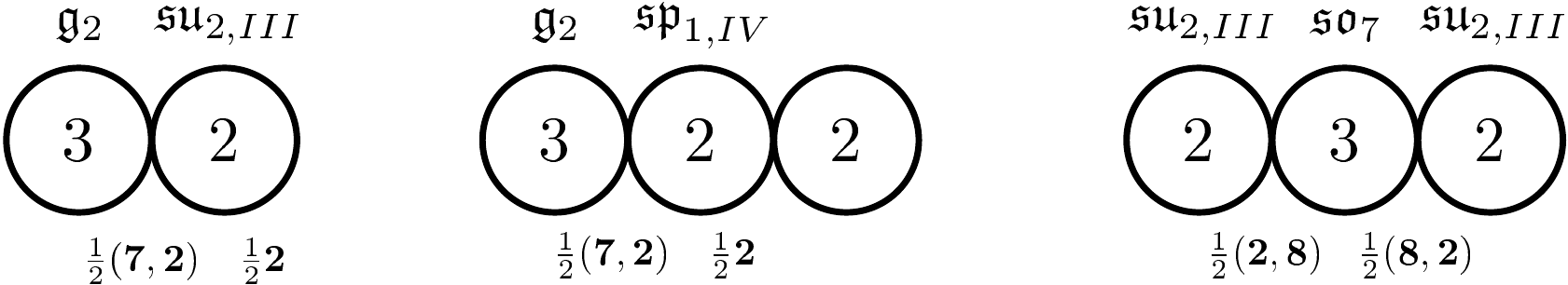}
\caption{Gauge algebra and matter content for multiple curve NHCs.}
\label{fig:NHC_multiple}
\end{figure}

Similarly, such non-Higgsable gauge theories can appear on multiple intersecting curves $C_i$ of negative self-intersection where neighboring curves intersect exactly once, as summarized in Figure~\ref{fig:NHC_multiple}. This can be deduced in a similar manner as above by finding the minimal number of times the curves with negative self-intersection are contained in the divisors associated to $f$, $g$, and $\Delta$. For the details of the derivation we refer to the original work in \cite{Morrison:2012np}. Our notation is such that the compact curves are given by $\{ u = 0 \}$, $\{ v = 0 \}$, and $\{ w = 0 \}$ from left to right. We further introduce the non-compact base divisors $\{ x_u = 0 \}$, $\{ x_v = 0 \}$, and $\{ x_w = 0 \}$ that intersect only the indicated compact curve exactly once. Sections of $-nK$ are then to leading order given by
\begin{align}
- n K \sim u^{k_1} v^{k_2} w^{k_3} x_u^{r_1} x_v^{r_2} x_w^{r_3} \,,
\end{align}
with higher order terms containing larger powers in $u$, $v$, and $w$.

We see that some of the NHCs above actually do have matter states, albeit in half-hypermultiplets. Since one cannot give a D-flat vacuum expectation value to a single half-hyper, the gauge theories can nevertheless not be Higgsed (geometrically) while preserving supersymmetry to a subgroup.

\subsection{Non-Higgsable Clusters With Torsion}
\label{subsec:NHCtorsion}

For models with non-trivial Mordell-Weil torsion, the Weierstrass coefficients have certain factorization properties given in \eqref{eq:torsionfrak}. The corresponding coefficients $a_n, b_n, c_n$ transform as section of $- n K$ and we can apply the same logic as above. Indeed one finds that the configurations of curves with non-trivial implications for the fiber degeneration are identical to the discussion above. Note that while the original NHCs do not possess any flavor symmetries, the presence of torsional sections often enforces flavor factors. We highlight them in the following discussions.

\subsection*{Single Curve NHCs With Torsion}
In the case of single curve NHCs we compute the leading vanishing order in $z$ for the coefficients in models with non-trivial Mordell-Weil torsion, cf.\ Table~\ref{tab:torsion_NHCs}. This can be used to compute the NHC gauge algebras and matter contents for models with Mordell-Weil-torsion.

\begin{table}
\centering
\renewcommand{\arraystretch}{1.2}
\begin{tabular}{| c || c | c | c | c | c | c | c |}
\hline
$m$  & $a_1$ & $a_2$ & $a_3$ & $a_4$ & $b_1$ & $b_2$ & $c_2$ \\ \hline \hline
$3$  & $z P_2$ & $z P_1$ & $z P_0$ & $z^2 P'_2$ & $z \tilde{P}_2$ & $z \tilde{P}_1$ & $z \tilde{P}'_1$ \\ \hline
$4$  & $z P_2$ & $z P_0$ & $z^2 P'_2$ & $z^2 P'_0$ & $z \tilde{P}_2$ & $z \tilde{P}_0$ & $z \tilde{P}'_0$ \\ \hline
$5$  & $z P_2$ & $z^2 P_4$ & $z^2 P_1$ & $z^3 P_3$ & $z \tilde{P}_2$ & $z^2 \tilde{P}_4$ & $z^2 \tilde{P}'_4$ \\ \hline
$6$  & $z P_2$ & $z^2 P_4$ & $z^2 P_0$ & $z^3 P'_2$ & $z \tilde{P}_2$ & $z^2 \tilde{P}_4$ & $z^2 \tilde{P}'_4$ \\ \hline
$7$  & $z P_2$ & $z^2 P_4$ & $z^3 P_6$ & $z^3 P_1$ & $z \tilde{P}_2$ & $z^2 \tilde{P}_4$ & $z^2 \tilde{P}'_4$ \\ \hline
$8$  & $z P_2$ & $z^2 P_4$ & $z^3 P_6$ & $z^3 P_0$ & $z \tilde{P}_2$ & $z^2 \tilde{P}_4$ & $z^2 \tilde{P}'_4$ \\ \hline
$12$ & $z P_2$ & $z^2 P_4$ & $z^3 P_6$ & $z^4 P_8$ & $z \tilde{P}_2$ & $z^2 \tilde{P}_4$ & $z^2 \tilde{P}'_4$ \\ \hline
\end{tabular}
\caption{Leading behavior of the sections $a_i, b_i, c_i$ in $z$ on a curve $C$ with $C\cdot C=-m$.}
\label{tab:torsion_NHCs}
\end{table}

Let us illustrate this procedure for the curve with self-intersection $(-3)$ and $\mathbb{Z}_2$ Mordell-Weil torsion. From Table~\ref{tab:torsion_NHCs} we read off the leading order behavior for $a_2$ and $a_4$
\begin{align}
a_2 = z P_1 \,, \quad a_4 = z^2 P_2 \,,
\end{align}
from which we find, cf.\ \eqref{eq:WSFTuning_Z2},
\begin{align}
f = z^2 \big( P_2 - \tfrac{1}{3} P_1^2 \big) \,, \quad g = \tfrac{1}{27} z^3 P_1 (2 P_1^2 - 9 P_2) \,, \quad \Delta = z^6 P_2^2 (4 P_2 - P_1^2) \,.
\end{align}
This indicates a fiber degeneration of type $I_0^*$. In order to determine the monodromy type we consider the monodromy cover, \cite{Grassi:2011hq},
\begin{align}
\begin{split}
\mu (\psi) & = \psi^3 + \Big( \frac{f}{z^2} \Big)\Big|_{z=0} \psi + \Big( \frac{g}{z^3} \Big)\Big|_{z=0} = \psi^3 + \big( P_2 - \tfrac{1}{3} P_1^2 \big) \psi + \tfrac{1}{27} P_1 \big( 2 P_1^2 - 9 P_2 \big)  \\
& = \tfrac{1}{27} \big( 3 \psi - P_1 \big) \big( 9 \psi^2 + 3 P_1 \psi + 9 P_2 - 2 P_1^2 \big) \,,
\end{split}
\end{align}
which fixes the fiber to be $I_0^{*, \text{ss}}$ with gauge algebra $\mathfrak{so}_7$. The discriminant of the corresponding monodromy cover is given by
\begin{align}
\Delta_0 = \frac{\Delta}{z^6} \Big|_{z = 0} = P_2^2 (4 P_2 - P_1^2) \,.
\end{align}
There are hypermultiplets in the spinor representation of $\mathfrak{so}_7$ located at the zeros of $P_2$ and one correspondingly finds two hypermultiplets $\mathbf{8}$. The remaining factor $4 P_2 - P_1^2$ defines a curve of the monodromy cover that can give rise to non-localized matter in the vector representation. The number of vector multiplets is given by the genus of the curve. However, since the genus in the present case is zero\footnote{This can be computed using $C\cdot (K + C) = 2 g - 2$ for the genus $g$ of a curve $C$.}, there is no non-local matter transforming as $\mathbf{7}$.

The given matter spectrum solves the anomaly condition for a $\mathfrak{so}_7$ theory on a $(-3)$ curve. The anomaly polynomial receives contributions from the Green-Schwarz term as well as from vector- and hypermultiplets. For $n_{\mathbf{8}}$ hypermultiplets in the spinor representation and $n_{\mathbf{7}}$ in the vector representation, one finds the anomaly polynomial 
\begin{align}
\mathcal{I}_8 = \mathcal{I}_8^{\text{GS}} + \mathcal{I}_8^{\text{v}} + \mathcal{I}_8^{\text{h}} = \tfrac{1}{32} \big( 1 - \tfrac{1}{2} n_{\mathbf{8}} \big) \big( \text{tr} F^2 \big)^2 - \tfrac{1}{24} \big( 1 + n_{\mathbf{7}} - \tfrac{1}{2} n_{\mathbf{8}} \big) \text{tr}F^4 \,,
\end{align}
which vanishes identically for $n_{\mathbf{8}}=2$ and $n_{\mathbf{7}}=0$. 

Taken by itself, this matter content would be incompatible with $\mathbb{Z}_2$ torsion, since the spinor representation is not invariant with respect to the $\mathbb{Z}_2$ center of a $\text{Spin}(7)$ gauge group. However, the geometry with $\mathbb{Z}_2$ Mordell-Weil torsion automatically contains two additional $I_2$ fibers over the two roots of $P_2$, leading to a $\mathfrak{su}_2 \oplus \mathfrak{su}_2$ flavor algebra. Therefore, we find that the $\mathfrak{so}_7$ matter is contained in two half-hypermultiplets transforming in the $(\mathbf{1}, \mathbf{8}, \mathbf{2})\oplus(\mathbf{2}, \mathbf{8}, \mathbf{1})$ representation (note that $\mathbf{2}$ is pseudo-real and $\mathbf{8}$ is real), which makes the matter content consistent with a symmetry group
\begin{align}
\mathcal{G} = \frac{\text{SU}(2) \times \text{Spin}(7) \times \text{SU}(2)}{\mathbb{Z}_2} \,.
\end{align}
Giving a vacuum expectation value to the matter states one breaks the gauge as well as the flavor symmetries. This in turn violates the restriction imposed by the $\mathbb{Z}_2$ Mordell-Weil torsion as well. An explicit geometric realization of this breaking in terms of a deformation is given by
\begin{align}
g \rightarrow g + \epsilon z^2 \,.
\end{align}
However, within  the class of models respecting the $\mathbb{Z}_2$ torsion, the theory has to preserve the flavor symmetries, and the model cannot be Higgsed.

\subsection*{Multi-Curve NHCs With Torsion}
Next, we consider base configurations with several intersecting curves of negative self-intersection. Again, the only relevant configurations are the ones already appearing for the generic NHCs, i.e.\, one configuration with two curves which intersect once and which have self-intersections $(-3,-2)$, as well as two chains of three curves of self-intersections $(-3, -2,-2)$ and $(-2,-3,-2)$.

In the case of two curves, which we describe by $\{u = 0\}$ and $\{v = 0\}$, one arrives at the form for the anti-canonical class, which necessarily contains the curves with negative self-intersections, given by
\begin{align}
- n K \sim u^{k_1} v^{k_2} x_u^{r_1} x_v^{r_2} \,,
\end{align}
from which we deduce the leading order behavior of the coefficients with Mordell-Weil torsion
\begin{align}
\begin{array}{| c | c | c | c | c | c | c |}
\hline
a_1 & a_2 & a_3 & a_4 & b_1 & b_2 & c_2 \\ \hline \hline
a_{1,0} \, u v x_u x_v & a_{2,0} \, u v x_v & a_{3,0} u^2 v x_u^2  & a_{4,0} u^2 v x_u & b_{1,0} \, u v x_u x_v & b_{2,0} \, u v x_v & a_{2,0} \, u v x_v \\ \hline
\end{array}
\label{tab:twonodeorders}
\end{align}
Again, $a_n,b_n,c_n$ are sections of $-nK$ and $a_{n,0},b_{n,0},c_{n,0}$ are complex constants. The same exercise can be repeated for three intersecting curves, leading to
\begin{align}
\begin{array}{| c | c | c | c | c |}
\hline
\text{configuration} & a_1 & a_2 & a_3 & a_4 \\ \hline \hline
(2,3,2) & a_{1,0} \, u v w x_u x_w & a_{2,0} \, u v^2 w x_v^2 & a_{3,0} \, u v^2 w x_v & a_{4,0} \, u v^2 w \\ \hline
(3,2,2) & a_{1,0} \, u v w x_u x_w & a_{2,0} \, u v w x_u^2 x_w & a_{3,0} \, u^2 v^2 w x_u x_v & a_{4,0} \, u^2 v^2 w x_v \\ \hline
\end{array}
\label{tab:threenodeorders}
\end{align}

In many situations the intersection of the compact curves with each other or with a non-compact divisor leads to non-minimal fiber singularities, i.e.\ $\text{ord}(f,g,\Delta) \geq (4,6,12)$. In these cases, one has to blow-up the intersection point, thus reducing the self-intersection of the central curve. In the description below we also include the gauge algebras on the blow-up divisors. This is very similar to what happens in the standard non-Higgsable clusters for $m \in \{ 9, 10, 11\}$. It can also happen that the singularity on the compact curve itself has $\text{ord}(f,g,\Delta) \geq (4,6,12)$ in which case there is no resolution and we denote the model as non-minimal.

\subsection*{$\boldsymbol{\mathbb{Z}_2}$ Torsion}
The single curve configurations with $\mathbb{Z}_2$ torsion are given by 
\begin{align}
\begin{array}{c}
\includegraphics[scale=0.6]{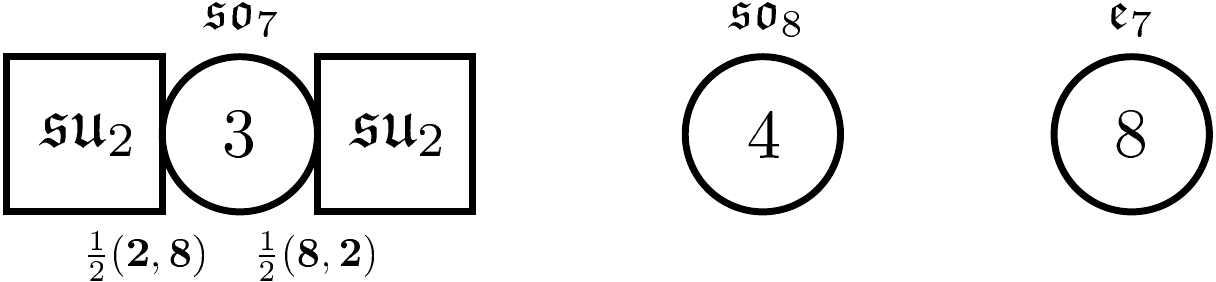}\end{array}
\label{eq:singleNHCZ2}
\end{align}
Curves with self-intersection smaller than $(-8)$ become too singular for a crepant resolution. For configurations with curves of self-intersection $(-5)$, $(-6)$, and $(-7)$ one encounters instead non-minimal singularities in codimension two which need to be blown up. The resulting configurations are given by
\begin{align}
\begin{array}{c}
\includegraphics[scale=0.6]{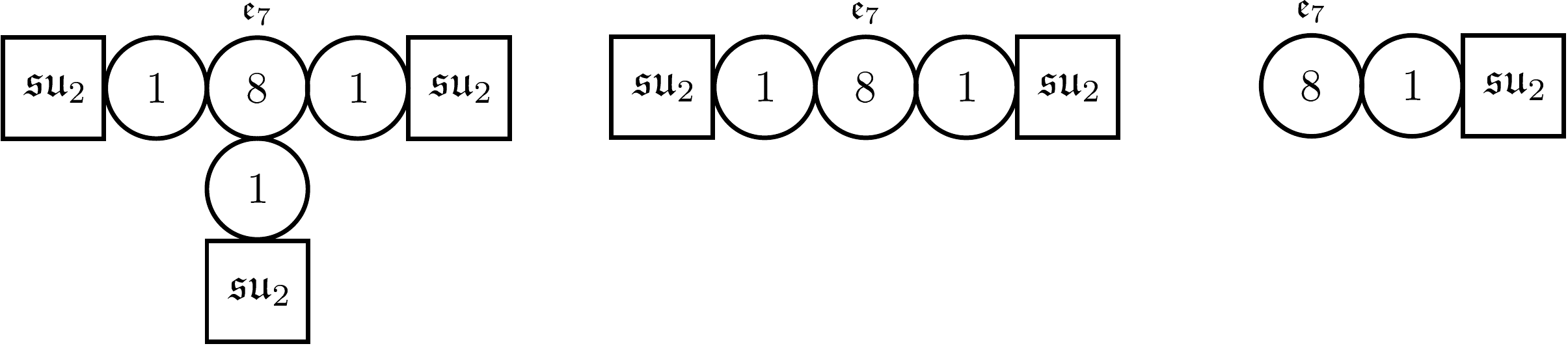}\end{array}
\end{align}
In all cases the final geometry contains a $(-8)$ curve with an $\mathfrak{e}_7$ algebra.

The configurations with multiple curves are
\begin{align}
\begin{array}{c}
\includegraphics[scale=0.6]{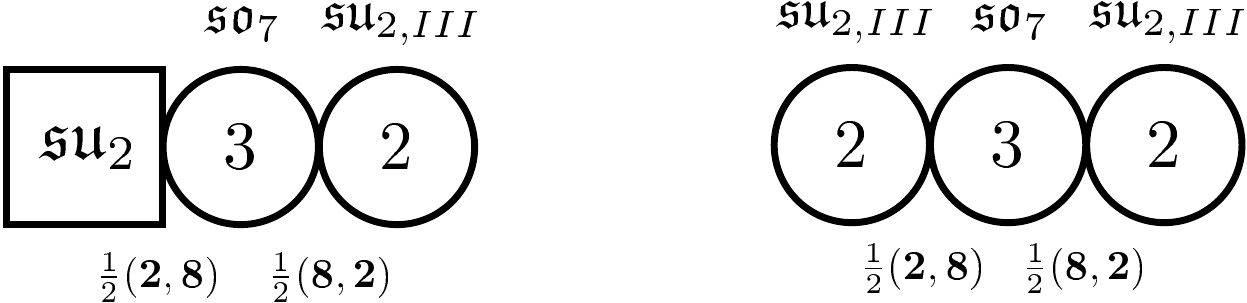}\end{array}
\end{align}
The remaining multi-curve NHC requires a blow-up an the resulting theory is given by
\begin{align}
\begin{array}{c}
\includegraphics[scale=0.6]{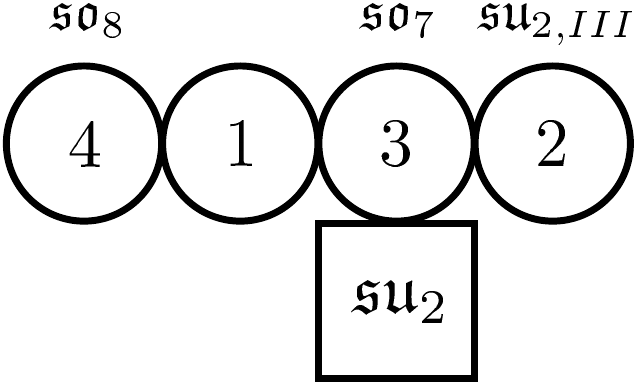}\end{array}
\label{eq:3nodeZ2}
\end{align}
The matter is identical to the two-curve cluster above and we see that essentially this results in the two-curve cluster connected to the NHC on a $(-4)$ curve by a $(-1)$ curve.

In all the cases above, the group structure is given by the simply-connected group induced from the algebras modded out by the discrete torsion group, i.e.\ for two-curve cluster it is given by
\begin{align}
\mathcal{G} = \frac{\widehat{\text{SU}}(2) \times \text{Spin}(7) \times \text{SU}(2)}{\mathbb{Z}_2} \,,
\end{align}
where we distinguish the flavor symmetry by a hat.

\subsection*{$\boldsymbol{\mathbb{Z}_3}$ Torsion}

With the orders for $a_1$ and $a_3$ given in Table~\ref{tab:torsion_NHCs}, we find for the single curve case\footnote{The starting configuration is a single curve with negative self-intersection. Several of these models, however, require further blow-ups leading to additional compact curves.}
\begin{align}
\begin{array}{c}
\includegraphics[scale=0.6]{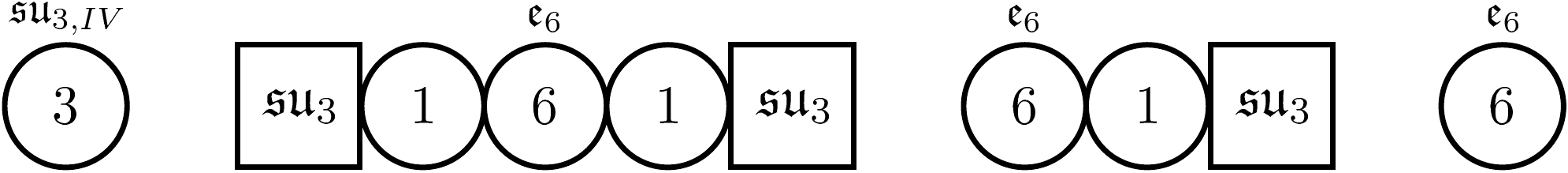}\end{array}
\end{align}
where we already performed the necessary blow-ups for $m = 4,5$. Beyond $m = 6$ one has non-minimal singularities already in codimension one.

All clusters with multiple curves require blow-ups and the final result is given by
\begin{align}
\begin{array}{c}
\includegraphics[scale=0.6]{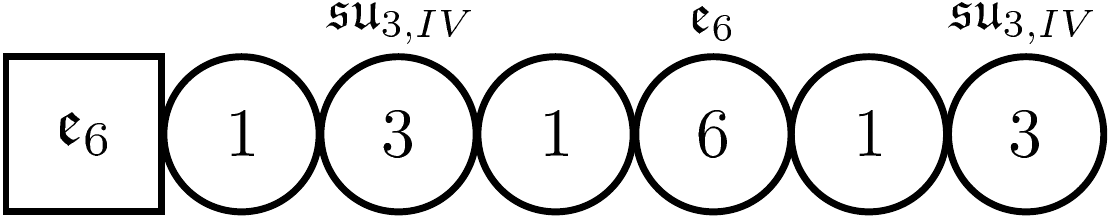}\end{array}
\end{align}
for two curves and
\begin{align}
\begin{array}{c}
\includegraphics[scale=0.6]{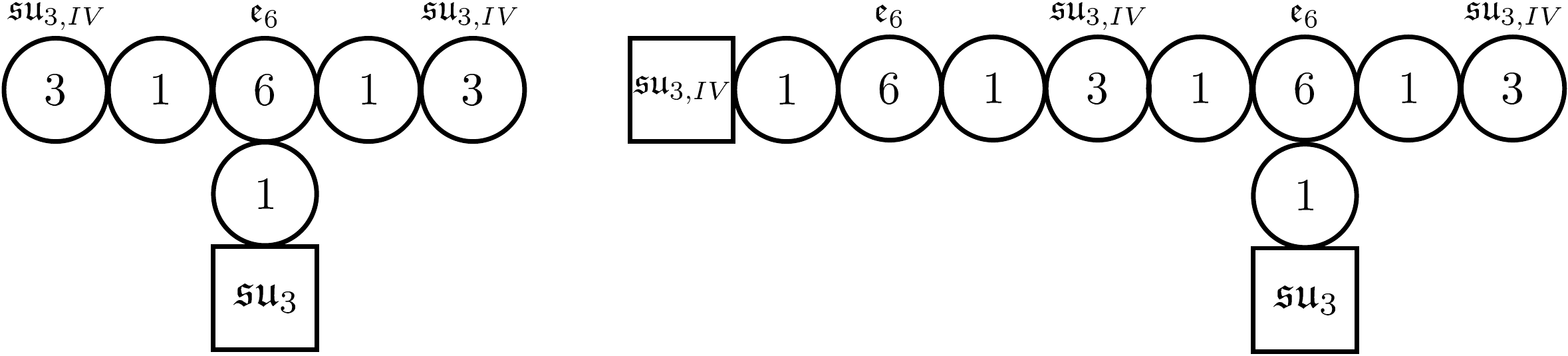}\end{array}
\end{align}
for three. All of these are variations of the $\mathfrak{e}_6$ superconformal matter as we will see later.

In all clusters, the group structure is uniquely defined. It is given by the product of all gauge and flavor group factors modded out by $\mathbb{Z}_3$.

\subsection*{$\boldsymbol{\mathbb{Z}_4}$ Torsion}
For $\mathbb{Z}_4$ torsion the single curve clusters are given by
\begin{align}
\begin{array}{c}
\includegraphics[scale=0.6]{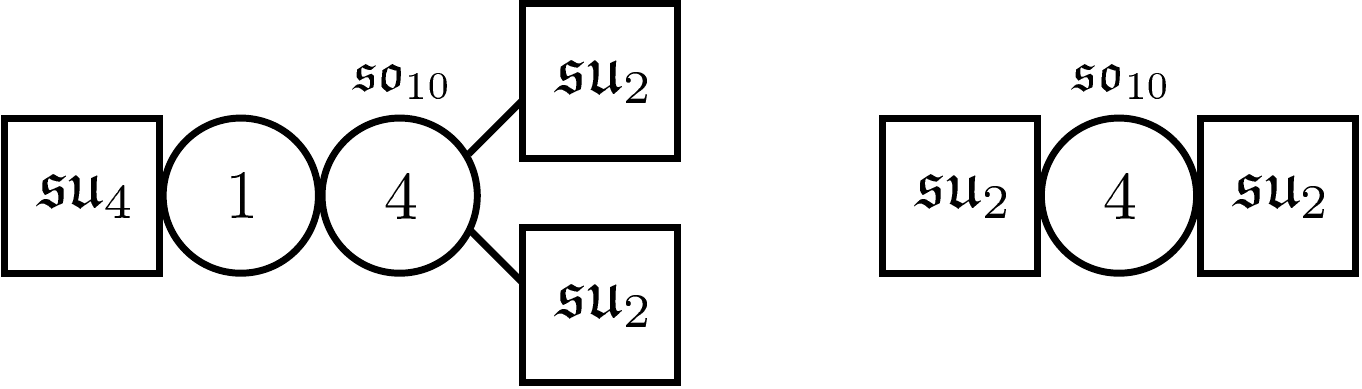}\end{array}
\end{align}
originating from $m = 3,4$. There are half-hypers in representation $(\mathbf{2}, \mathbf{10})$ between the $\mathfrak{so}_{10}$ and $\mathfrak{su}_2$'s. The group structure is given by the $\mathbb{Z}_4$ quotient of the simply-connected realization.

As for the example in section \ref{subsec:NHCtorsion}, these two theories can be Higgsed to the classical NHC on $(-4)$ curve with $\mathfrak{so}_8$ algebra, breaking the $\mathbb{Z}_4$ torsion structure alongside the flavor symmetries. Note that $\mathbb{Z}_4$ torsion does not allow for NHCs on multiple curves or configurations with $m>4$; these turn out to have no crepant resolution due to $(4,6,12)$ singularities in codimension one.

\subsection*{$\boldsymbol{\mathbb{Z}_2 \times \mathbb{Z}_2}$ Torsion}
As for $\mathbb{Z}_4$ torsion, there are only single curve clusters with $m = 3,4$, given by
\begin{align}
\label{eq:nhc3z2z2}
\begin{array}{c}
\includegraphics[scale=0.6]{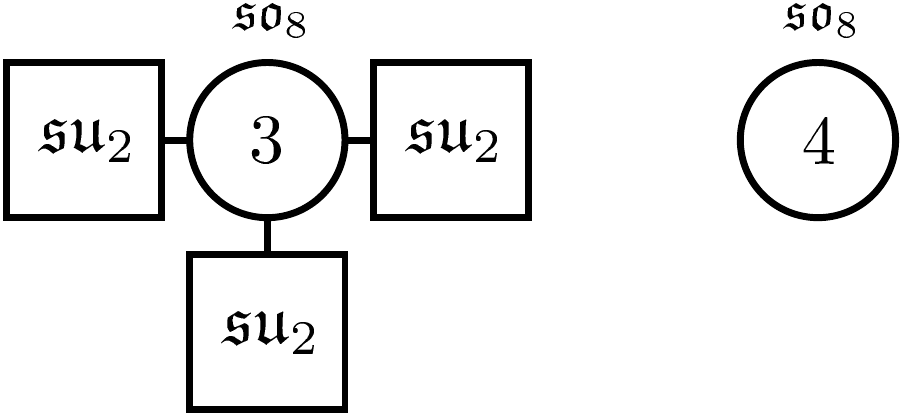}\end{array}
\end{align}
The $(-4)$ curve does not have any matter. The matter for the $(-3)$ curve is given by
\begin{align}
\tfrac{1}{2} (\mathbf{2}, \mathbf{1},\mathbf{1}, \mathbf{8}_{\text{v}}) \oplus \tfrac{1}{2} (\mathbf{1}, \mathbf{2}, \mathbf{1}, \mathbf{8}_{\text{s}}) \oplus \tfrac{1}{2} (\mathbf{1}, \mathbf{1}, \mathbf{2}, \mathbf{8}_{\text{co}}) \,,
\end{align}
where the subscripts (v, s, co) indicate the vector, spinor, and co-spinor representation of $\mathfrak{so}_8$
Interestingly, if one tries to tune two of the intersections with the $\mathfrak{su}_2$ flavor factors on top of each other, the third $\mathfrak{su}_2$ factor automatically intersects at the same point as well. Simultaneously, this intersection point enhances to $(f, g, \Delta) = (4, 6, 12)$ and one has to perform a blow-up, thus reducing the self-intersection of the curve to $(-4)$ and getting rid of all the matter.

In the case with multiple curves only the two curve case leads to a valid configuration
\begin{align}
\begin{array}{c}
\includegraphics[scale=0.6]{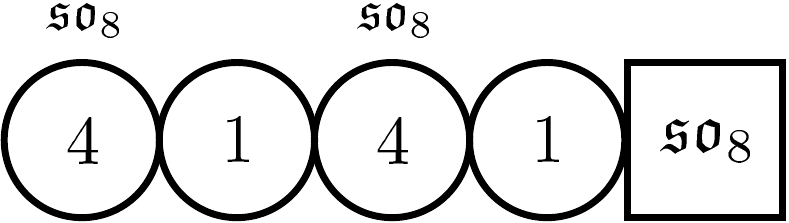}\end{array}
\end{align}
The configurations with three curves do not have a crepant resolution.

\subsection*{Remaining Cases}
The remaining possibilities for the Mordell-Weil torsion, i.e.\ $\mathbb{Z}_5$, $\mathbb{Z}_6$, $\mathbb{Z}_4 \times \mathbb{Z}_2$ and $\mathbb{Z}_3 \times \mathbb{Z}_3$, immediately lead to $(4,6,12)$ singularities in codimension one.

\section{Non-Simply-Connected SCM}
\label{sec:NSCFTs}

The next main ingredient in our discussion is going to be non-simply connected superconformal matter. On an intuitive level it is clear why superconformal matter must be revisited: Since a non-simply-connected (gauge-)group comes with a modified charge lattice that restricts certain matter representations, one expects that a similar logic should hold for their non-perturbative extension, the superconformal matter.  Indeed, in Section~\ref{sec:SCFT} we have already seen that the presence of torsion can lead to a breaking of the $\text{E}_8$ flavor symmetry of the E-string theory to a subgroup consistent with the global quotient. Before we discuss the various superconformal matter theories for each quotient factor in detail, we want to comment on the following three generic features that appear in these models:
\begin{itemize}
\item \textbf{Classical SCM}, as discussed e.g.\ in \cite{DelZotto:2014hpa}, often already comes with compatible centers, and therefore a non-simply-connected flavor group, as anticipated in \cite{Ohmori:2018ona}. In Section~\ref{ssec:Classic} we show that these models indeed admit torsional sections that are compatible with the respective centers.  
\item \textbf{Discrete jumps in the ramp of gauge groups} appear on the tensor branch. This feature results from the restricted monodromy, which forbids various gauge group and matter factors and hence heavily modifies the tensor branch, as we show in simple examples in Section~\ref{ssec:Flavor}. 
\item \textbf{Additional singular $\boldsymbol{{I}_1}$ loci} intersect one or multiple flavor branes in a single point. As we discuss in Section~\ref{ssec:DiscComp}, these curves are often singular themselves, which severely affects the tensor branch of these theories.
\end{itemize}

\subsection{Torsion and SCM} 
\label{ssec:Classic}
Some of the superconformal matter theories discussed in the literature (see e.g.\ \cite{DelZotto:2014hpa}) in fact admit a non-simply-connected flavor group. This has already been anticipated by the authors of \cite{Ohmori:2018ona}, who noted that in fact superconformal matter of type $(G,G)$ is changed to be of type $(G,G)/\text{center}(G)$. In this section, we present the geometric realization of this statement, which is due to the presence of an  $n$-torsional section in the respective Weierstrass model, with $\mathbb{Z}_n$ the center of the simply-connected cover $G^*$. We will mainly consider the $\mathfrak{e} \times \mathfrak{e}$ and $\mathfrak{so} \times \mathfrak{so}$ type superconformal matter.

We start with $\mathbb{Z}_2$. The $\mathfrak{e}_7 \times \mathfrak{e}_7$ SCM can be directly engineered in the $\mathbb{Z}_2$ torsion model of~\eqref{eq:WSFTuning_Z2} by setting $a_2 = u^2 v^2\,,~a_4 = u^3 v^3$. The tensor branch is given by
\begin{align}
\begin{array}{c}
\includegraphics[scale=0.7]{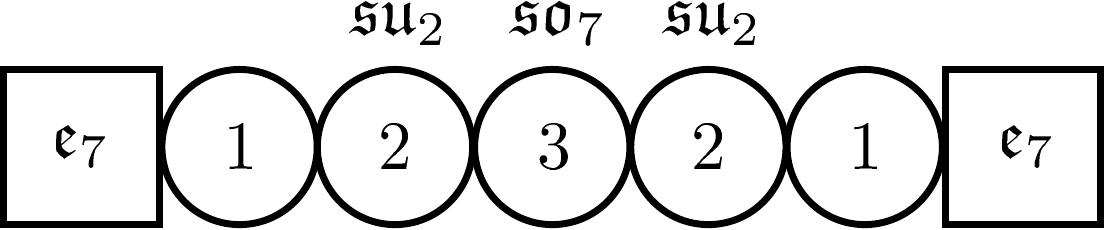} \end{array}
\end{align} 
where each individual gauge factor has a $\mathbb{Z}_2$ center, as enforced geometrically by the Weierstrass model. Note also that the $(2,3,2)$ NHC which appears here is unmodified when $\mathbb{Z}_2$ torsion is present, cf.~\eqref{eq:singleNHCZ2}.
 
Similarly, the $\mathfrak{e}_7 \times \mathfrak{su}_{2,III}$ collision\footnote{In order to distinguish the $\mathfrak{su}_2$ algebras arising from type $I_2$ and type $III$, singularities, we add a subscript for the latter. We proceed similarly for the $\mathfrak{su}_{3}$ of type $I_3$ vs type $IV$.} is obtained from $a_2 = u^2 v\,,~a_4 = u^3 v$ and does in fact correspond to one of the $\mathbb{Z}_2$ discrete holonomy instanton theories as discussed in Section~\ref{sec:SCFT}. Moreover, superconformal matter of type $\mathfrak{so}_{8+4n} \times \mathfrak{so}_{8+4m}$ can be engineered by factorizing $a_2 = u v\,,~a_4 = u^{2+n} v^{2+m}$. The tensor branches of those theories are
\begin{align}
\begin{array}{c}
\label{eq:sonsomZ2SCM}
\includegraphics[scale=0.7]{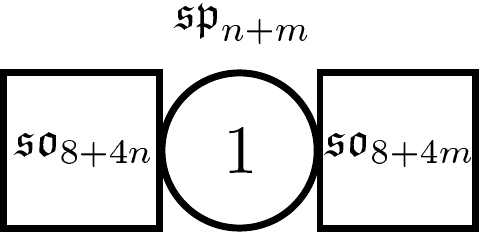} \end{array}
\end{align}
with matter transforming in the bi-fundamental representation of two adjacent algebras. In cases where either $n$ or $m$ are zero (but not both), the $\mathfrak{so}_8$ flavor factor is reduced to $\mathfrak{so}_7$.

We note that the torsion never allows collisions of type $\mathfrak{so}_{8+4n}\times \mathfrak{so}_{10+4m}$. This can be understood from the fact that anomaly cancelation on the tensor branch requires the presence of a single fundamental hypermultiplet of the $\mathfrak{sp}$ gauge group. However, this is incompatible with the $\mathbb{Z}_2$ torsion. Geometrically, this is ensured in the $\mathbb{Z}_2$ torsion model by an automatic enhancement of the flavor symmetry. A similar effect appears for $[\text{Spin}(10+4n) \times \text{Spin}(10+4m)]/\mathbb{Z}_4$ models.\footnote{Notably, the Sp group on the tensor branch does not have a $\mathbb{Z}_4$ center. As we shall see later, often only a subgroup of the full torsion group is modded out.}

We show in Section~\ref{sec:Z4TorsionSCM} that, even when two extended flavor factors have compatible centers, it is not guaranteed that a modding by that group exists: The modding can be obstructed by the presence of incompatible gauge group factors or representations on the tensor branch. An example for this is superconformal matter of type $\mathfrak{e}_7 \times \mathfrak{so}$ which includes a single $\mathfrak{sp}$ fundamental, that is not allowed by the $\mathbb{Z}_2$ embedding. In Section~\ref{ssec:Z2SCM}, we will see that $(\text{E}_7 \times \text{SO}(2n))/\mathbb{Z}_2 $ superconformal matter is possible, but it comes with a modified tensor branch that is consistent with the global $\mathbb{Z}_2$.

The $\mathbb{Z}_3$ case can be handled analogously. For example, $\mathfrak{e}_6 \times \mathfrak{e}_6$ type superconformal matter can be engineered from~\eqref{eq:WSFTuning_Z3} with $a_1=0$ (which sets $f\equiv0$), and $a_3=u^2 v^2$.  The tensor branch is given by
 \begin{align}
 \label{eq:e6e6Z3}
\begin{array}{c}
  \includegraphics[scale=0.7]{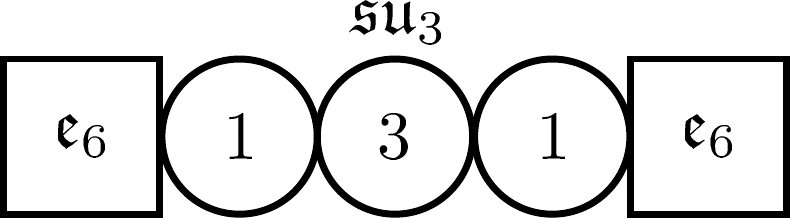} \end{array}
 \end{align}
without any matter, which is trivially consistent with the the global $\mathbb{Z}_3$. Also here we note that the non-Higgsable cluster of a $(-3)$ curve is unmodified by the $\mathbb{Z}_3$ torsion.

To summarize, we find that indeed some superconformal matter theories that have already been constructed in the literature admit a non-simply-connected flavor group, which is geometrically manifest by Mordell-Weil torsion in the Weierstrass model. In general, this is an important lesson that needs to be taken into account, when constructing superconformal matter within a certain torsion model. When starting with a given torsion, specific flavor configurations might further enhance the Mordell-Weil group beyond the starting configuration. Such an enhancement e.g.\ occurs when one chooses $a_2=u v$ and $a_4=u^2 v^2$ in an $\mathbb{Z}_2$ torsion model, which engineers an $\mathfrak{so}_8 \times \mathfrak{so}_8$ flavor algebra. After a coordinate shift, one easily sees that this model in fact admits a $\mathbb{Z}_2 \times \mathbb{Z}_2$ torsion, which is what one expects from the center of $\mathfrak{so}_8$. In Appendix~\ref{sec:EnhancedWSFs}, we show the conditions under which such a torsion enhancement is possible. Note that this can also be read in reverse, showing which deformations preserve possible torsion subgroups.
  
\subsection{Example: Flavor Groups vs Algebras}
\label{ssec:Flavor}

After having illustrated that some superconformal matter theories in fact already admit a torsion factor, we want to compare two superconformal matter theories that admit the same flavor algebra, but different flavor groups and hence different tensor branches. We pick the  $(\text{E}_7 \times \text{SU}(2n))/\mathbb{Z}_2$ theory which we compare with its simply-connected cover. This type of theory is engineered in the $\mathbb{Z}_2$ torsion model of~\eqref{eq:WSFTuning_Z2} by setting $a_2= u^2$ and $a_4 = u^3 v^n$, resulting in the Weierstrass functions
\begin{align}
f =   \tfrac13 (-u + 3 v^n) u^3 \,, \quad g=\tfrac{1}{27}  (  2 u-9 v^n)u^5 \,, \quad 
\Delta =    u^9 v^{ 2 n } (-u + 4 v^n) \, .
\label{eq:E7dicinst}
\end{align}
For $n>0$, the collision leads to a $(4,6,10+2n)$ non-minimal singularity. Note that there is also an $I_1$ component intersecting the origin as well. Performing the first resolution by blowing up $u=v=0$ with an exceptional divisor $e_1$ and taking the proper transform results in the Weierstrass functions
\begin{align} 
f =  \tfrac13 (-\tilde{u} + 3 \tilde{v}^n e_1^{(n-1)}) \tilde{u}^3 \,, \quad \tfrac{1}{27}  (  2 \tilde{u}-9 \tilde{v}^n e_1^{(n-1)})\tilde{u}^5 \, ,  \quad
\Delta =   \tilde{u}^9 \tilde{v}^{ 2 n } e_1^{ 2(n-1)} (-\tilde{u} + 4 \tilde{v}^n) \, .
\end{align}
Over $e_1$, we thus find an $\mathfrak{su}_{2n-2}$ gauge algebra factor, as needed from compatibility with the prescribed center. Continuing this process $n-1$ times results in the chain
\begin{align}
\label{eq:e7su2n_Z2}
\begin{array}{c}
\includegraphics[scale=0.7]{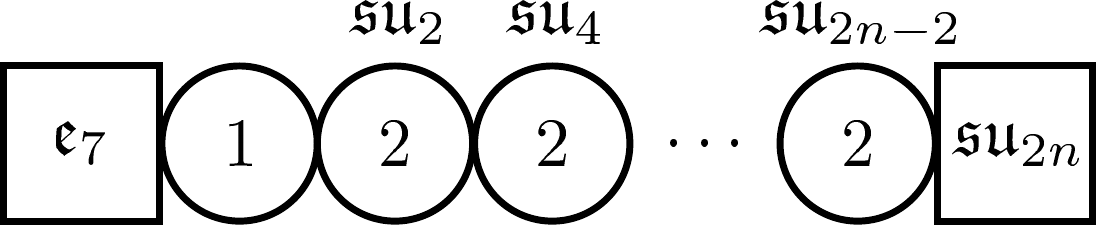} \end{array}
\end{align}
All gauge algebra  and the bi-fundamental $(\boldsymbol{2k},\overline{\boldsymbol{2k+2}})$ matter factors are compatible with the $\mathbb{Z}_2$ center that is modded out in addition to gauge anomaly cancellation. Note that the chain above ends on one of the $\mathbb{Z}_2$ discrete holonomy instanton theory.  

Let us compare this to the $\text{E}_7 \times \text{SU}(2n)$ theory constructed in a similar fashion, but in the most general Weierstrass model without any torsion. The resulting tensor branch is given by \cite{DelZotto:2014hpa} 
\begin{align}
\begin{array}{c}
\includegraphics[scale=0.7]{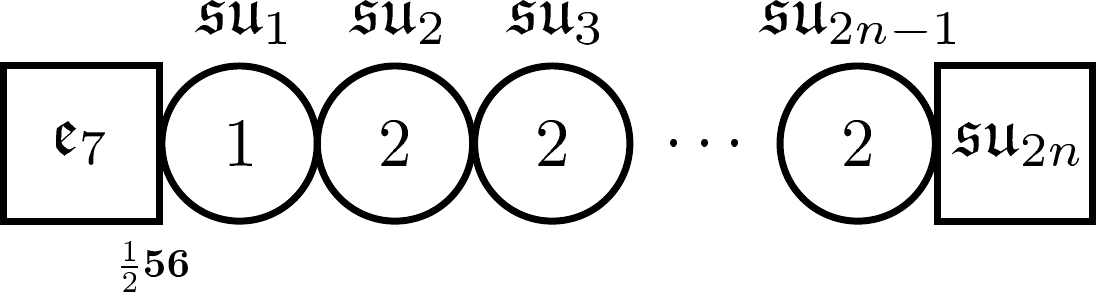} \end{array}
\end{align}
 The direct comparison shows how the non-simply-connected version of the $\mathfrak{e}_7 \times \mathfrak{su}_{2n}$ theory differs from its cover. We find that every algebra factor and representation that is incompatible with the $\mathbb{Z}_2$ modding, is indeed missing in~\eqref{eq:e7su2n_Z2}. This includes the $\mathfrak{su}_{2k+1}$ factors as well as the $\mathbf{56}$ half-hyper. This results in a ramp of $\mathfrak{su}_{2k}$ gauge factors that runs towards $\mathfrak{e}_7$ twice as fast as without the quotient, which requires less blow-ups and consequently tensor multiplets. We present an overview of the two theories with and without torsion in Table~\ref{tab:e7_su2n_comparison}. As already pointed out in \cite{Ohmori:2018ona}, a characteristic feature of these theories is a modified matter spectrum and a jump in the rank of the gauge groups in the ramp between the flavor factors.
 
\begin{table}
\renewcommand{\arraystretch}{1.4}
\centering
\small
\begin{tabular}{|c|c|c||c|}\cline{2-4}
\multicolumn{1}{c|}{} &    $\text{E}_7 \times \text{SU}(2n)$ 			&   $[\text{E}_7 \times \text{SU}(2n)]/\mathbb{Z}_2$&  Difference  \\ \hline
 Tensors 			&  $2n-1$ 																				& $n$ 												& $n-1$ \\ \hline
 rank($G$) 		& $(n-1)(2n-1)$																		& $(n-1)^2$ 									& $n(n-1)$  \\ \hline  
 \# Vectors 		& $\frac13 (n-1) ( 2 n-1) (4 n + 3)$ 										& $\frac13 (n-1) ( 4 n^2-2n-3)$ 	& $\frac43 n (n^2-1)$ \\ \hline 
 \# Hypers 		& $\frac23 n ( 4 n^2-1)+ \frac12 56$ 									& $\frac43 n ( n^2-1)$ 					& $\frac23 n(1+2 n^2)+\frac12 56$  \\ \hline 
 5d  dim($C$) 	& $n (2 n-1)$ 																			& $n+ ( n-1)^2$ 							&  $n^2 - 1$ \\ \hline
\end{tabular}
\caption{Comparison of the $\mathfrak{e}_7 \times \mathfrak{su}_{2n}$ SCM theories with and without torsion. $G$ refers to the gauge algebras. In the last column we give the dimension of the Coulomb branch dim$(C)$ of the 5D circle reduced theory.}
\label{tab:e7_su2n_comparison}
\end{table} 

In a very similar fashion, theories of type $(\text{E}_6 \times \text{SU}(3n))/\mathbb{Z}_3$ differ from their simply-connected version by omitting all $\mathfrak{su}_{3k \pm 1}$ algebra factors (and the corresponding tensors and hypermultiplets, see Section~\ref{sssec:Z3SCM}). 
    
\subsection{Singular Discriminant Components and Resolutions}
\label{ssec:DiscComp}
A distinctive feature of Weierstrass models with torsion points is that the Weierstrass functions $f$ and $g$ are highly tuned as compared to the standard Weierstrass form. Recall for example the $\mathbb{Z}_2$ torsion model and its Weierstrass functions
\begin{align}
f = a_4 - \tfrac{1}{3} a_2^2 \,, \quad g = \tfrac{1}{27} a_2 (2 a_2^2 - 9 a_4) \,, \quad \Delta = a_4^2 (4 a_4 - a_2^2) \,, 
\end{align}
Note that in addition to the $\mathfrak{su}_2$ locus, there is an $I_1$ component that includes the $a_4$  polynomial of the respective $\mathfrak{su}_2$. When tuning this model, we therefore often obtain an $I_1$ divisor with a double-, triple- or higher-point singularity. For example for the model above with $a_2 = u^n\,,~a_4 = v^m$, the $I_1$ component becomes singular, too. Moreover, care needs to be taken when $m$ is even, since then the $I_1$ divisor becomes reducible and can be split into two (possibly singular) divisors. Perturbative examples for this phenomenon are given e.g.\ by choosing $n=1$ and $m=3$, which results in the discriminant
\begin{align}
\Delta =  v^6 (-u^2 + 4 v^3) \,.
\end{align}
This corresponds to an $\mathfrak{sp}_3$ gauge algebra over $v=0$ and an $I_1$ locus with a double-point singularity at $u=v=0$. At this locus, the Weierstrass model enhances to an $(2,3,8)$ singularity with a local $\mathfrak{so}_{12}$ algebra contributing a hypermultiplet in the two-fold antisymmetric representation $\mathbf{14}$ of $\mathfrak{sp}_3$, compatible with the center factor. 

Similarly, for $n=2\,,~m=3$, the discriminant is given by
\begin{align}
\Delta= v^6 (-u^4 + 4 v^3) \,.
\end{align}
Now, the $\mathfrak{sp}_3$ fiber becomes split and supports an $\mathfrak{su}_6$ algebra, and the $I_1$ locus has a triple-point singularity at the origin. At the intersection with the $\mathfrak{su}_6$, this results in an enhancement to a $(3,5,9)$ singularity, i.e.\ that of an $\mathfrak{e}_7$ algebra, which gives again rise to the two-fold antisymmetric $\mathbf{15}$ of $\mathfrak{su}_6$.\footnote{This can be seen from the decomposition: $\mathfrak{e}_7 \rightarrow \mathfrak{su}_3 \times \mathfrak{su}_6$, where the adjoint is decomposed as $\mathbf{133}\rightarrow (\mathbf{8},\mathbf{1}) + (\mathbf{1},\mathbf{35}) + (\boldsymbol{\overline{3}},\boldsymbol{15}) + (\boldsymbol{3},\boldsymbol{\overline{15}})$.} 

In a similar vein, tuning the $I_1$ locus can lead to extra superconformal matter, e.g.\ when choosing $n=2$ and $m=4$, in which case the discriminant becomes
\begin{align}
 \Delta =v^8 (-u^2 + 2 v^2) (u^2 + 2 v^2)=v^8(u  - \sqrt{2} v) (u  + \sqrt{2} v)(u  - i\sqrt{2} v) (u  + i\sqrt{2} v)  \, .
\end{align}
This leads to an $\mathfrak{su}_8$ over $v=0$ and four $I_1$ loci all intersecting at the origin. In fact, all five curves meet at the origin, producing a $(4,6,12)$ singularity. This configuration is an E-string theory whose flavor group got broken to $\text{SU}(8)/\mathbb{Z}_2$. Upon blowing up, the singularity can be removed leading to a $(-1)$ curve intersected by the four $I_1$ curves at different points. The corresponding diagram is given by
\begin{align}
\label{eq:su8_I14_1}
\begin{array}{c}
\includegraphics[scale=0.5]{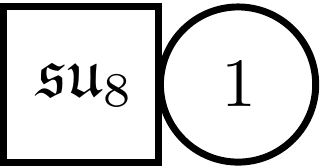} \end{array}
\end{align} 
Degenerating the $I_1$ locus further by setting $n=4$, the discriminant becomes
\begin{align}
\begin{split}
\Delta &= v^8 (-u^4 + 2 v^2) (u^4 + 2 v^2)\\ &= -v^8 (u^2  - \sqrt{2} v) (u^2  + \sqrt{2} v)(u^2  - i\sqrt{2} v) (u^2  + i\sqrt{2} v)\, .
\end{split}
\end{align}
Now, there is still the $\mathfrak{su}_8$ factor over $v=0$ but the $I_1$ curves have slightly changed. After two blow-ups in the base, this results in the chain 
\begin{align}
\label{eq:su8_I14_2}
\begin{array}{c}
  \includegraphics[scale=0.5]{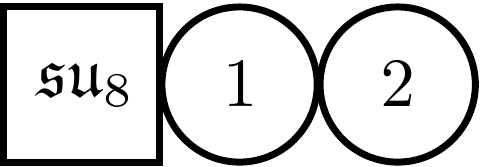} \end{array} 
\end{align} 
which is simply the same $\text{SU}(8)/\mathbb{Z}_2$ theory as before, but with additional unpaired tensors. Note that~\eqref{eq:su8_I14_1} can be reached from~\eqref{eq:su8_I14_2} along an RG flow that decompactifies the $(-2)$ curve.

Similarly, when taking the factorization $a_2 = u v^{l+1}\, ,~ a_4 = u ^4 v $,
\begin{align}
\begin{split}
f=&\frac13 u^2 v (3 u^2 - v^{(1 + 2 r)}) \, , \qquad g= \frac{1}{27}u^3 v^{(
 2 + r)} (-9 u^2 + 2 v^{(1 + 2 r)}) \\   \Delta=&  u^{10} v^3 (4 u^2 - v^{(1 + 2 r)}) \, .
 \end{split}
\end{align}
one obtains a $\mathfrak{su}_{2,III} \times \mathfrak{so}_{16}$ flavor group, where $l$ only affects the $I_1$ locus and its intersection at the origin. For $l=0$, a single resolution with a $(-1)$ curve suffices, resulting in
\begin{align}
\begin{array}{c}
\includegraphics[scale=0.5]{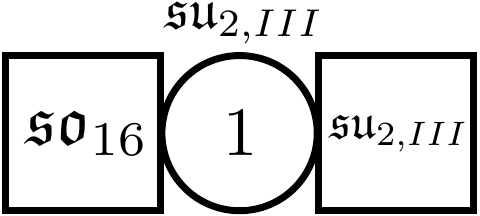} \end{array}  
\end{align}
However, for $l>0$ the $I_1$ locus develops a double-point singularity at the origin and one needs $l$ additional blow-ups that host $\mathfrak{su}_{2,III}$ factors to get a fully non-singular model,
\begin{align}
\begin{array}{c}
\includegraphics[scale=0.5]{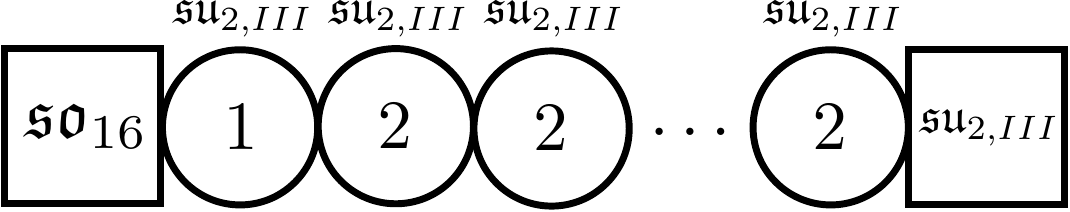} \end{array}  
\end{align}
Hence, we find a chain that repeats $l$ times an $\mathfrak{su}_2$ gauge factor from the right, similar to the orbi-instanton discussed in Section \ref{sec:SCFT}. From another perspective, the above model can be interpreted as an $\mathfrak{su}_2$ flavor brane that collides with a discrete holonomy instanton at a $\mathbb{Z}_{l-1}$ orbifold singularity. In the Weierstrass model, this is all encoded in the structure of the $I_1$ locus. Features like these require taking extra care when analyzing the theory, since the $I_1$ loci can influence the structure of the tensor branch substantially. 

In the process of studying non-minimal singular Weierstrass models, we use that traverse intersections of two divisors that lead to singularities of vanishing order less than $(8,12,24)$ can be reduced by a sequence of blow-ups in the base. Since some of the the models we analyze can have non-transverse intersections, we need to be careful. In particular, it can happen that a singularity of type $(8,12,24)$ (or worse) only becomes apparent after performing some blow-ups. We demonstrate this with the $\mathbb{Z}_2$ torsion model where we tune the coefficients $a_2 = u^3\,,~a_4 = u^3 v^6$. This leads to the Weierstrass model
\begin{align}
\begin{split}
f=\tfrac13 u^3 (u^3 - 3 v^6)\, , \quad   g=\tfrac{1}{27} u^6 (2 u^3 - 9 v^6)\,,\quad \Delta= -u^9  v^{12} (u^3 - 4 v^6)\,.
\end{split}
\end{align}
At first glance this model seems to admit a crepant resolution: It has a simple  $\mathfrak{e}_7 \times \mathfrak{sp}_6 $ collision, but also an $I_1$ curve with a triple-point singularity at the origin. The overall vanishing orders at the origin sum up to $(6,9,24)$ which is just below the bound of a non-crepant resolution. Upon performing the first blow-up, one obtains a model with
\begin{align}
\begin{split}
f=\tfrac13 e_1^2 u^3 (-u^3 + 3 e_1^3 v^6) \, , \quad  g= \tfrac{1}{27} e_1^3 u^6 (2 u^3 - 9 e_1^3 v^6)\,,\quad \Delta= e_1^{12} u^9 v^{12} (-u^3 + 4 e_1^3 v^6)\,.
\end{split}
\end{align}
This has an $I_{6}^{*,\text{s}}$ fiber along the blowup divisor $\{e_1=0\}$, which corresponds to an $\mathfrak{so}_{20}$ gauge algebra. However, now we see a $(8,12,24)$ singularity at the intersection of the $\mathfrak{so}_{20}$ divisor $\{ e_1 = 0 \}$ and the $\mathfrak{e}_7$ divisor $\{ u = 0 \}$, revealing that no crepant resolution is possible. Cases like these are relatively generic and complicate a systematic study of the non-simply-connected superconformal matter theories.

\section{Non-Simply-Connected Conformal Matter Zoo}
\label{sec:zoo}

In this section we discuss superconformal matter theories with various torsion factors, as well as their tensor branches, in more detail. The theories are constructed by systematically engineering singularities in the restricted Weierstrass models compatible with the Mordell-Weil torsion factors. In order to obtain superconformal matter theories, the singularities are engineered such that the vanishing orders are below $(4,6,12)$ in codimension one, and between $(4,6,12)$ and $(8,12,24)$ in codimension two. These conditions mean that a crepant resolution by blowing up the base is possible. This allows us to determine all gauge group factors and matter representations of each theory. In order to present our results in a systematic way, we discuss each torsion theory using the following approach:
\begin{enumerate}
\item For each torsion factor, consider all E$_8$ breaking patterns of a discrete holonomy instanton,  which can be inferred from Table~\ref{tab:E8Broke}. Subsequently, consider all further breakings that can be engineered in a torsion-preserving way (see e.g.\ \cite{Aspinwall:1998xj}).
\item From this, we obtain non-compact flavor branes at the coordinates of the $\mathbb{C}^2$ base ($u=0$ and $v=0$), which host the E-string theory at the origin.
\item For each flavor group over a codimension one locus, we engineer higher and higher singularities, until we hit the limit of crepantly resolvable theories. 
\item We perform all resolutions in the base, which allows us to deduce the 6d tensor branch. We then determine the gauge groups and matter content, check anomaly cancellation, as well as compatibility with torsion.
\end{enumerate}
Note that some theories cannot be reached via this procedure of starting from a discrete holonomy instanton and engineering further singularities. We call these ``outlier theories''. We present a few interesting outliers in the following sections, but the bulk of them is discussed in Appendix~\ref{app:outlier}.

Let us briefly recap our notational conventions: 
\begin{itemize}
\item Square nodes denote flavor algebras and circle nodes denote gauge algebras.
\item We specify these algebras inside the squares and above the circles. We follow the convention $\mathfrak{su}_2\simeq\mathfrak{sp}_1$, $\mathfrak{su}_1=\mathfrak{su}_0=\emptyset$. Likewise, nodes without a gauge algebra above them signal trivial algebras. 
\item The number in the circle denotes the negative self-intersection of the corresponding divisor.
\item Unless specified otherwise, all matter transforms in the bi-fundamental representation of two adjacent nodes. For groups other than $\mathfrak{su}_{n}$, the ``fundamental'' is the lowest-dimensional non-trivial irreducible representation.\footnote{Lines connecting the nodes are only used in order to unclutter the notation and do not indicate any special kind of matter.}
\end{itemize}

\subsection[Single Factors Theories: \texorpdfstring{$\mathbb{Z}_{n}\,,  n =2 \ldots 6$}{Zn, n=2...6}]{Single Factors Theories: \texorpdfstring{$\boldsymbol{\mathbb{Z}_{n}\,,  n =2 \ldots 6}$}{Zn, n=2...6}.} 
We start by considering theories with Mordell-Weil torsion given by a single $\mathbb{Z}_n$ factor and employ the strategy outlined above. It turns out that $\mathbb{Z}_2$ torsion is the most versatile in terms of different models, since a wide range of compatible gauge group factors exist. Typically, one obtains fewer possibilities when the order of the torsion is enhanced. However for non-prime torsion, there is the possibility that the gauge and flavor groups are only affected by a subgroup of the full torsion, which in turn increases the number of possibilities again. 

\subsubsection{$\boldsymbol{\mathbb{Z}_2}$ Torsion: Conformal Matter}
\label{ssec:Z2SCM}
Superconformal matter with $\mathbb{Z}_2$ Mordell-Weil torsion is described by the Weierstrass model in~\eqref{eq:WSFTuning_Z2}. Note that the model generically has an $\mathfrak{su}_2$ singularity at the zeros of $a_4$. The possible breakings of E-string theories by a $\mathbb{Z}_2$ discrete holonomy instanton are\footnote{As mentioned in Section~\ref{sec:rev}, the group structure is sometimes not uniquely specified by the indicated quotients. In these cases, the group action of $T$ can be deduced from the matter states that are present on the tensor branch of the theories.}
\begin{align}
\label{eq:z2Instantons}
\renewcommand{\arraystretch}{1.3}
\begin{array}{|c|c|c|c|} \cline{2-4}
\multicolumn{1}{c|}{}&\text{ Flavor Group} & a_2 & a_4 \\ \hline
1& [\text{E}_7 \times \text{SU}(2)] /\mathbb{Z}_2  &  u^2 &   u^3 v \\[0pt]
2& [\text{Spin}(8) \times \text{Spin}(8)] / [\mathbb{Z}_2 \times \mathbb{Z}_2]  & u v  & u^2 v^2 \\[0pt]
3&  [\text{Spin}(16)] /\mathbb{Z}_2 & u v & u^4 \\ \hline 
4& \text{SU}(8)/\mathbb{Z}_2 & u^2 & v^4 \\[0pt]     
5& [\text{SU}(6) \times \text{SU}(2)_{III}] / \mathbb{Z}_2 & u^2 & u v^3 \\[0pt] 
6& [\text{Spin}(8) \times \text{SU}(4)]/\mathbb{Z}_2  & u^2 & u^2 v^2 \\[0pt]
7& [\text{Spin}(12) \times \text{SU}(2)_{III}] /\mathbb{Z}_2 & u v & u v^3 \\ \hline \hline
8& \text{Sp}(4)/\mathbb{Z}_2  & u^3 & v^4 \\[0pt] 
9& [\text{Sp}(3) \times \text{SU}(2)_{III}]/\mathbb{Z}_2  & u^3 & u v^3 \\[0pt]     
10& [\text{Spin}(8) \times \text{Sp}(2)]/\mathbb{Z}_2  & u^3 & u^2 v^2 \\ \hline
\end{array}
\end{align} 
For the first three theories, the E$_8$ flavor group is broken to one of its maximal subgroups, while theories 4-7 have rank lower than 8. The last three theories have an additional monodromy that folds the flavor group to a non-simply laced group. 

In the following, we will discuss the first seven of these theories and defer the last three to the Appendix. Starting with theory 1, we can enhance the $\mathfrak{su}_2$ flavor symmetry over $v=0$ to an $\mathfrak{su}_{2n}$ factor simply by replacing $u^3v\rightarrow u^3v^n$ in $a_4$. This example has already been discussed in Section~\ref{ssec:Flavor}, and its tensor branch is summarized in \eqref{eq:e7su2n_Z2}. The same is true for theory~2, for which we can further enhance the ranks of the $\mathfrak{so}$ algebras by setting $a_4=u^{2+n} v^{2+m}$, leading to well known theories~\cite{DelZotto:2014hpa} whose tensor branches have already been discussed in \eqref{eq:sonsomZ2SCM}.

We continue our discussion with theory 3, which is an $\mathfrak{so}_{16}$ theory intersected by an $I_1$ component. Enhancing its flavor algebra leads to theories with tensor branch
\begin{align}
\label{eq:Z2Theory4}
\begin{array}{c}
\includegraphics[scale=0.7]{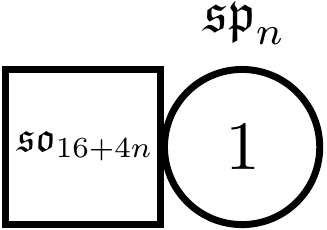} \end{array} 
\end{align}
with-half hypermultiplets in the bi-fundamental representation. This cancels the $\mathfrak{sp}_n$ gauge anomalies.

Moving on to theory 4, we have the E-string theory from an $\mathfrak{su}_8 \times I_1^4$ collision. We can enhance the $\mathfrak{su}_8$ factor to $\mathfrak{su}_{8+2n}$ by changing $a_4$ in the Weierstrass form to $a_4 = u^{4+n}$. Note that for these choices the $I_1$ locus is irreducible when $n$ is odd, factors into two components if $n$ is divisible by 2, and into 4 components if $n$ is divisible by 4. The form of tensor branch depends on $2n$ mod $8$, which specifies the rank of the gauge algebra on the final $(-1)$ curve,
\begin{align}
\label{eq:SU8nZ2}
\begin{array}{c}
\includegraphics[scale=0.7]{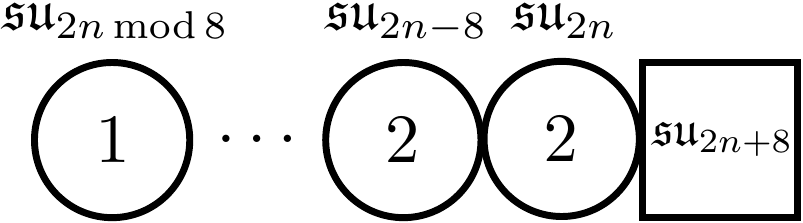} \end{array} 
\end{align} 
Note that the $(-1)$ curve does not carry any gauge algebra if $n$ is a multiple of 4. All matter multiplets are bi-fundamental. However, in the case that $n=3$ mod $4$, the chain ends on an $\mathfrak{su}_6$ over the $(-1)$ curve and the $I_1$ curves do not split. In such a case, one finds another antisymmetric $\mathbf{15}$-plet representation at the collision with the $I_1$ curve. The $\mathbf{15}$-plet can be seen to arise from an enhancement to an $(3,5,9)$ singularity in codimension two, encoding an $\mathfrak{e}_7$ fiber (note that an $\mathfrak{e}_7$ enhancement is consistent with the restricted $\mathbb{Z}_2$ monodromy). The matter representation can then be inferred from decomposing the adjoint $\mathbf{133}$ \cite{Katz:1996xe} into irreducible $\mathfrak{su}_6$ representations, which gives rise to a $\mathbf{15}$-plet. In the case $n=2$ mod $4$, on the other hand, the chain ends on an $\mathfrak{su}_4$ over the $(-1)$ curve which intersects two $I_1$ components. The $\mathfrak{su}_4 $ comes with fundamentals, but also with two $\mathbf{6}$-plet half-hypermultiplets (this is the two-fold antisymmetric representation), originating from the intersections with the two $I_1$ components. This is again confirmed by noting that the $(-1)$ curve intersects the $I_1$ curve in an enhanced $(2,3,6)$  singularity, which yields a codimension two enhancement to $\mathfrak{so}_{8}$. This local enhancement is compatible with the desired center in its simply-connected realization as one expects from the $\mathbb{Z}_2$ torsion. By using the Katz-Vafa rule and branching the adjoint $\mathbf{28}$ of $\mathfrak{so}_8$ into $\mathfrak{su}_4$ representations, one finds the required $\mathbf{6}$-plets as necessary for anomaly cancellation.

Next we consider theory 5, which is an $\mathfrak{su}_6 \times \mathfrak{su}_{2,III}$ broken E-string theory, and enhance the $\mathfrak{su}_6$ factor to $\mathfrak{su}_{2n+6}$ by setting $a_4=u v^{3+n}$. This results in the chain
\begin{align}
\begin{array}{c}
\includegraphics[scale=0.7]{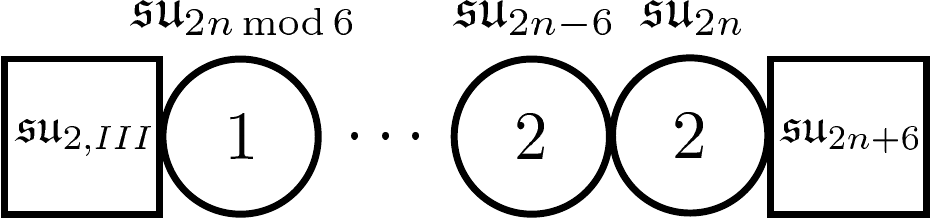} \end{array} 
\end{align}
Note that the rank of the $\mathfrak{su}$ factors above jumps by $6$ in the quiver.  Again, if the chain ends on an $\mathfrak{su}_4$ on the last $(-1)$ curve, there are additional states to the usual bi-fundamentals. Indeed, we find a half-hypermultiplet in the $(\mathbf{6,2})$ representation, but also a full hypermultiplet in the $(\mathbf{4,2})$ representation at the intersection with the $\mathfrak{su}_2$ flavor brane. This can be explicitly verified from its intersection with the $\mathfrak{su}_2$, which enhances to an $\mathfrak{e}_7$ singularity in codimension two, from which we get the two-fold antisymmetric representation required for anomaly cancellation.

We can also enhance the $\mathfrak{su}_2$ to an $\mathfrak{so}_8$, and additionally keep enhancing the $\mathfrak{su}_{2n+6}$ factors further, which gives the chain 
\begin{align}
\label{eq:susoz2}
\begin{array}{c}
\includegraphics[scale=0.7]{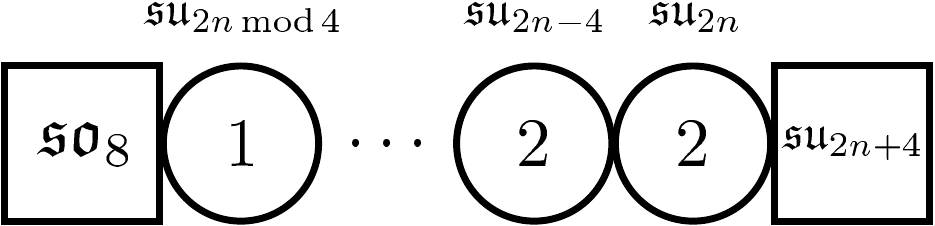} \end{array}
\end{align}
Depending on $n$, the above theory can either end on the broken E-string theory 6 (for $n$ even), or on a theory with a gauged $\mathfrak{su}_2$ on the $(-1)$ curve (for $n$ odd). In the latter case, there are $(\mathbf{2 ,6})$ bi-fundamentals hypermultiplets and $(\mathbf{8,2})$ half-hypermultiplets under the flavor $\mathfrak{so}_8$, as required by anomalies. As noted, the above chain also includes the enhanced version of the $[\text{SO}(8) \times \text{SU}(4)]/\mathbb{Z}_2$ of theory 6.

We continue with theory 7, where we enhance the $\mathfrak{so}_{12}$ factor by setting $a_4=u v^{3 + n}$. This results in the tensor branch
\begin{align}
\begin{array}{c}
\includegraphics[scale=0.7]{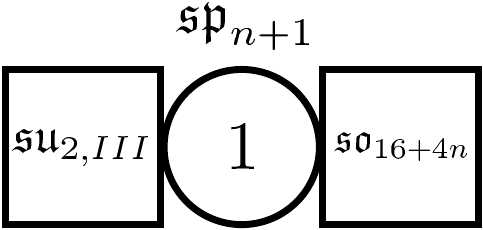} \end{array} 
\end{align} 
Enhancing the other flavor group, i.e.\ the $\mathfrak{su}_2$ (which is actually a type $III$ singularity) to an $\mathfrak{so}_{8+4n}$, basically results in a theory we have already discussed before in \eqref{eq:sonsomZ2SCM}. Remember, however, that it is often impossible to enhance multiple flavor factors to arbitrarily high vanishing order, as this typically leads to non-minimal singularities. A couple of theories where multiple flavor factors can be enhanced simultaneously (but not necessarily both indefinitely), are presented in Appendix~\ref{app:outlier}.

Another class of models can be obtained by taking theory 1 and enhancing the $\mathfrak{su}_2$ factor to an $\mathfrak{so}_{8+4n}$ group. Concretely this is achieved by setting $a_2 = u^2 v  \, ,~a_4 = u^3  v^{4+n}$. This branch is relatively close to the standard conformal matter, but there is also an $I_1$ component intersecting the origin as well, changing the tensor branch in a subtle way,
\begin{align}
\begin{array}{c}
\includegraphics[scale=0.7]{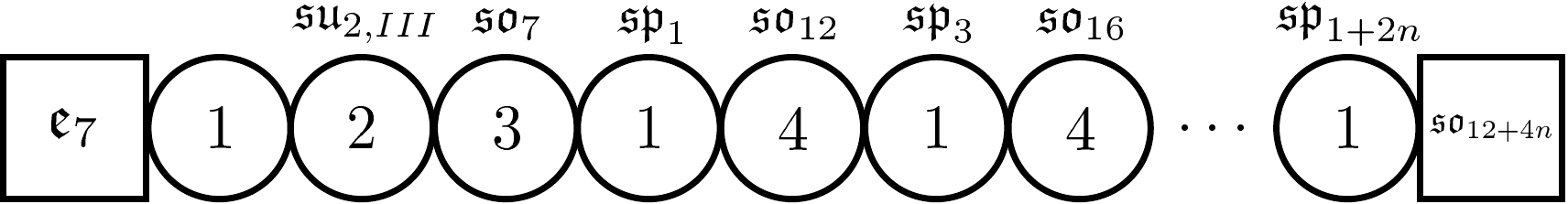} \end{array} 
\end{align}
Here, the $\mathfrak{su}_{2}$ over the $(-2)$ curve is of type $III$. Note that for $n=0$ the chain terminates at the $\mathfrak{so}_{12}$ factor and for higher $n$ multiple chains of $(-1)\,(-4)$ curves are appended, as shown above. It is important to point out that the above type of theories differ from the standard superconformal matter (see e.g.~\cite{DelZotto:2014hpa}), which has $\mathfrak{so}_{2n+1}$ gauge algebras over the $(-4)$ curves. The $\mathbb{Z}_2$ factor furthermore changes the spectrum with respect to the non-torsion case, such that with torsion only bi-fundamental matter is present. Note that the last ``bi-fundamental'' between the $\mathfrak{so}_7$ and the $\mathfrak{su}_2$  factor is in the eight-dimensional spinor representation of $\mathfrak{so}_7$ as for the $\mathbb{Z}_2$ NHCs above.  

Before we continue our discussion with other torsion groups, we want to emphasize that the tensor branch strongly depends on the form of the additional $I_1$ component. To illustrate this, consider theory 4 and set $a_2=v^4$. This does not change the flavor algebra at $v=0$. Taking $n=3$ or $n=4$, corresponding to $a_4 = u^6$ or $a_4= u^7$. From the perspective of the flavor brane, these simply look like an $\mathfrak{su}_{12}$ or $\mathfrak{su}_{14}$ flavor symmetry intersecting the $I_1$ component and from that perspective one might have expected theories like those presented in \eqref{eq:SU8nZ2}. However, with this modification, the $I_1$ locus develops a triple-point singularity for $n=3$ and a septuple-point singularity for $n=4$ at the origin. As a consequence, the tensor branches are actually quite different from~\eqref{eq:Z2Theory4},
\begin{align}
&\begin{array}{c}
\includegraphics[scale=0.7]{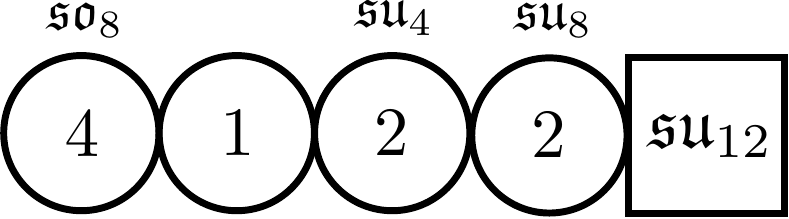} \end{array}\\   
&\begin{array}{c}
\includegraphics[scale=0.7]{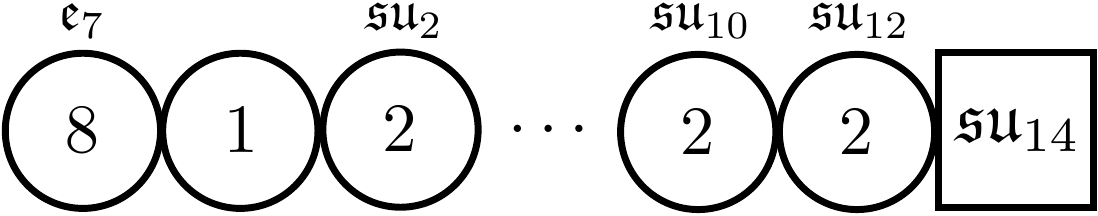} \end{array}  
\end{align}
Indeed, the resulting tensor branches are closer to theories like~\eqref{eq:susoz2} and \eqref{eq:e7su2n_Z2}, where the $\mathfrak{so}_8$ or $\mathfrak{e}_7$ flavor algebra is gauged.  More theories of similar type can be found in the Appendix~\ref{app:outlier}.  

\subsubsection{$\boldsymbol{\mathbb{Z}_3}$ Torsion: Conformal Matter}
\label{sssec:Z3SCM}
F-theory models with  $\mathbb{Z}_3$ torsion are constructed from the Weierstrass model~\eqref{eq:WSFTuning_Z3}. It is easy to see that this model only allows fibers of type $I_{3n}$, $IV$, and $IV^{*,\text{s}}$, which are consistent with the restricted monodromy.

We start with presenting the $\mathbb{Z}_3$ discrete holonomy instanton theories:
\begin{align}
\renewcommand{\arraystretch}{1.3}
\begin{array}{|c|c|c|c|} \cline{2-4}
\multicolumn{1}{c|}{}&\text{ Flavor } & a_1 & a_3 \\ \hline 
1&[\text{E}_6 \times \text{SU}(3)] /\mathbb{Z}_3  &  u &   u^2 v \\[0pt] 
2&\text{SU}(9)/\mathbb{Z}_3  & u & v^3 \\[0pt] 
3&[\text{E}_6 \times \text{SU}(3)_{IV}] /\mathbb{Z}_3  & u v & u^2 v \\ \hline
4& [\text{SU}(6) \times \text{SU}(3)_{IV}]/\mathbb{Z}_3 & u & u v^2 \\  \hline
\end{array}
\end{align}
Note that the last entry above is not a maximal commutant in $\text{E}_8$, but can instead be viewed as originating from a deformed $ \mathfrak{e}_6 \times \mathfrak{su}_{3}$ theory. 

Theory 1 is somewhat analogous to theory 1 of the $\mathbb{Z}_2$ case. New types of conformal matter arise from colliding an $\mathfrak{e}_6$ with an $\mathfrak{su}_{3n}$ algebra. Setting $a_1 = u$ and $a_3 = u^2 v^n$, in theory 1 leads to theories with a tensor branch
\begin{align}  
\begin{array}{c}
\includegraphics[scale=0.7]{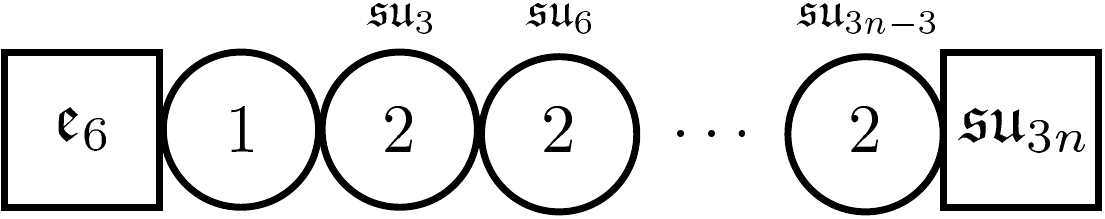} \end{array} 
\label{eq:Z3chain}
\end{align}
This is similar to the classic $\mathfrak{e}_6 \times \mathfrak{su}$ collision, but the ranks of the gauge algebras on the tensor branch increase in steps of three. This necessarily had to happen, since they have to be compatible with the $\mathbb{Z}_3$ torsion factor.
 
Let us move on to theory 2 in the table above and enhance the $I_9$ fiber to an $I_{3n}$ fiber by setting $a_1 =  u\,,~ a_3 = v^n$. From the Weierstrass model
\begin{align}
\begin{split}
f =& \tfrac{1}{48} \kappa_1 u (-\kappa_1^3 u^3 + 24 v^n)\, ,\qquad g= \tfrac{1}{864} (\kappa_1^6 u^6 - 36 \kappa_1^3 u^3 v^n + 216 v^{2 n}) , \\ \Delta=& \tfrac{1}{16} v^{3 n} (-\kappa_1^3 u^3 + 27 v^n)
\end{split}
\end{align}
we find that the $I_1$ locus develops a triple-point singularity at the origin for $n>2$. Each blow-up in the base reduces the rank of the gauge algebra over the blow-up divisor by $9$, which results in the tensor branch 
\begin{align}
\label{eq:Su3nZ3}
\begin{array}{c}
\includegraphics[scale=0.7]{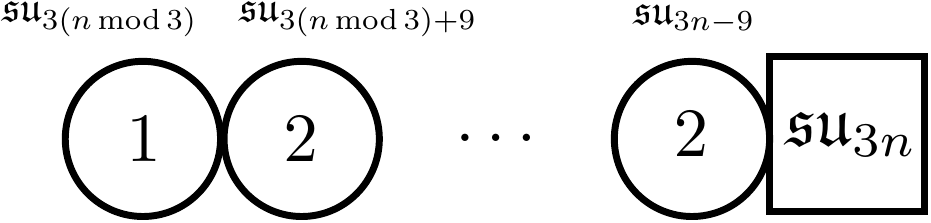} \end{array}  
\end{align}  
The matter sector consists of bi-fundamentals. If there is an $\mathfrak{su}_6$ factor over the final $(-1)$ curve, the intersection with the $I_1$ component gives an additional half-hyper in the $\mathbf{20}$-dimensional triple-antisymmetric representation. To see this note that the $\mathfrak{su}_6$ divisor intersects the $I_1$ direction in an $\mathfrak{e}_6$ point. Similarly, for $n=0$~mod~$3$, the theory has no gauge group over the $(-1)$ curve, which we identify as the E-string theory where the $\mathfrak{su}_9$ flavor factor is enhanced further. The three types of theories defined by $n=0,1,2$ mod $3$ are in precise agreement with the constructions found in~\cite{Ohmori:2018ona}.

Let us continue with theory 3. The flavor algebra at the $\mathfrak{su}_3$ end can only be enhanced once by setting $a_3=u^2 v^{n}$ with $n=0,1$. Higher values of $n$ lead to non-minimal singularities. For $n=1$, we obtain the classical $\mathfrak{e}_6 \times \mathfrak{e}_6$ conformal matter theory discussed in~\eqref{eq:e6e6Z3}. 
Let us stress at this point that tuning the $I_1$ fiber of the generic model decides whether we have perturbative matter, or whether the collision of the flavor branes leads to superconformal matter theories.  For example, for $a_1 =\kappa$ with $\kappa\in\mathbb{C}$ and $a_3 = u^n v^m$, we can engineer theories with $\mathfrak{su}_{3n} \times \mathfrak{su}_{3m}$ groups. The codimension-two vanishing order at their intersection $u=v=0$ is benign (i.e.\ below $(4,6,12)$) and hence leads to ordinary, perturbative theories rather than superconformal matter theories. However, by choosing $a_1 = \kappa_1 u + \kappa_2 v$ and $a_3 = u^n v^m$ the discriminant becomes
\begin{align}
\label{eq:nonPertenhanceZ3}
\Delta= u^{3 n} v^{3 m} (-\kappa_1^3 u^3 - 3 \kappa_1^2 \kappa_2 u^2 v - 3 \kappa_1 \kappa_2^2 u v^2 - \kappa_2^3 v^3 + 27 u^n v^m)
\end{align} 
where the collision of the $\mathfrak{su}_{3n} \times \mathfrak{su}_{3m}$ branes appears at the triple-point singularity of the $I_1$ factor. There, the singularity enhances to $(4,6,3(m+n+1))$ if $m+n > 2$.  Blowing up the intersection point introduces a $(-1)$ curve with an $\mathfrak{su}_{3(n+m-1)}$ gauge algebra that is intersected three times by the $I_1$ curve. The tensor branch is given by
\begin{align} 
\begin{array}{c}
\includegraphics[scale=0.7]{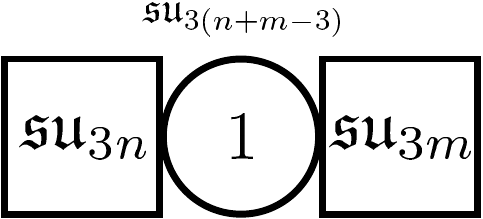} \end{array}  
\end{align}  
Notably, in the case where the central node has an $\mathfrak{su}_6$ gauge factor (i.e.\ for $n+m=5$), more matter is required to cancel the local gauge anomaly. In total we require $15$ hypermultiplets in the fundamental $\mathbf{6}$ of $\mathfrak{su}_6$, which arise naturally as bi-fundamentals for the three distinct combinations of $n$ and $m$, which have $\mathfrak{su}_{15} \times I_1$, $\mathfrak{su}_{12} \times \mathfrak{su}_3$ and $\mathfrak{su}_{9} \times \mathfrak{su}_6$ flavor groups. In addition, anomaly cancellation requires a half-hyper in the triple-antisymmetric $\mathbf{20}$ is required. This can indeed be found by analyzing the discriminant carefully and noting that over $\{ \kappa_1 u + \kappa_2 v = 0 \}$ there is a collision with the exceptional divisor, which enhances the $I_6$ to an $\mathfrak{e}_6$ singularity. From the branching of the $\mathfrak{e}_6$ adjoint one finds the triple-antisymmetric representation as expected. Note that this is the only type of enhancement that can produce the required representation while being consistent with the restricted monodromy. For $n+m>5$, further blow-ups are required that lead to the chain
\begin{align}
\label{eq:SU_SU_SU_Z3}
\begin{array}{c}
\includegraphics[scale=0.7]{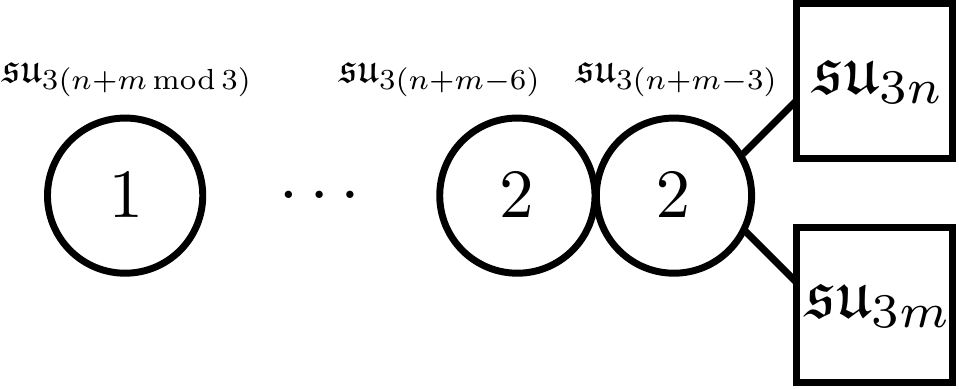}\end{array}
\end{align}
Notably, this theory has a similar tensor branch structure to the one given in~\eqref{eq:Su3nZ3}. Hence it appears plausible that the theory~\eqref{eq:SU_SU_SU_Z3} is a deformation of~\eqref{eq:Su3nZ3} which splits the $\mathfrak{su}_{3(n+m)}$ flavor algebras into $\mathfrak{su}_{3m}\times\mathfrak{su}_{3n}$.

Let us move on to theory 4. While a collision of an $I_{3m}$ with an $I_{3n}$ fiber leads to a perturbative theory for any $n$ and $m$, we get superconformal theories if we replace one of the $I_{3m}$ by an $\mathfrak{su}_3$ of type $IV$, which takes us to theory 4. Enhancing the other $I_{3m}$ factor by setting $a_3=u v^n$, we obtain the Weierstrass function
\begin{align}
\begin{split}
f=& \tfrac{1}{48} u^2 (-u^2 + 24 v^n) \,,\qquad g= \tfrac{1}{864} u^2 (u^4 - 36 u^2 v^n + 216 v^{2 n}) \\ \Delta=&  \tfrac{1}{16}  u^4 v^{3 n} (-u^2 + 27 v^n)\,,
\end{split}
\end{align}
where the $I_1$ component develops a double point singularity at the origin. For $n>2$ and even, the tensor branch is given by 
\begin{align} 
\begin{array}{c}
\label{eq:su3su3n_even}
\includegraphics[scale=0.7]{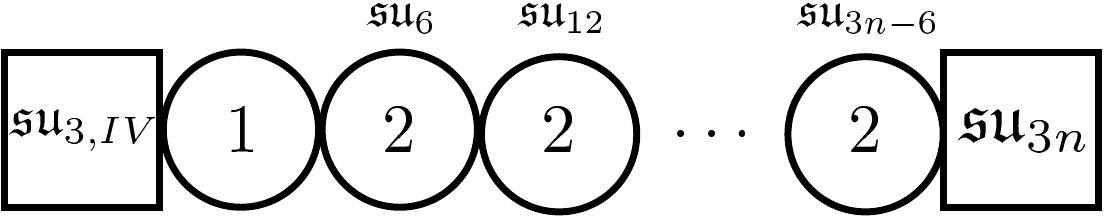} \end{array}  
\end{align}
where the left-most part is again the $\mathbb{Z}_3$ discrete holonomy instanton theory. For $n>2$ and odd, one has
\begin{align} 
\begin{array}{c}
\label{eq:su3su3n_odd}
\includegraphics[scale=0.7]{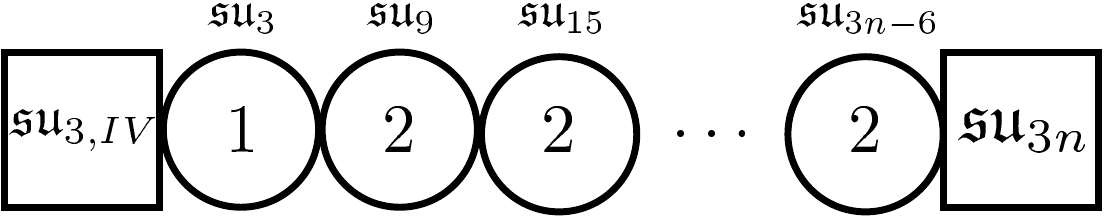} \end{array}  
\end{align}
Note that the two theories~\eqref{eq:su3su3n_even} and~\eqref{eq:su3su3n_odd} could also been obtained by enhancing~\eqref{eq:nonPertenhanceZ3}, setting $n=1$, and deforming to $\kappa_2 =0$. Various other outlier theories can be found in Appendix~\ref{app:outlier}.

\subsubsection{$\boldsymbol{\mathbb{Z}_4}$ Torsion: Conformal Matter}
\label{sec:Z4TorsionSCM}
For $\mathbb{Z}_4$ conformal matter we analyze theories whose Weierstrass model is given in \eqref{eq:WSFTuning_Z4}. One could have expected to find only groups with a $\mathbb{Z}_4$ center. However, this is not the case: it is also possible to find groups with a $\mathbb{Z}_2$ subcenter of the full $\mathbb{Z}_4$ as already being evident for the most generic model, which has an $\mathfrak{su}_4 \times \mathfrak{su}_2$ algebra. Group factors with $\mathbb{Z}_4$ and $\mathbb{Z}_2$ center also appear when considering the three different discrete holonomy instanton configurations
\begin{align}
\label{eq:Z4DHInst}
\renewcommand{\arraystretch}{1.3}
\begin{array}{|c|c|c|c|} \cline{2-4}
\multicolumn{1}{c|}{}&\text{ Flavor } & a_1 & a_2 \\ \hline 
1&[\text{SU}(8) \times \text{SU}(2)]/\mathbb{Z}_4  & u & v^2 \\[0pt] 
2&[\text{Spin}(10) \times \text {SU}(4)]/\mathbb{Z}_4 &  u &   u v \\[0pt]  \hline
3& [\text{SU}(4) \times \text{SU}(4) \times \text{SU}(2)]/\mathbb{Z}_4 &  u+v &   u v \\[0pt]  \hline
\end{array}
\end{align}

We start our discussion by enhancing the $\mathfrak{su}_2$ flavor algebra in the first discrete holonomy instanton theory by setting $a_1 = u^n$. From this, we obtain the following chain of theories
\begin{align}
\begin{array}{c}
\includegraphics[scale=0.7]{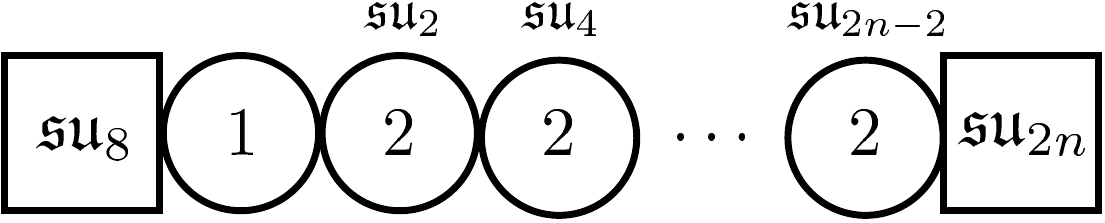} \end{array}  
\end{align}  
If we instead further enhance the $\mathfrak{su}_8$ flavor symmetry by setting $a_2  = v^{2n}$, we get
\begin{align}
\begin{array}{c}
\includegraphics[scale=0.7]{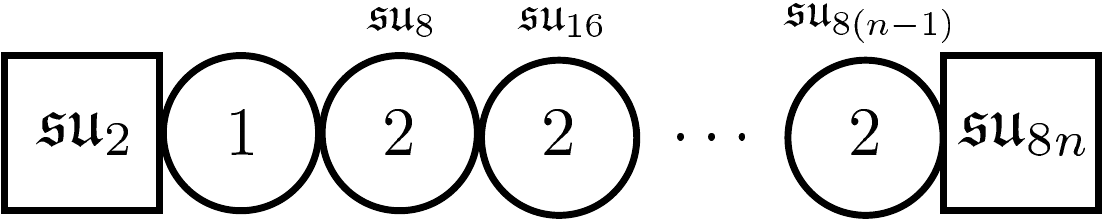} \end{array}  
\end{align}
As in the examples above, the jump in the gauge group rank proceeds along the gauge factor coming from the (broken) flavor theory of the E-string theory. Note that a similar theory exists when the power of $v$ in $a_2$ is odd, $a_2=v^{2n+1}$, which leads to
\begin{align}
\begin{array}{c}
\includegraphics[scale=0.7]{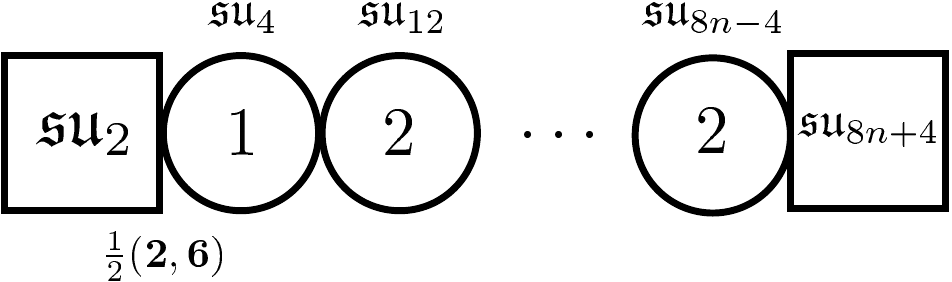} \end{array}  
\end{align} 
The $\mathfrak{su}_4$ above the $(-1)$ curve next to the $\mathfrak{su}_2$ flavor algebra comes with a half-hypermultiplet in the $(\mathbf{2,6})$ representation, as required by anomaly cancellation and consistent with a $\mathbb{Z}_2$ center charge. Geometrically, the representation arises from a $(2,3,7)$ singularity at the collision point, which corresponds to an $\mathfrak{so}_{10}$ enhancement. This is also consistent with the full $\mathbb{Z}_4$ center. Branching its adjoint into $\mathfrak{su}_2 \times \mathfrak{su}_4$ representations indeed gives rise to the required representation.

For the second theory in~\eqref{eq:Z4DHInst}, one can enhance the $\mathfrak{su}_4$ factor to $\mathfrak{su}_{4n}$ by setting $a_2 = u v^n$, which results in the chain
\begin{align}
\begin{array}{c}
\includegraphics[scale=0.7]{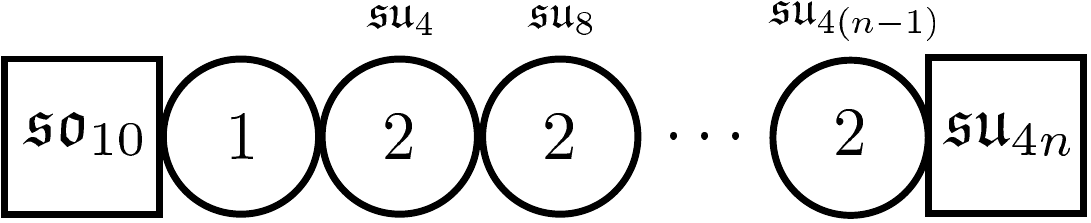} \end{array}  
\end{align} 
Note the order 4 jump in between the $\mathfrak{su}_{4k}$ factors. Enhancing the $\mathfrak{so}_{10}$ side on the other hand (by setting $a_1=u^n$) leads to
\begin{align}
\begin{array}{c}
\includegraphics[scale=0.7]{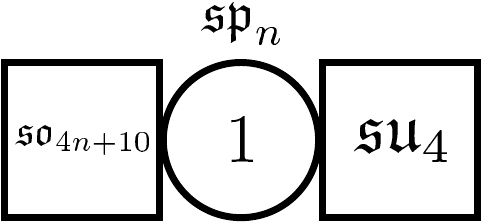} \end{array}  
\end{align} 

Finally, we consider the third theory in~\eqref{eq:Z4DHInst}, which represents the only non-maximal $\mathfrak{e}_8$ subgroup with $\mathfrak{su}_4^2 \times \mathfrak{su}_2$ flavor group. Note that this theory can in fact be seen as a deformation obtained from the $\mathbb{Z}_4 \times \mathbb{Z}_2$ discrete holonomy instanton case that we consider in Section~\ref{sssec:Z2Z4SCM}, which reduces the global flavor group by an $\mathfrak{su}_2$ factor. Starting from that theory, there is the possibility to either enhance the $\mathfrak{su}_4$ factors or the $\mathfrak{su}_2$ factor. The former case is obtained by setting $a_2=u^n v^m$ and results in
\begin{align}
\begin{array}{c}
\includegraphics[scale=0.7]{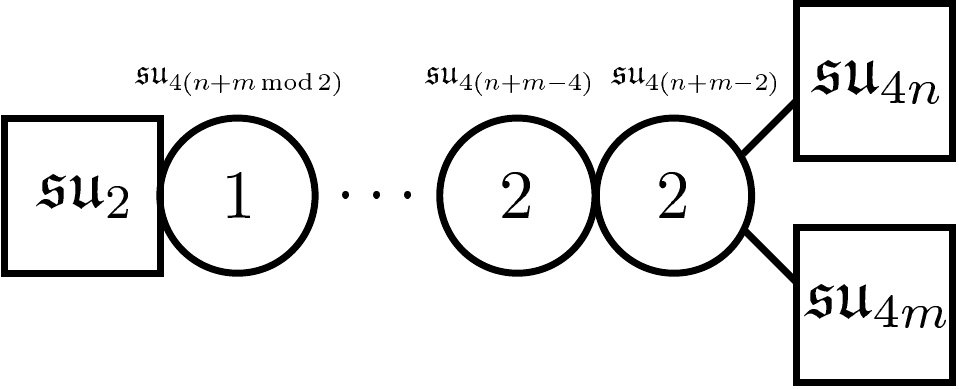} \end{array}   
\end{align}
From a technical point of view, the above chain can be obtained by performing a coordinate shift after the first blow-up. Setting $w = u-v$ puts the $\mathfrak{su}_2$ over the toric locus $w=0$. For $(n+m)=1$~mod~$2$, there is again an $\mathfrak{su}_4$ attached to the $\mathfrak{su}_2$ flavor factor, with half-hypers in the $(\mathbf{2,6})$ representation, as required by anomaly cancellation. This is directly seen by noting that the intersection locus is a $(2,3,7)$ singularity and hence yields the desired representations from the decomposition of the adjoint as argued before.
For the case $n+m=3$, one obtains a single $\mathfrak{su}_4$ gauge algebra on the $(-1)$ curve
\begin{align}
\begin{array}{c}
\includegraphics[scale=0.7]{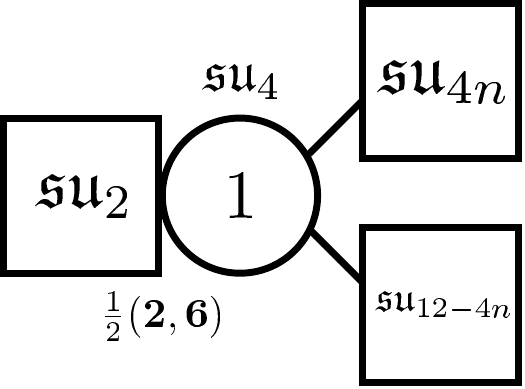} \end{array}   
\end{align}  
We can also enhance the non-toric $\mathfrak{su}_2$ flavor group to $\mathfrak{su}_{2k}$ by setting $a_1=(u+v)^k$. Note that the $\mathfrak{su}_2$ flavor is only modded by a $\mathbb{Z}_2$ subgroup of the full $\mathbb{Z}_4$. If we enhance this flavor group further, we obtain the following tensor branch:
\begin{align}
\begin{array}{c}
\includegraphics[scale=0.7]{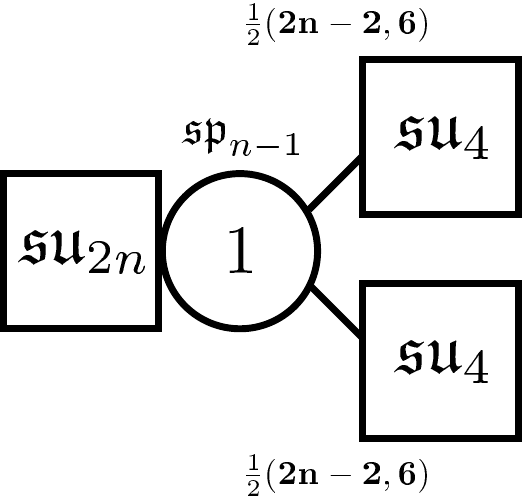} \end{array}  
\end{align}
Here, we find collisions of orders $(2,3,3+2n )$ where the $\mathfrak{sp}_{n-1}$ intersects the $\mathfrak{su}_4$ flavor factors. Hence, we expect half-hypermultiplets that transform in the two-fold antisymmetric of $\mathfrak{su}_4$, consistent with anomaly cancellation. In order to keep the appearing singularities crepantly resolvable, one cannot enhance the non-toric $\mathfrak{su}_2$ and the other $\mathfrak{su}_4$ factors arbitrarily. In fact, the only consistent configuration where one enhances multiple sides is by setting $a_1 = (u+v)^2$ and $a_2 = u v^2$ which results in
\begin{align} 
\begin{array}{c}
\includegraphics[scale=0.7]{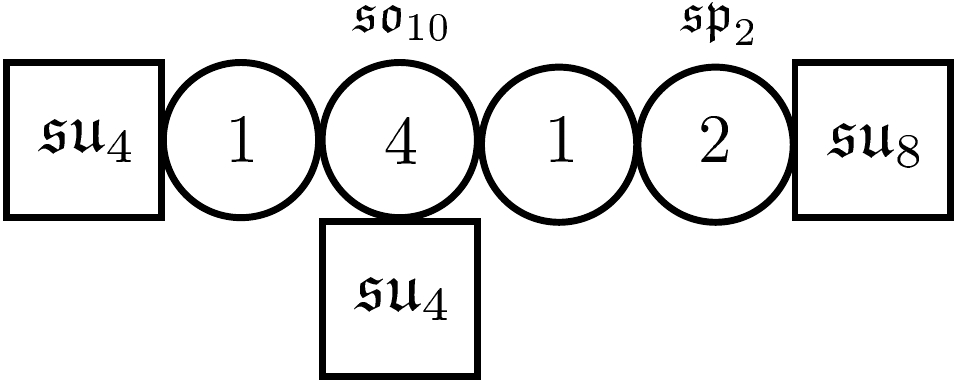} \end{array}  
\end{align}
Note that one can interpret the above configuration in part as theory 2 glued to another instanton configuration along the $\mathfrak{so}_{10}$ algebra, and further enhancing the $\mathfrak{su}_4$. 

Finally we present the possibility to engineer an $\mathfrak{so}_{10+4n} \times \mathfrak{so}_{10+4m}$ superconformal matter collision which appeared already in \cite{DelZotto:2014hpa}. Our geometric construction shows that the groups will be modded by a $\mathbb{Z}_4$ quotient factor. The theory is obtained by setting $a_1=u^{1+n} v^{1+m}$ and $a_2=u v$ and reads
\begin{align}
\begin{array}{c}
\includegraphics[scale=0.7]{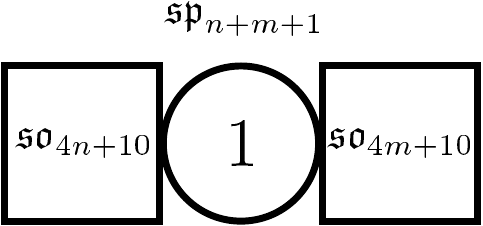} \end{array}  
\end{align} 
This concludes the simple theories which can be obtained by enhancing flavor groups of $\mathbb{Z}_4$ discrete holonomy instanton theories. Various other outlier theories can be found in Appendix~\ref{app:outlier}.
  
\subsubsection{$\boldsymbol{\mathbb{Z}_5}$ Torsion: Conformal Matter}

Next, we focus on the $\mathbb{Z}_5$ torsion model. Monodromies of this type only allow for fibers of type $I_{5n}$, i.e.\ $\mathfrak{su}_{5n}$ algebras, as one expects from the field theory side. The Weierstrass model is given in Eqn.~\eqref{eq:WSFTuning_Z5}. The only discrete holonomy instanton theory is
\begin{align}
\renewcommand{\arraystretch}{1.3}
\begin{array}{|c|c|c|} \hline
\text{Flavor Group} & a_1 & b_1 \\ \hline 
[\text{SU}(5) \times \text{SU}(5)]/\mathbb{Z}_5  & u & v  \\ \hline
\end{array}
\end{align}  
Enhancing one of the $\mathfrak{su}_5$ flavor factor is the only non-trivial possibility for superconformal matter of this kind (enhancing both factors simultaneously leads to non-minimal models), which is obtained by setting $a_1 = u\, ,~ b_1 = u^n$. This results in a theory with an $[\text{SU}(5) \times \text{SU}(5n)]/\mathbb{Z}_5$ flavor group. This class is the analog of $\mathfrak{e}_8 \times \mathfrak{su}$ superconformal matter restricted to geometries with a 5-torsion section. Its tensor branch reads
\begin{align}
\begin{array}{c}
\includegraphics[scale=0.7]{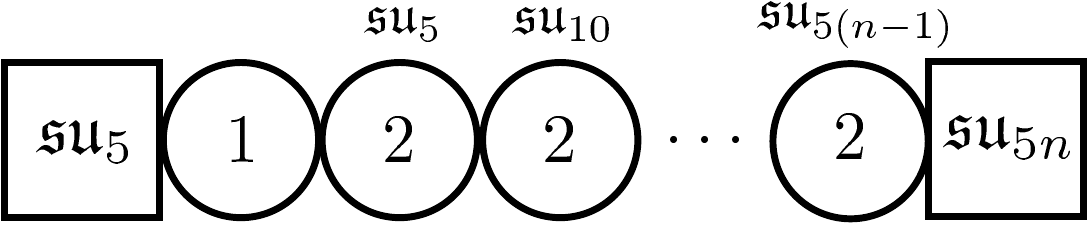} \end{array}  
\end{align}  
At the end of the chain, we have the regular $\mathbb{Z}_5$ discrete holonomy instanton theory. This class of theories has also been constructed from a field theory perspective in~\cite{Ohmori:2018ona}.

\subsubsection{$\boldsymbol{\mathbb{Z}_6}$ Torsion: Conformal Matter}

Finally, the $\mathbb{Z}_6$ restricted model given in~\eqref{eq:WSFTuning_Z6} demands a similarly strong tuning as the $\mathbb{Z}_5$ case, which leads to a single consistent discrete holonomy instanton theory,
\begin{align}
\renewcommand{\arraystretch}{1.3}
\begin{array}{|c|c|c|} \hline
\text{Flavor Group} & a_1 & b_1 \\ \hline 
[\text{SU}(6) \times \text{SU}(3) \times \text{SU}(2)]/\mathbb{Z}_6  & u & v  \\ \hline
\end{array}
\end{align} 
Note that there are groups with centers $\mathbb{Z}_2$, $\mathbb{Z}_3$, and $\mathbb{Z}_6$. Since $\mathbb{Z}_6\simeq\mathbb{Z}_2\times\mathbb{Z}_3$, this means that also subgroups of $\mathbb{Z}_6$ appear as centers.

We have the choice to extend the flavor groups of the E-string theory from several directions, which leaves a chain with jumps in their respective orders. Technically, this is best achieved by performing a coordinate shift in the above Weierstrass model and then increasing the vanishing order of the shifted coordinate. Remember that
\begin{align}
 \Delta = \tfrac{1}{2^{24}} (3a_1 - 5 b_1) (3 a_1 + b_1)^2 (a_1 + b_1)^3 (a_1 - b_1)^6\,.
\end{align}
Setting $w=a_1-b_1=(u-v)$ and increasing the vanishing order to $w^n$ allows us to construct the chain
\begin{align}
\begin{array}{c}
\includegraphics[scale=0.7]{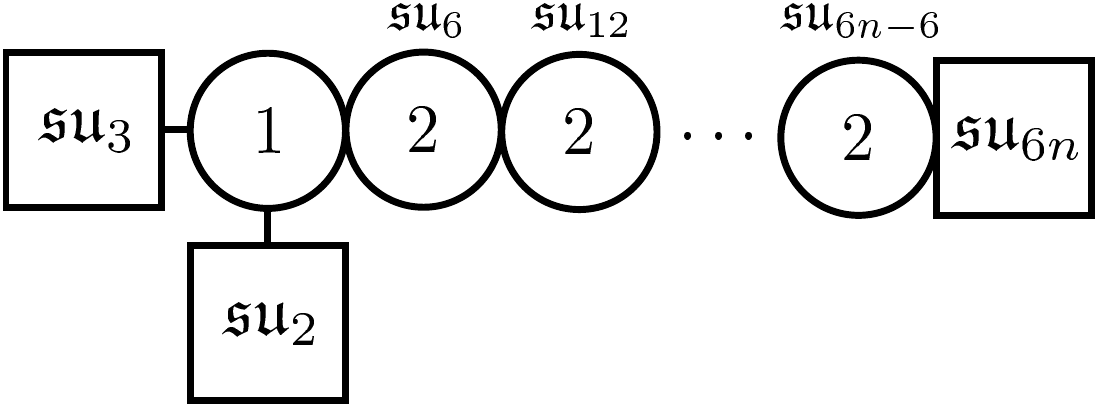} \end{array}   
\end{align}
as was also deduced in \cite{Ohmori:2018ona} from field theory arguments. However, we can also construct extended chains along other directions: Setting $w=a_1+b_1=(u+v)$ and then sending $w\to w^n$ leads to
\begin{align}
\begin{array}{c}
\includegraphics[scale=0.7]{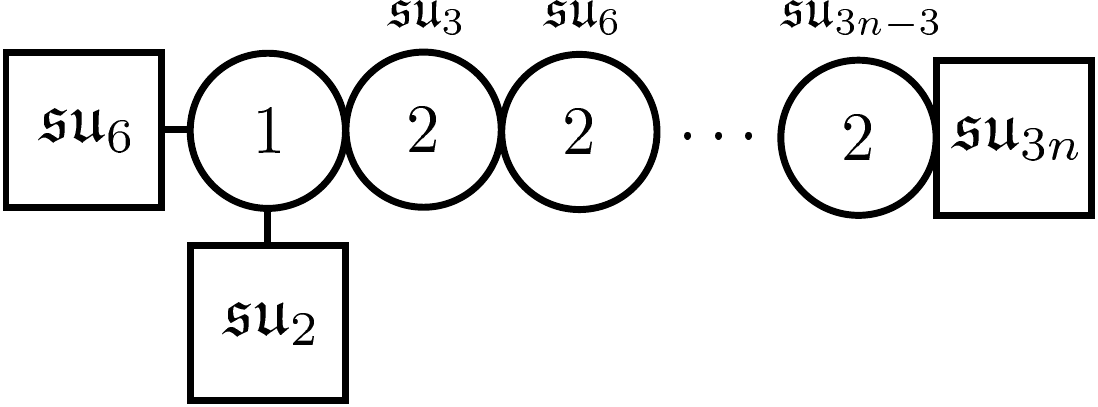} \end{array} 
\end{align}
while setting $w=3a_1+b_1=(3u+v)$ and then increasing the vanishing order to $w^n$ gives
\begin{align}
\begin{array}{c}
\includegraphics[scale=0.7]{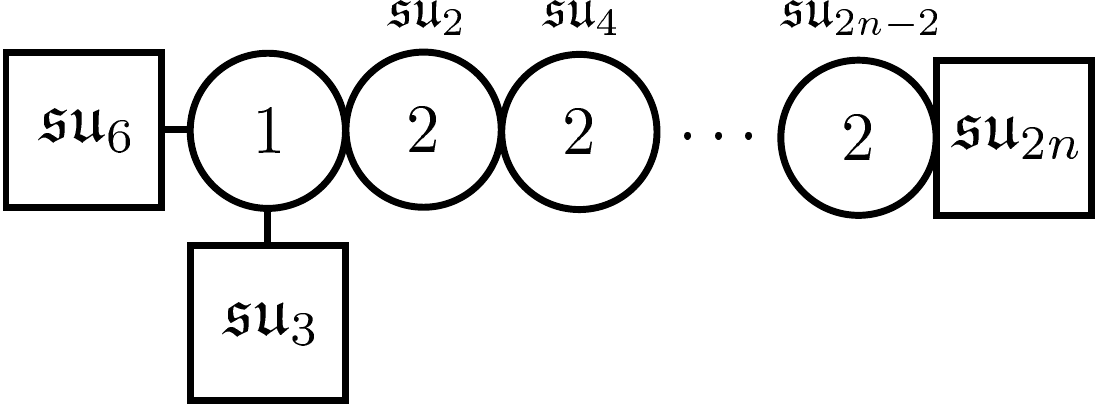} \end{array} 
\end{align}
Simultaneous enhancement of two gauge factors will result in non-minimal singularities.
  
\subsection[Double Factors Theories: \texorpdfstring{$\mathbb{Z}_2 \times \mathbb{Z}_2 \,,~\mathbb{Z}_2 \times \mathbb{Z}_4 \,,~\mathbb{Z}_3 \times \mathbb{Z}_3$}{Z2xZ2, Z2xZ4, Z3xZ3}]{Double Factors Theories: \texorpdfstring{$\boldsymbol{\mathbb{Z}_2 \times \mathbb{Z}_2 \,,~\mathbb{Z}_2 \times \mathbb{Z}_4\,,~\mathbb{Z}_3 \times \mathbb{Z}_3}$}{Z2xZ2, Z2xZ4, Z3xZ3}}

This section continues the investigation of superconformal matter theories with non-simply-connected groups that admit two quotient factors. These theories have similar features to the ones we have already discussed. However, tuning the torsion points severely restricts the possible breaking patterns of the E-string theory by discrete holonomy instantons. Moreover, as e.g.\ in the $\mathbb{Z}_4$ case, the center of some of the group factors is only modded out by a subgroup of the full torsion group.

\subsubsection{$\boldsymbol{\mathbb{Z}_2 \times \mathbb{Z}_2}$ Torsion: Conformal Matter}
Here we summarize the general models with $\mathbb{Z}_2 \times \mathbb{Z}_2$ torsion given by the Weierstrass models spelled out in Eqn.~\eqref{eq:WSFTuning_Z2xZ2}. These models admit a couple of $\text{E}_8$ breaking patterns:
\begin{align}
\renewcommand{\arraystretch}{1.3}
\begin{array}{|c|c|c|c|} \cline{2-4}
\multicolumn{1}{c|}{}&\text{ Flavor Group } & b_2 & c_2 \\ \hline 
1& [\text{SU}(4)^2 \times \text{SU}(2)^2]/ [\mathbb{Z}_2 \times \mathbb{Z}_2]  & u^2 & v^2  \\  \hline 
2& [\text{Spin}(12) \times \text{SU}(2)^2] / [\mathbb{Z}_2 \times \mathbb{Z}_2]   & u v & v^2  \\  \hline
3& [\text{Spin}(8) \times \text{Spin}(8)] / [\mathbb{Z}_2 \times \mathbb{Z}_2]   & u v & \frac12 u v   \\  \hline
\end{array}
\end{align} 
Again, the rank of the resulting gauge algebra factors can be further increased by tuning higher vanishing orders. Note that the first broken instanton theory carries a $\mathbb{Z}_4 \times \mathbb{Z}_2$ torsion factor which we discuss later.\footnote{Indeed, by shifting the coordinates as $ \widetilde{u} = 2 (u - v), ~ \widetilde{v} = \frac12 (u + v)$, one obtains the same model from the $\mathbb{Z}_2 \times \mathbb{Z}_4$ torsion model, cf.~\eqref{eq:E8z2z4}.} 

Enhancing one of the $\mathfrak{su}_4$ factors to $\mathfrak{su}_{4n}$ by setting $b_2=u^{2n}$ (or $c_2=v^{2m}$), we get the tensor branch
\begin{align}
\begin{array}{c}
\includegraphics[scale=0.7]{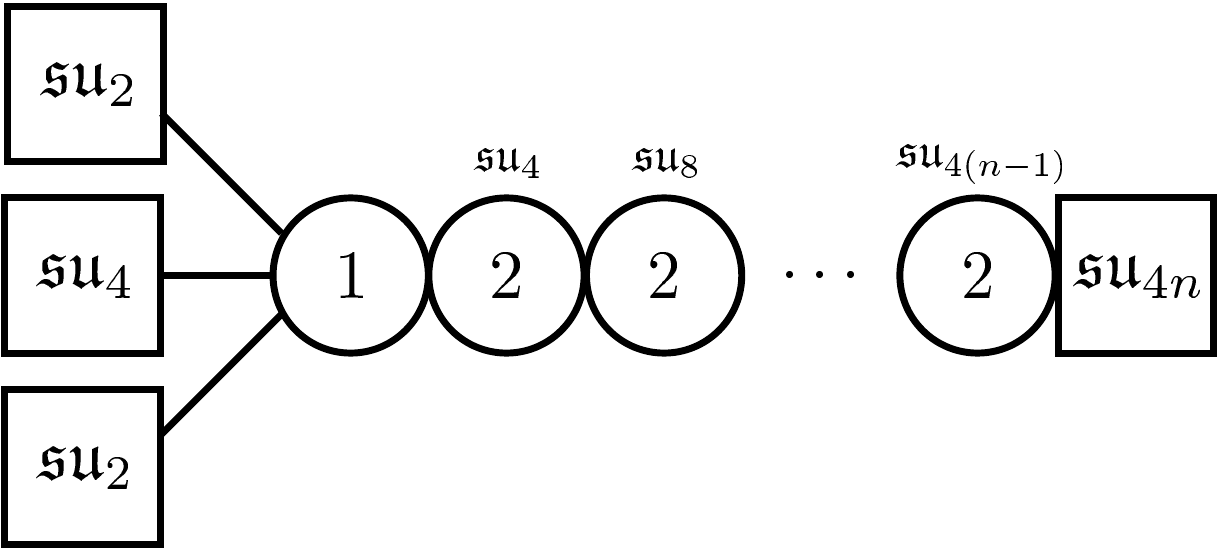} \end{array} 
\end{align}
This enhancement also preserves the $\mathbb{Z}_4$ torsion point, which is why the group symmetry is  modded by the full $\mathbb{Z}_4 \times \mathbb{Z}_2$. 

This, however, is not the case when enhancing the $\mathfrak{su}_4$ to an $\mathfrak{su}_{4n+2}$ by setting $b_2=u^{2n-1}$. In that case, the outer $\mathfrak{su}_2$ factors do not split anymore but become a single $\mathfrak{su}_2$ flavor factor. Moreover, the $\mathfrak{su}_4$ is folded to an $\mathfrak{sp}_2$. The rank of the gauge groups on the tensor branch is still going to jump by steps of four for the $\mathfrak{su}$ factors, but it ends with an $\mathfrak{su}_2$ on a $(-1)$ curve intersecting the $\mathfrak{su}_2$ and $\mathfrak{sp}_2$ flavor factor at a double-point singularity. The full tensor branch reads
\begin{align}
\begin{array}{c}
\includegraphics[scale=0.7]{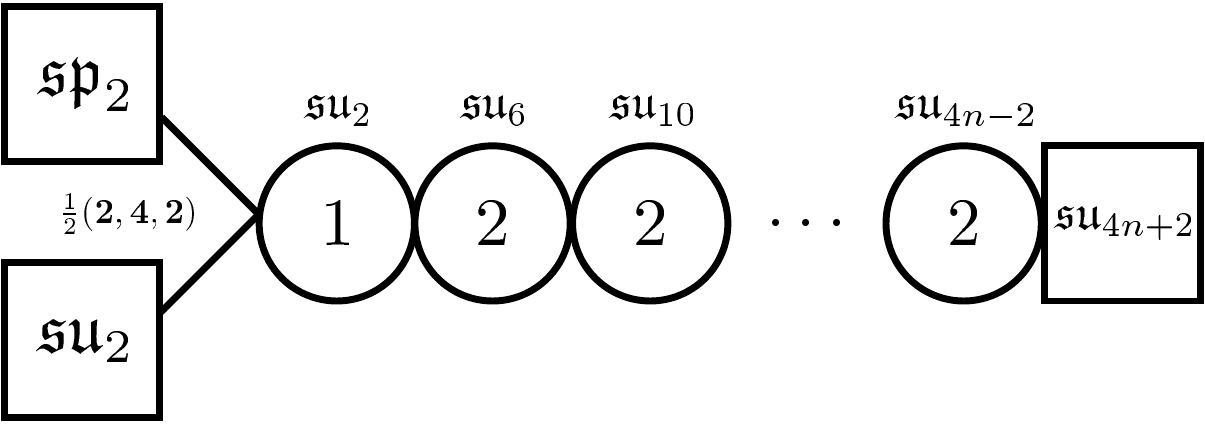} \end{array} 
\end{align}
Note that similarly, we can enhance the original $\mathfrak{su}_2$ factors, which yields a similar tensor branch but only jumps by two in the rank of the $\mathfrak{su}$ gauge factors.

Turning to the second discrete holonomy instanton theory above, we can further enhance the $\mathfrak{so}$ factor by setting $c_2=v^{2+n}$, which results in theories of the type
\begin{align}
\begin{array}{c}
\includegraphics[scale=0.7]{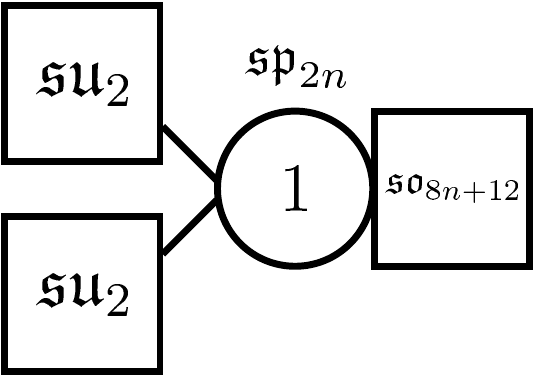} \end{array} 
\end{align}   
as well as 
\begin{align}
\begin{array}{c}
\includegraphics[scale=0.7]{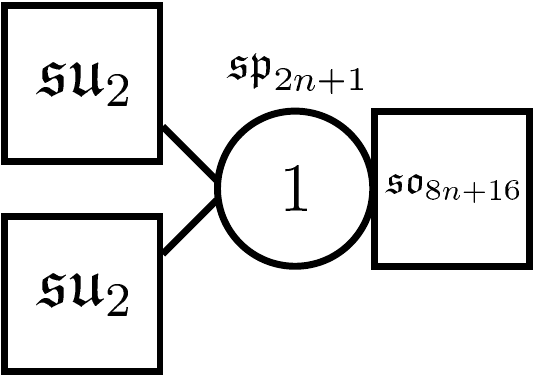} \end{array} 
\end{align} 

Alternatively, we can enhance an $\mathfrak{su}_2$ factor on the other end and obtain a chain very similar to that of $\mathbb{Z}_2$ matter with modified end part:
\begin{align}
\begin{array}{c}
\includegraphics[scale=0.7]{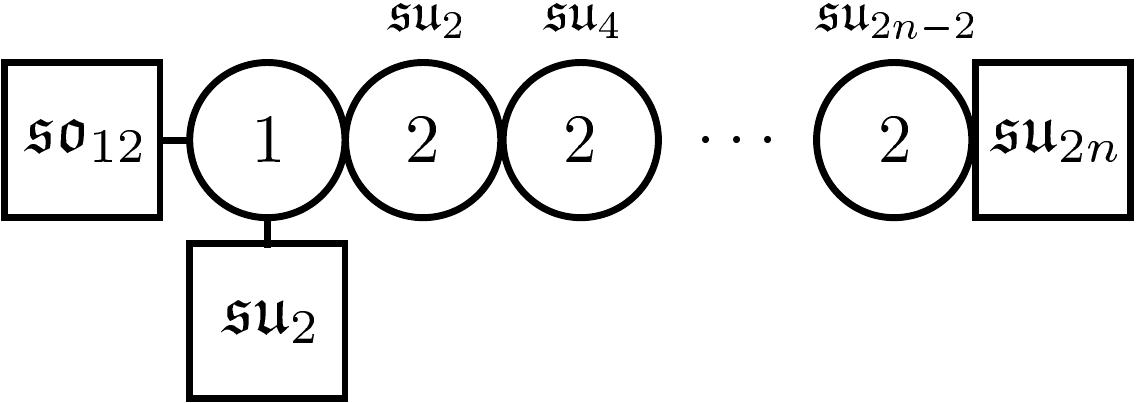} \end{array} 
\end{align}

A tuning that sits in between the second and first E-string theory can be obtained by using $\mathfrak{su}_4\simeq\mathfrak{so}_6$ and enhance it to an $\mathfrak{so}_{12} \times \mathfrak{so}_{8+4n}\times \mathfrak{su}_2$ collision by setting $b_2 = u^2 v^1 \, , \, c_2 = u^k v^n$. This, however, cannot be done beyond $k=1$, since one eventually ends up with $(4,6,12)$ singularities in codimension one (or $(8,12,24)$ singularities in codimension two). However, one is free to send $n$ to any other value which results in the chain
\begin{align}
\begin{array}{c}
\includegraphics[scale=0.7]{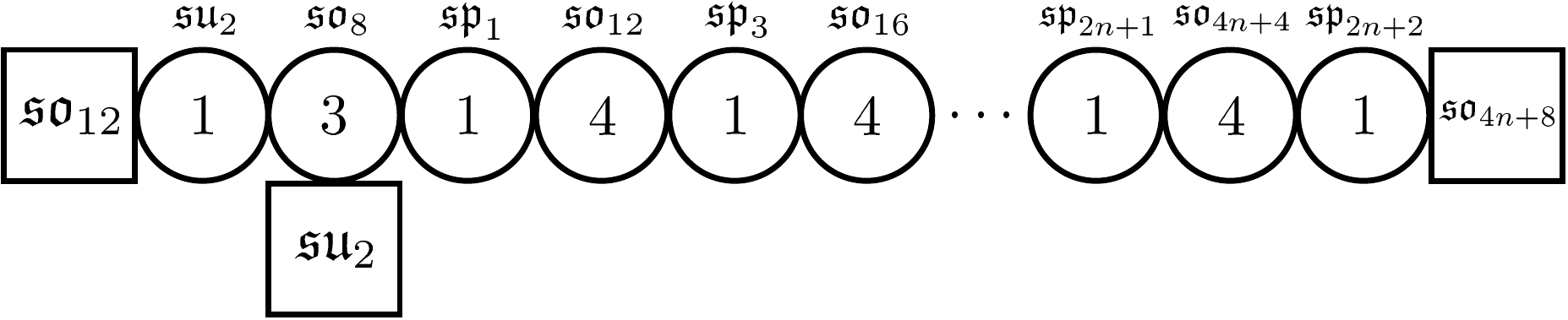} \end{array}
\end{align}
Note that for $n \leq 1$, the additional $\mathfrak{su}_2$ flavor factor decouples. The $\mathfrak{so}_8$ on the $(-3)$ curve is just the $\mathbb{Z}_2 \times \mathbb{Z}_2$ NHC as given in \eqref{eq:nhc3z2z2}, with three half-hypers in the bi-fundamental representation of vector, spinor and co-spinor representation, as required by anomaly cancellation. The rest of the matter are half-hypers in the bi-fundamental representations.

\subsubsection{$ \boldsymbol{\mathbb{Z}_2 \times \mathbb{Z}_4}$ Torsion: Conformal Matter}
\label{sssec:Z2Z4SCM}
The $\mathbb{Z}_2 \times \mathbb{Z}_4$ model breaks $\text{E}_8$ to $[ \text{SU}(4)^2 \times \text{SU}(2)^2] / [\mathbb{Z}_2 \times \mathbb{Z}_4]$ generically, which can be observed in the Weierstrass model given in Eqn.~\eqref{eq:WSFTuning_Z2xZ4}. This means that the simplest discrete holonomy instanton theory is
\begin{align}
\label{eq:E8z2z4}
\renewcommand{\arraystretch}{1.3}
\begin{array}{|c|c|c|}\hline
$Flavor Group$ & a_1 & b_1  \\ \hline
[\text{SU}(4)^2 \times \text{SU}(2)^2] / [\mathbb{Z}_2 \times \mathbb{Z}_4] & u & v \\ \hline
\end{array} 
\end{align}
We can enhance one of the  $\mathfrak{su}_2$ flavor factors by e.g.\ setting $b_1 = v^n$, which results in the tensor branch
\begin{align}
\label{eq:SU2n_Z4Z2}
\begin{array}{c}
\includegraphics[scale=0.7]{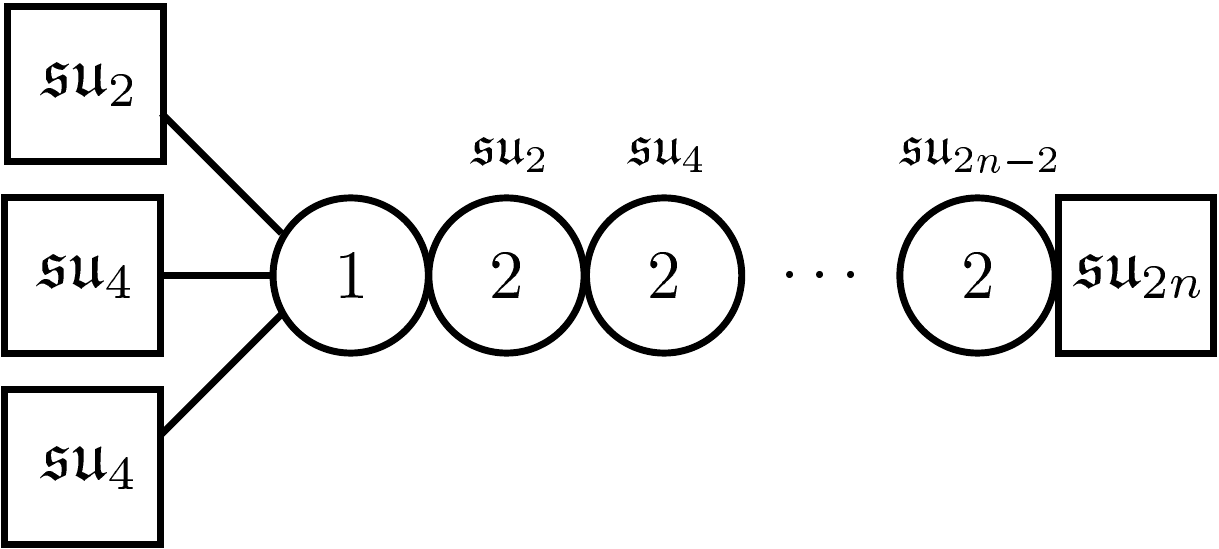} \end{array} 
\end{align} 
Note that we are not able to simultaneously enhance the second $\mathfrak{su}_2$ factor in the model without introducing a codimension-two $(8,12,24)$ point at the origin. Alternatively we can enhance the $\mathfrak{su}_4$ factor by setting $a_1 = u\,,~ b_1 = \frac14 (v^n+u)$, which gives an analogous chain
\begin{align}
\begin{array}{c}
\includegraphics[scale=0.7]{SU4n_Z4Z2.pdf} \end{array} 
\end{align} 

\subsubsection{$\boldsymbol{\mathbb{Z}_3 \times \mathbb{Z}_3}$ Torsion: Conformal Matter}

The final model we are discussing preserves two 3-torsion sections and has the restricted Weierstrass model given in \eqref{eq:WSFTuning_Z3xZ3}. The simplest factorization is also its generic gauge group in a compact setup, which is identical to the associated discrete holonomy instanton theory
\begin{align}
\renewcommand{\arraystretch}{1.3}
\begin{array}{|c|c|c|}\hline
$Flavor Group$ & a_1 & b_1 \\ \hline
\text{SU}(3)^4/[\mathbb{Z}_3 \times \mathbb{Z}_3] & u & v \\ \hline
\end{array}
\end{align} 
We can now increase the rank of any of the $\mathfrak{su}_3$ algebras. Since the $b_1$ locus is toric, it is easiest to set $b_1 = v^n$. From this, we obtain the analog of the $\text{E}_8 - \text{A}$ superconformal matter theory for this restricted monodromy, with tensor branch
\begin{align}
\begin{array}{c}
\includegraphics[scale=0.7]{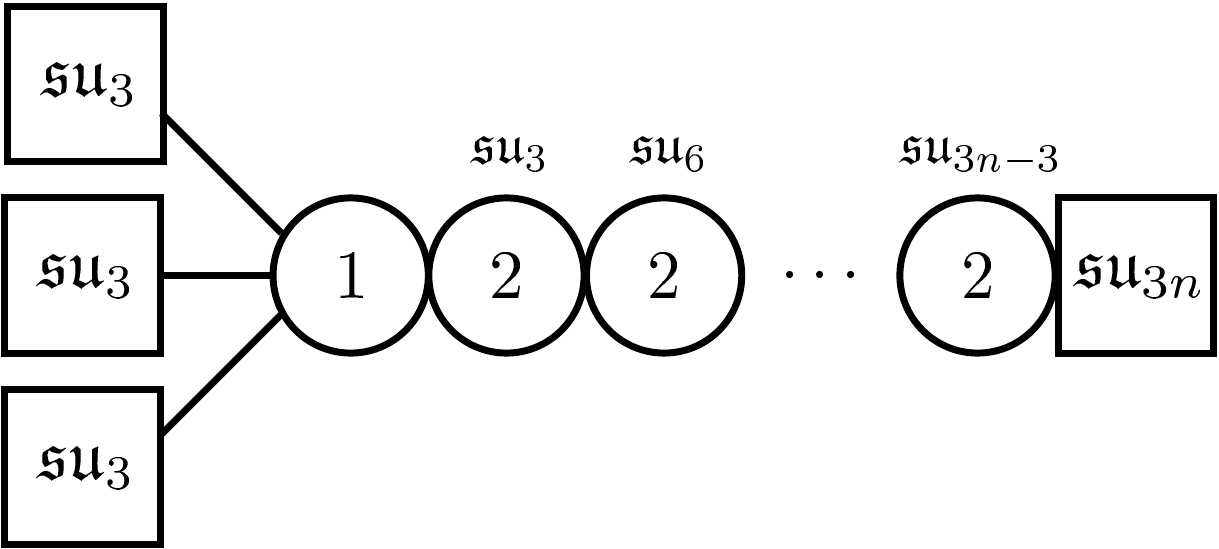} \end{array} 
\label{eq:Z3Z3model}
\end{align}
Enhancing more than one $\mathfrak{su}_3$ factor leads to non-minimal singularities.

\subsection{Construction of General SCFTs with Torsion}
\label{subsec:SCFTconstr}

With the above building blocks at hand we can outline the construction of general 6d SCFTs with Mordell-Weil torsion along the lines of \cite{Heckman:2018pqx}\footnote{In contrast to the torsion models, progenitor theories have an $\text{E}_8$ flavor group factor with trivial center.}. This constructive approach proceeds in two steps. 

First, one performs a fission process for which one switches on relevant deformations that trigger an RG flow. These deformations can have two effects: 
\begin{enumerate}
\item They deform the gauge and flavor algebras on the tensor branch. \label{itm:Deformation1}
\item They can lead to a flow in which one of the curves of negative self-intersection. decompactifies.\label{itm:Deformation2}
\end{enumerate}
Concerning effect~\ref{itm:Deformation1}, we have already seen in Section~\ref{sec:SCFT} that the deformations which modify the group structure are restricted to a subclass by a non-trivial Mordell-Weil torsion. Effect~\ref{itm:Deformation2} might split the compact part of the geometry in two pieces and is not problematic from the point of view of the global group structure: the resulting disconnected parts all respect the restricted Weierstrass form and consequently the action of $T$ is preserved.

Second, one performs a fusion operation which connects two previously disconnected parts, either via a $(-1)$ curve or by gauging a common flavor group factor in the UV. This step is more problematic in the presence of Mordell-Weil torsion, since the gauging has to be performed in a way that is compatible with the global group structure. Having multiple parts with respective torsional groups $T_i$, a fusion operation can only preserve the maximal common subgroup. Even if there is a non-trivial common subgroup, the fusion process still has to be performed in a torsion-preserving way, since otherwise the torsional part is lost in the process.
This can be easily understood in terms of the restricted monodromies in models with Mordell-Weil torsion. The individual pieces before the fusion process only allow for monodromies in certain subgroups of $\text{SL}(2,\mathbb{Z})$. After the fusion process, all monodromies appear in the same configuration and one might find a different subgroup of $\text{SL}(2,\mathbb{Z})$ or even the full duality group. The torsion of the fusion product accordingly preserves or destroys the torsional sections of the individual pieces.

We want to demonstrate the above considerations in two relatively simple examples. The first possibility for a fusion process is to connect two distinct pieces via a curve with self-intersection $(-1)$. For that consider for example the $\mathbb{Z}_2 \times \mathbb{Z}_2$ NHC on a $(-4)$ curve and the $\mathbb{Z}_2$ two-curve cluster. Connecting the two pieces via the exceptional curve will at least break the torsion group to the maximal common subgroup $\mathbb{Z}_2$. Indeed, we see that the fusion product can actually be identified with the resolved $\mathbb{Z}_2$ three curve cluster in \eqref{eq:3nodeZ2}.
\begin{align}
\begin{array}{c}
\includegraphics[scale=0.6]{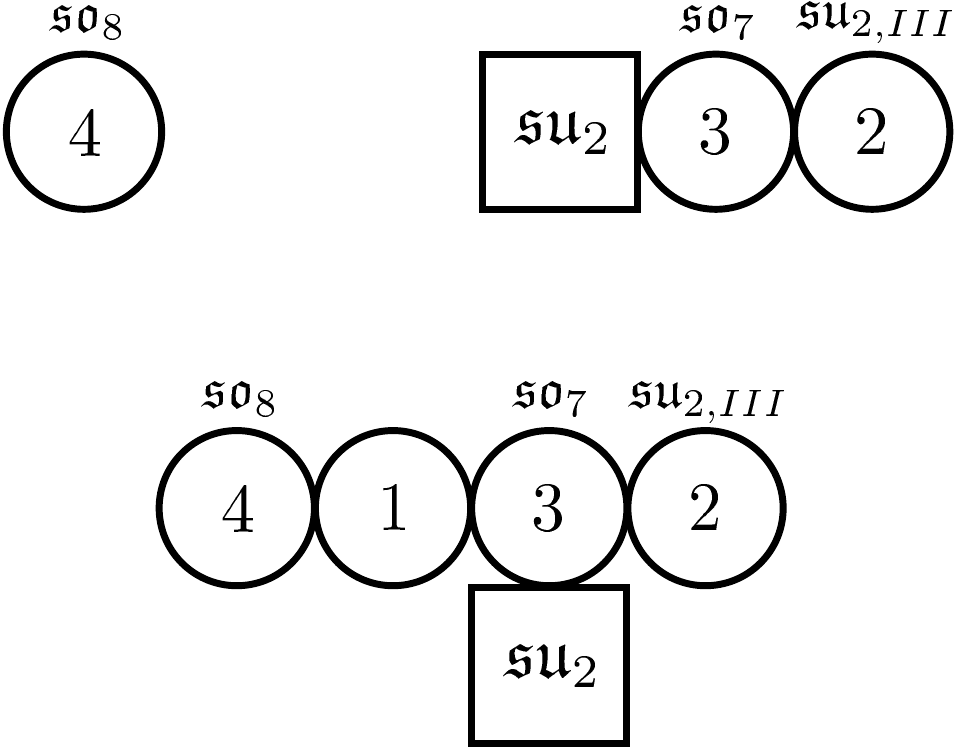}\end{array}
\end{align}
This shows that the fusion process can in fact preserve the $\mathbb{Z}_2$ torsion.

The second way to connect disconnected theories is by gauging a common flavor symmetry. If the individual parts do contain compatible torsion sections, this gauging of the flavor symmetry can be performed in a way to retain at least part of the torsion group. As an example, consider a fission product of the $\mathbb{Z}_3 \times \mathbb{Z}_3$ theory given in \eqref{eq:Z3Z3model} derived from decompactifying the first $(-2)$ curve and the $\mathbb{Z}_3$ discrete holonomy instanton theory. Gauging the two $\mathfrak{su}_3$ flavor algebras one finds
\begin{align}
\begin{array}{c}
\includegraphics[scale=0.6]{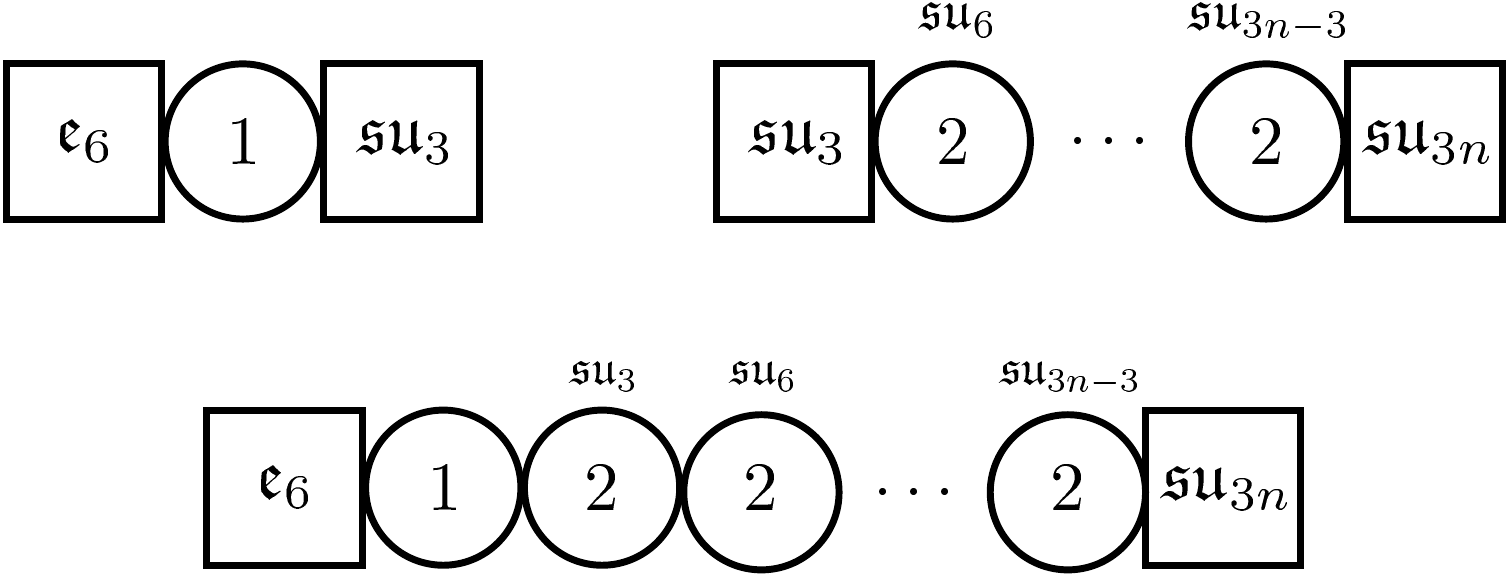} \end{array}
\end{align}
which is identical to the $\mathbb{Z}_3$ theory \eqref{eq:Z3chain} and shows that a torsional $\mathbb{Z}_3$ subgroup can be preserved in the fusion process.

\section{Conclusion and Outlook}	
\label{sec:concl}

In this article we construct a wide range of 6D SCFTs with non-simply connected non-Abelian flavor groups by tuning the geometry in an F-theory setup. The tensor branch of these theories inherits this non-simply-connected structure.

In more detail, we proceed by analyzing singular non-compact Calabi-Yau 3-folds which are elliptically-fibered with extra torsional sections. These models feature a non-trivial, finite Mordell-Weil group $T$. The construction restricts the full $\text{SL}(2,\mathbb{Z})$ duality of type IIB string theory to a congruence subgroup. As a consequence, not all $(p,q)$ 7-branes are allowed and the flavor and gauge groups of the models are constrained. Moreover, the Mordell-Weil torsion mods out a part of the center of the simply-connected cover $\mathcal{G} = \mathcal{G}^*/T$. This restricts the possible flavor and gauge groups to those Lie groups whose center admit an action of $T$, or a subgroup thereof.

Within this set of restricted Weierstrass models, we construct the essential building blocks for 6d SCFTs and their tensor branches for all possible $T$. We start by analyzing non-Higgsable clusters with Mordell-Weil torsion. Many of the original gauge algebras of~\cite{Morrison:2012np} on curves with negative self-intersection need to be enhanced in order to respect the global group structure. This enhancement is automatic once we restrict the geometry such that the elliptic fibration has Mordell-Weil torsion. 

We then turn to the E-string theory in the presence of Mordell-Weil torsion, which can be interpreted as theories with discrete holonomy instanton in the heterotic/M-theory framework. These theories result in flavor groups which are subgroups of $\text{E}_8$ with an associated action of $T$ as described above. 

Finally, we investigate the collision of two non-compact components of the discriminant locus which lead to non-minimal singularities in codimension two, also known as superconformal matter theories. The non-trivial group $T$ imposes strong restrictions on the realization of the tensor branch as well as the deformations of this class of theories. This results in important differences for theories with the same flavor algebra but different flavor group. We further confirm some field theoretic constructions in \cite{Ohmori:2018ona} as a subclass of models derived in this way and complement the explicit geometric construction. We observe that singularities of vanishing order $(8,12,24)$ (or worse) in codimension two often appear when trying to enhance flavor group factors. If blown up, such singularities lead to $(4,6,12)$ singularities in codimension one over the blow-up divisor. Interestingly, in many cases colliding two flavor groups looks benign. However, after blowing up the collision, we find that $(8,12,24)$ singularities appear between the blown up divisor and one of the flavor group factors.

There are several interesting avenues to pursue further. In Section~\ref{sec:SCFT} we already hinted at the modifications of allowed deformations and RG flows in the presence of Mordell-Weil torsion. It would be interesting to explore these ideas and obtain a complete network of 6d SCFTs with torsion. Beside the techniques employed in \cite{Heckman:2015ola, Heckman:2016ssk, Heckman:2018pqx}, an alternative approach to classify the deformations associated to T-brane data is via $(p,q)$ string junctions, see e.g.\ \cite{Hassler:2019eso}, which can also be utilized in geometries with restricted monodromies, i.e.\ Mordell-Weil torsion.

We also suggested an M-theory interpretation of models with non-trivial torsion in Section~\ref{sec:SCFT}. Exploring this further would not only shed some light on the M-theory picture of the group structure in 6d SCFTs, but on M-theory models in general. The appearance of fractional fluxes might further elucidate the interplay between the gauge group of the 7d super-Yang-Mills theory on an ADE-singularity and the worldvolume theory of M5-branes probing it, with possible effects in holographic descriptions. Compactifying the resulting 6d theories leads to additional possibilities in the flavor backgrounds due to the global group structure. An investigation involving these might reveal further structure in the lower-dimensional superconformal field theories, as discussed e.g.\ in \cite{Ohmori:2018ona}.

Of course another way to investigate the implications of the global group structure in M-theory is to use M-/F-theory duality. For that one compactifies the 6d model F-theory model on a circle (including possible twists), and deforms the 5d theory onto the Coulomb branch \cite{Bhardwaj:2018yhy, Bhardwaj:2018vuu,Bhardwaj:2019fzv}. It is suggestive that the modification of the group structure in the 6d setup also affects the allowed twists and resulting 5d theories. 

Beyond this, a pure 5d approach to M-theory on elliptically-fibered Calabi-Yau 3-fold with torsion, similar to \cite{Apruzzi:2019vpe, Apruzzi:2019opn, Apruzzi:2019enx}, could be used to study possible effects of restrictions imposed by the presence of torsional sections. In this way, one can fully benefit from the powerful geometrical F-theory construction of 6d SCFTs acting as seeds for lower-dimensional theories via compactification. 

Furthermore, the explicit geometries of this work can serve as a starting point for the constructions of various other theories in six and lower dimensions. One promising direction, presented in \cite{Anderson:2018heq,Anderson:2019kmx} is to combine the fiber translation encoded in the finite Mordell-Weil group with an automorphism in the base. This allows to combine quotients in the base, such as in \cite{Apruzzi:2017iqe}, with a non-trivial action on the fiber to obtain new smooth local Calabi-Yau quotients, with new F- and M-theory duals.

Finally, a full classification of non-simply-connected flavor groups also includes additional $\text{U}(1)$ factors in the flavor sector. In \cite{Apruzzi:2020eqi}, it was shown that these do not necessarily originate from the free part of the Mordell-Weil group, but can descend from deformed non-Abelian symmetries. In our analysis we often found additional $I_1$ loci whose tuning strongly influenced the overall tensor branch structure. It is tempting to connect these $I_1$ singularities of the fiber, or combinations thereof dictated by their ABJ anomalies, to the additional Abelian flavor factors. Since these also contribute to the group structure, their inclusion might increase the allowed possibilities. Moreover, they impact the embedding of the discrete torsion group $T$ into $\mathcal{G}^*$.
 
 \subsection*{Acknowledgments}
We thank Jonathan Heckman, Thorsten Schimannek, Gianluca Zoccarato for valuable discussions. We further thank Jonathan Heckman for his comments on the draft. The work of M.D. is supported by the individual DFG grant DI 2527/1-1. The work of P.K.O.\ is supported by a Swedish grant of the Carl Trygger Foundation for Scientific Research.
 
\appendix
\section{Enhanced Weierstrass Models}
\label{sec:EnhancedWSFs}

In this appendix we list the specific factorizations of the Weierstrass coefficients that ensure the presence of torsional sections. We then further discuss the possibility to enhance the Mordell-Weil torsion by further tunings.

\subsection{Weierstrass Models with Mordell-Weil Torsion}

The enhanced Weierstrass models for all allowed torsion groups are given by
\begingroup
\allowdisplaybreaks
\begin{subequations}
\label{eq:torsionfrak}
\begin{align}
\renewcommand{\arraystretch}{1.5}
\mathbb{Z}_2:
& \quad f = a_4 - \tfrac{1}{3} a_2^2 \,, 
\quad g = \tfrac{1}{27} a_2 (2 a_2^2 - 9 a_4) \,, 
\quad \Delta = a_4^2 (4 a_4 - a_2^2) \,, \label{eq:WSFTuning_Z2}\\[6pt]
\mathbb{Z}_3: & \quad f = \tfrac{1}{2} a_1 a_3 - \tfrac{1}{48} a_1^4 \,, \quad g = \tfrac{1}{4} a_3^2 + \tfrac{1}{864} a_1^6 - \tfrac{1}{24} a_1^3 a_3 \,,\nonumber\\
&\quad \Delta = \tfrac{1}{16} a_3^3 (27 a_3 - a_1^3) \,, \label{eq:WSFTuning_Z3}\\[6pt]
\mathbb{Z}_4: 
& \quad f = - \tfrac{1}{48} a_1^4 + \tfrac{1}{3} a_1^2 a_2 - \tfrac{1}{3} a_2^2 \,, 
\quad g = \tfrac{1}{864} \big( a_1^2 - 8 a_2 \big) \big( a_1^4 - 16 a_1^2 a_2 - 8 a_2^2 \big) \,, \nonumber\\
& \quad \Delta = - \tfrac{1}{16} a_1^2 a_2^4 \big( a_1^2 - 16 a_2 \big) \,, \label{eq:WSFTuning_Z4}\\[6pt]
\mathbb{Z}_5: 
& \quad f = \tfrac{1}{48} (-a_1^4-8 a_1^3 b_1+16 a_2^2 b_1^2+8 a_1 b_1^3-16 b_1^4)\,,\nonumber\\ 
&\quad g = \tfrac{1}{864} (a_1^2 - 2 a_1 b_1 + 2 b_1^2) (a_1^4 + 14 a_1^3 b_1 + 26 a_1^2 b_1^2 - 116 a_1 b_1^3 + 76 b_1^4)\,,\nonumber\\ 
& \quad \Delta = \tfrac{1}{16} (a_1 - b_1)^5 b_1^5 (a_1^2 + 9 a_1 b_1 - 11 b_1^2)\,,\label{eq:WSFTuning_Z5}\\[6pt]
\mathbb{Z}_6: 
& \quad  f = \tfrac{1}{192} b_1 (3 a_1^3 - 3 a_1^2 b_1 - 3 a_1 b_1^2 - b_1^3)\,,\nonumber\\
&\quad  g = \tfrac{1}{2^{12} 3^3} (3 a_1^2 - 6 a_1 b_1 - b_1^2) (9 a_1^4 - 6 a_1^2 b_1^2 - 24 a_1 b_1^3 - 11 b_1^4)\,,\nonumber\\
&\quad  \Delta = \tfrac{1}{2^{24}} (3a_1 - 5 b_1) (3 a_1 + b_1)^2 (a_1 + b_1)^3 (a_1 - b_1)^6\,,\label{eq:WSFTuning_Z6}\\[6pt]
\mathbb{Z}_2 \times \mathbb{Z}_2: 
& \quad  f = \tfrac13 (b_2 c_2 - b_2^2 - c_2^2)\,,
\quad  g = -\tfrac{1}{27} (b_2 + c_2) (b_2 - 2 c_2) (2 b_2 - c_2)\,,\nonumber\\
& \quad  \Delta = -b_2^2 c_2^2 (b_2 - c_2)^2\,,\label{eq:WSFTuning_Z2xZ2}\\[6pt]
\mathbb{Z}_2 \times \mathbb{Z}_4: 
& \quad f = -\tfrac{1}{768} a_1^4 - 7/24 a_1^2 b_1^2 - 1/3 b_1^4\,,\nonumber\\
& \quad g = \tfrac{1}{2^{11}3^3}(a_1^2 + 16 b_1^2) (a_1^2 - 24 a_1 b_1 + 16 b_1^2) (a_1^2 + 24 a_1 b_1 + 16 b_1^2)\,,\nonumber\\
& \quad \Delta = -\tfrac{1}{2^{16}} a_1^2 b_1^2 (a_1 - 4 b_1)^4 (a_1 + 4 b_1)^4\,, \label{eq:WSFTuning_Z2xZ4}\\[6pt]
\mathbb{Z}_3 \times \mathbb{Z}_3:
& \quad f = -\tfrac{1}{48} a_1 (a_1 - 2 b_1) (a_1 - 2 \omega b_1) (a_1 - 2 \omega^2 b_1)\,,\nonumber\\
& \quad g = \tfrac{1}{864} (a_1^2 + 2 a_1 b_1 - 2 b_1^2) (a_1^2 + 2 \omega a_1 b_1 - 2 \omega^2 b_1^2) (a_1^2 + 2 \omega^2 a_1 b_1 - 2 \omega b_1^2)\,,\nonumber\\
& \quad \Delta = \tfrac{1}{432} (a_1 + b_1)^3 (a_1 + \omega b_1)^3 (a_1 + \omega^2 b_1)^3 b_1^3\,,\label{eq:WSFTuning_Z3xZ3}
\end{align}
\end{subequations}
\endgroup
where $\omega$ corresponds to a non-trivial cube-root of unity and the index of the coefficients $a_i, b_i, c_i$ indicates the degree in terms of the anti-canonical class of the base, i.e.\ $a_n \sim - n K$. 

\subsection{Torsion Enhancements}

Next, we discus factorizations that enhance the torsion in the enhanced Weierstrass models. In Table~\ref{fig:torsiontuning}, we have determined all tunings that further enhance the torsion. Read in reverse, this can also be understood as deformations (such as Higgsing) that preserve a certain subgroup of the initial torsion.
\begin{table}
\centering
\includegraphics{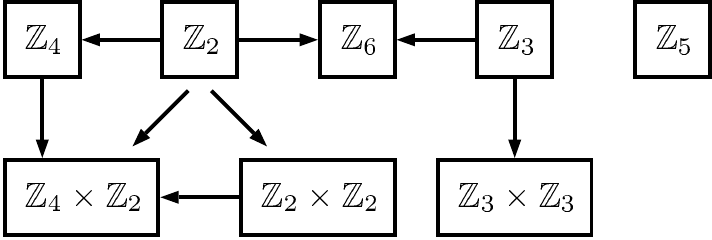}\\[12pt]
\renewcommand{\arraystretch}{1.3}
\begin{tabular}{|c|c|c|}\hline
 start torsion  &  tuned torsion  &  tuning  \\ \hline
$\mathbb{Z}_2 $ &$ \mathbb{Z}_4$ & $\begin{array}{l}a_{2} \rightarrow \frac{a_{1}^2}{4}-2 \tilde{a}_{2}\, , \\  a_4 \rightarrow \tilde{a}_{2}^2 \end{array}$ \\ \hline
$\mathbb{Z}_2 $ &$ \mathbb{Z}_6$ & $\begin{array}{l}a_2 \rightarrow \frac{1}{16} (b_1^2 + 6 a_1 b_1 - 3 a_1^2)\, , \\  a_4 \rightarrow \frac{1}{256} (a_1 - b_1)^3 (3 a_1 + b_1) \end{array}$ \\ \hline
$\mathbb{Z}_2 $ &$ \mathbb{Z}_2 \times \mathbb{Z}_2$ & $\begin{array}{l}a_2 \rightarrow -b_2 + 2c_2\, , \\  a_4 \rightarrow -b_2 c_2+c_2^2  \end{array}$ \\ \hline
$\mathbb{Z}_2 $ &$ \mathbb{Z}_2 \times \mathbb{Z}_4$ & $\begin{array}{l} a_2 \rightarrow  \frac{1}{16} (a_1^2-24 a_1 b_1+16 b_1^2)\,, \\  a_4 \rightarrow -\frac{1}{16} a_1 b_1 (a_1-4 b_1)^2 \end{array}  $ \\ \hline  \hline
$\mathbb{Z}_3 $ &$ \mathbb{Z}_6 $ &$\begin{array}{l} a_1 \rightarrow b_1 \\ a_3 \rightarrow \frac{1}{32} (\tilde{a}_1 - b_1)^2 (\tilde{a}_1 + b_1)  \end{array}$ \\ \hline 
$\mathbb{Z}_3 $ &$ \mathbb{Z}_3 \times \mathbb{Z}_3 $ &$\begin{array}{l} a_1 \rightarrow i \sqrt{3}\; \tilde{a}_1 \\ a_3 \rightarrow -\frac{i}{3 \sqrt{3}}(\tilde{a}_1^3+b_1^3)  \end{array}$ \\ \hline \hline 
$\mathbb{Z}_4 $ &$ \mathbb{Z}_2 \times \mathbb{Z}_4$ &$\begin{array}{l} a_1 \rightarrow -4i b_1 \\ a_2 \rightarrow \frac{1}{16} (a_1^2 - 16 b_1^2)  \end{array}$ \\ \hline 
$\mathbb{Z}_2 \times \mathbb{Z}_2 $ &$ \mathbb{Z}_2 \times \mathbb{Z}_4$ &$\begin{array}{l} b_2 \rightarrow a_1 b_1 \\ c_2 \rightarrow -\frac{1}{16} (a_1-4b_1)^2 \end{array}$ \\ \hline 
\end{tabular}
\caption{Chain of torsion enhancements and their explicit globally defined Weierstrass tunings.} 
\label{fig:torsiontuning} 
\end{table}

Models with $\mathbb{Z}_2$, $\mathbb{Z}_3$, and $\mathbb{Z}_2\times\mathbb{Z}_2$ torsion admit a tuning $f\equiv0$ or $g\equiv0$ while preserving the torsion points. Those configurations have a constant $J$-invariant and are hence strongly coupled. The allowed tunings are
\begin{align}
\renewcommand{\arraystretch}{1.3}
\begin{array}{|c|c|c|c|c|}\hline
\text{Factor} & \text{Tuning} &f & g & \Delta  \\ \hline 
\mathbb{Z}_2 & a_4 = a_2^2/3& 0&  -(a_2^3/27)&  a_2^6/27 \\ \hline
\mathbb{Z}_2 & a_4 = (2 a_2^2)/9 & -(a_2^2/9) & 0 & -((4 a_2^6)/729) \\ \hline \hline 
\mathbb{Z}_2 \times \mathbb{Z}_2 & b_2= -e^{(2\pi i/3)} c_2 & 0& -i/(3\sqrt{3}) c_2^3 & -c_2^6 \\ \hline   
\mathbb{Z}_2 \times \mathbb{Z}_2 & b_2 = -c_2 & -c_2^2 & 0 & -4 c_2^6 \\ \hline  \hline
\mathbb{Z}_3 & a_1 = 0 & 0 & a_3^2/4 & (27 a_3^4)/16 \\ \hline
\end{array}
\end{align}
Note that the models in line $1$ and $3$ become identical, as do the models in line $2$ and $4$.

The $\mathbb{Z}_3$ model has a type $IV$ fiber with vanishing orders $(\infty,2,4)$, whereas all other models have a type $I_0^{*,s}$ fiber. The latter are very restricted and only allow for very few further specializations. These include $\mathfrak{so}_8 \times \mathfrak{so}_8 $ collisions which can be resolved by a simple $(-1)$ curve, or by having a more general polynomial Ansatz 
\begin{align}
a_2=(u-\kappa_1 v)(u-\kappa_2 v)(u-\kappa_3 v)\,, 
\end{align}
with $\kappa\in\mathbb{C}$. This geometry has an $\mathfrak{so}_8^3$ flavor algebra. Resolving the collision requires three exceptional divisors resulting in
\begin{align}
\begin{array}{c}
\includegraphics[scale=0.5]{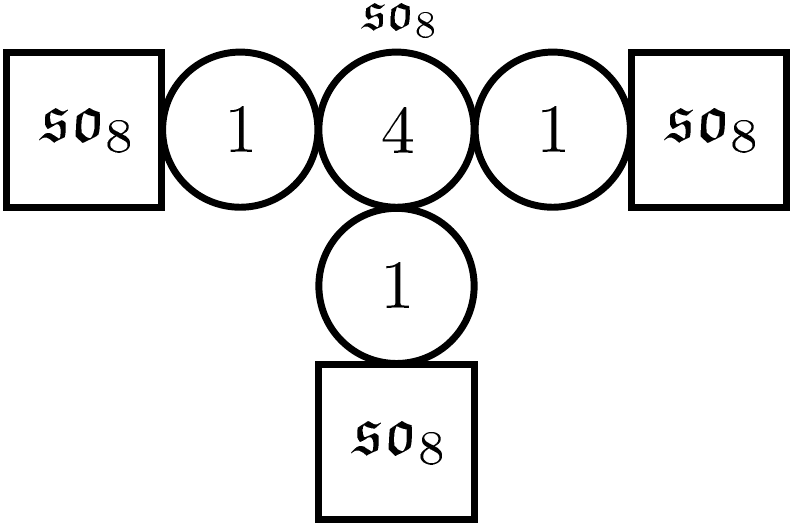} \end{array}
\end{align}

The $\mathbb{Z}_3$ case allows the collision of even more components of the discriminant locus at the origin. The maximal singularity is achieved for the collision of five type $IV$ fibers, for
\begin{align}
\label{eq:Z3IVtunings}
a_3 = \prod_i^5 (u-\kappa_i v)\,. 
\end{align}
The first blow-up yields an $\mathfrak{e}_6$ gauge factor. A smooth base requires five more blow-ups and the final configuration is given by
\begin{align}
\begin{array}{c}
\includegraphics[scale=0.5]{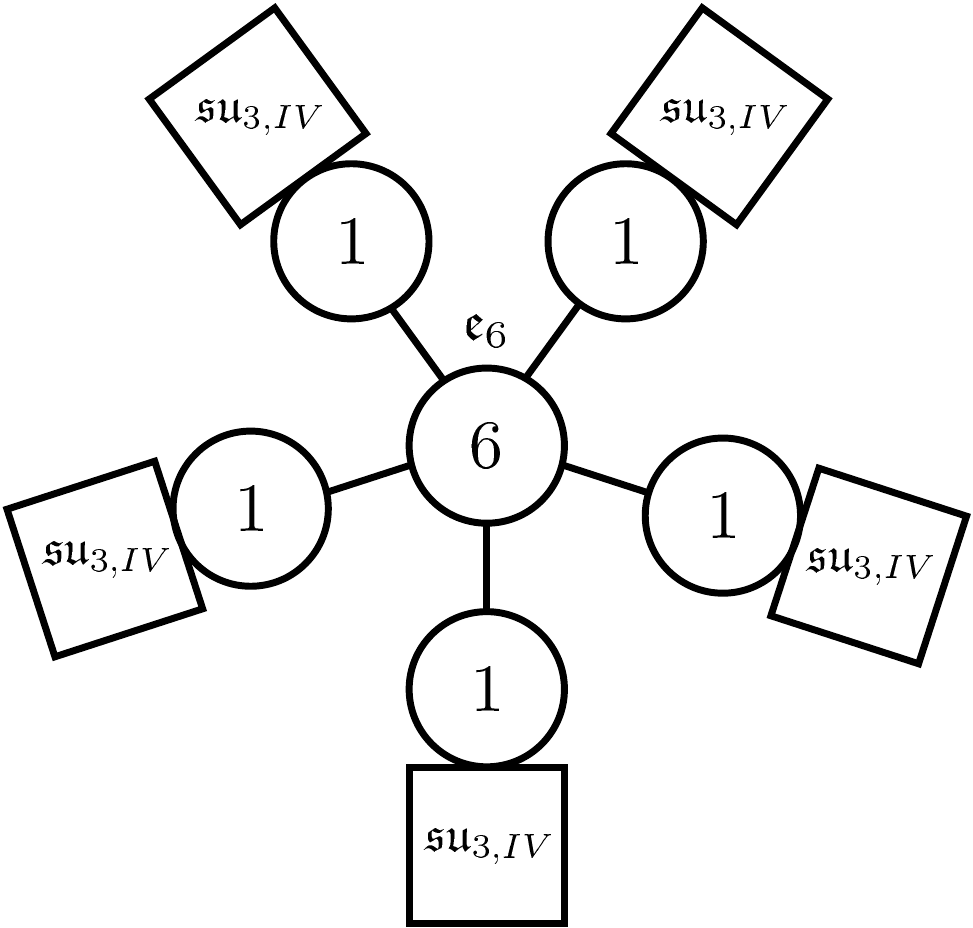} \end{array}
\end{align}
Alternatively, one can replace two of the $\mathfrak{su}_{3,IV}$ with one $\mathfrak{e}_6$ flavor algebra. Technically this is done by setting some of the parameters $\kappa_i$ in \eqref{eq:Z3IVtunings} to zero. Two possibles configurations can be generated in this way
\begin{align}
\begin{array}{c c}
\includegraphics[scale=0.5]{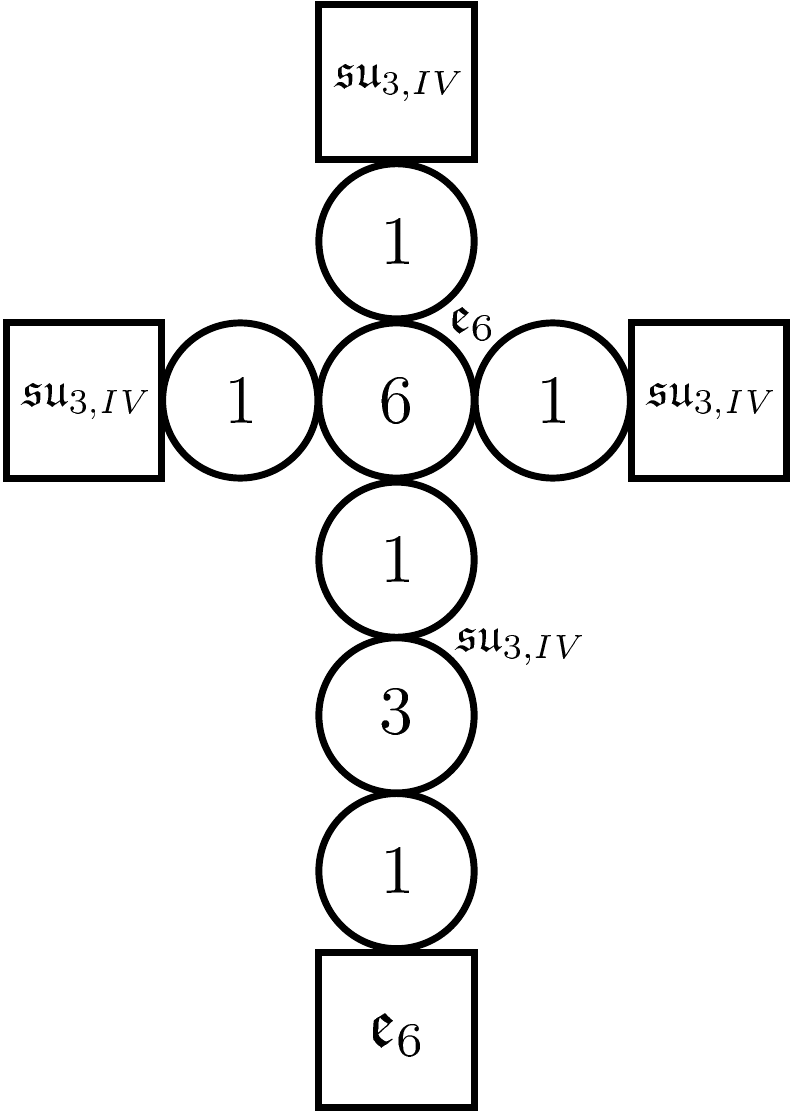} & \includegraphics[scale=0.5]{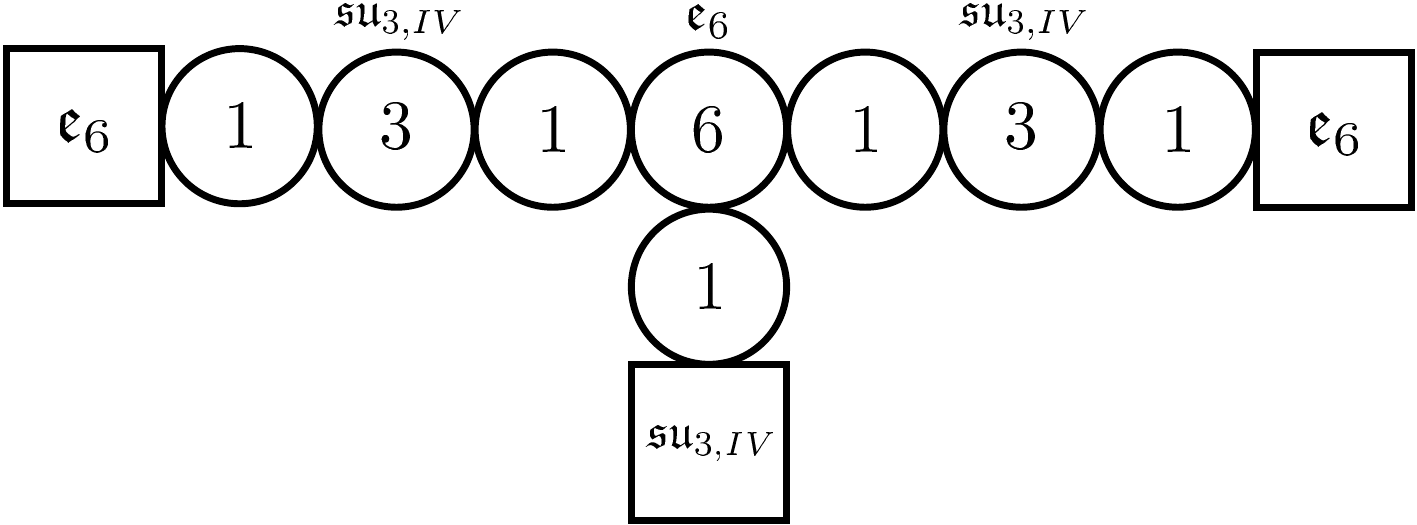} \end{array}
\end{align}
where the flavor symmetry contains one and two $\mathfrak{e}_6$ factors, respectively.

\section{Outlier Theories}
\label{app:outlier}

\subsubsection*{$\mathbb{Z}_2$ outlier theories}
A simple class of outlier theories can be constructed by tuning the $I_1$ locus such that it has a double- or triple-point singularity at the origin. This increases a regular collision to a singularity with vanishing order $\geq(4,6,12)$ in codimension two and hence enhances singularities that would have had perturbative matter to require a blow-up at the origin and hence lead to superconformal matter. This has already been encountered in~\eqref{eq:Su3nZ3} in Section~\ref{sssec:Z3SCM} for $\mathbb{Z}_3$ torsion models. Here, we use the tuning $a_4 = u^n v^m$ and $a_2 =\kappa$, which results in $\mathfrak{su}_{2n}\times \mathfrak{su}_{2m}$ flavor algebras with perturbative bi-fundamental matter at the origin. Setting $a_2 = \kappa_1 u+\kappa_2 v$, the $\mathfrak{su}_{2n}$ and $\mathfrak{su}_{2m}$ factors reduce to $\mathfrak{sp}_{n} \times \mathfrak{sp}_m$ and their intersection is of $\mathfrak{so}$ type i.e.~$(2,3,2(n+m+1))$; this still results in perturbative matter. However, setting $a_2$ to a generic quadratic polynomial, $a_2 = \kappa_1 u^2+\kappa_2 v^2+\kappa_3 u v$, leads back to an $\mathfrak{su}_n$ flavor group, but with a non-minimal collision. The tensor branch is given as
\begin{align}
\begin{array}{c}
\includegraphics[scale=0.7]{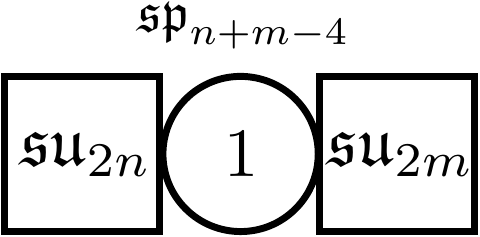}\end{array}
\end{align}
Notably, if $n+m=4$ the $(-1)$ curve is has no gauge algebra and we obtain a discrete holonomy instanton theory. Similarly, if we choose $a_2= \kappa_1 u^3+\kappa_2 v^3 + \kappa_3 u v^2 + \kappa_4 u^2 v$, the flavor factors become $\mathfrak{sp}_n$ factors again, but with a non-minimal collision giving rise to
\begin{align}
\begin{array}{c}
\includegraphics[scale=0.7]{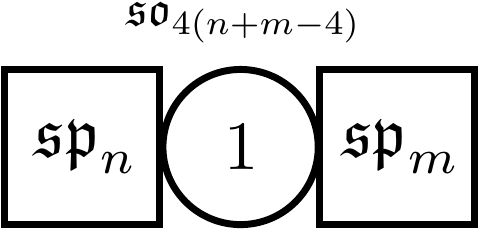}\end{array}
\end{align}
in cases with $n+m \geq7$. For the lower cases we find
\begin{align}
\begin{array}{|c|c|}\hline
n+m & $gauge symmetry$ \\ \hline
\leq 3 & \text{no blowup}\\ 
4 & \emptyset \\ 
5 & \mathfrak{su}_{2,III} \\
6 & \mathfrak{so}_7 \\ \hline
\end{array}
\end{align}  
In the following we continue the systematic enhancement of E-string theories. They all feature an additional monodromy that folds the outer flavor group to a non-simply laced version. 
We start with E-string theory 8 of Table~\eqref{eq:z2Instantons} with flavor group $\text{Sp}(4)/\mathbb{Z}_2$ and enhance it by setting $a_2=u^3$ and $a_4 = v^n$. For $n\in\{4,5,6,7\}$, this leads to the following tensor branches: 
\begin{align}
\begin{array}{c}
\includegraphics[scale=0.7]{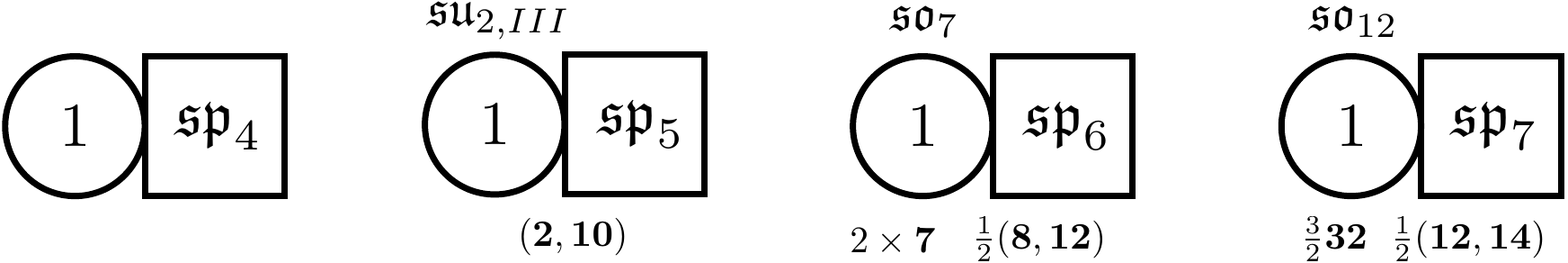} \end{array}
\end{align}
The first case is the discrete holonomy instanton theory with no gauge group. In the $\mathfrak{sp}_5$ case there are bi-fundamentals of the $\mathfrak{su}_2$. Also note that one would expect an additional $\mathfrak{sp}_2$ flavor factor \cite{Bertolini:2015bwa}, which we do not see geometrically realized. There are two singular $I_1$ components thought that intersect the $(-1)$ curve. In the last case, we find bi-fundamental half-hypermultiplets in the $(\mathbf{12,14})$ representation, but in addition we expect three half-hypers in the spinor representation of $\mathfrak{so}_{12}$. These arise from an $\mathfrak{e}_7$ enhancement of multiplicity three at the codimension two locus $u=e_1=0$. Note that this is consistent if one assumes that the gauge group is Spin$(12)/\mathbb{Z}_2$ where the $\mathbb{Z}_2$ is identified with the $\mathbb{Z}_2$ factor in the $\mathbb{Z}_2\times\mathbb{Z}_2$ center of Spin$(12)$ that acts trivially on the spinor representation.
      
The chain above can in fact be increased up to $\mathfrak{sp}_{11}$ factors by increasing $n$ up to $11$. However, at this point, the chain starts introducing multiple gauge factors:
\begin{align}
\begin{array}{c}
\includegraphics[scale=0.7]{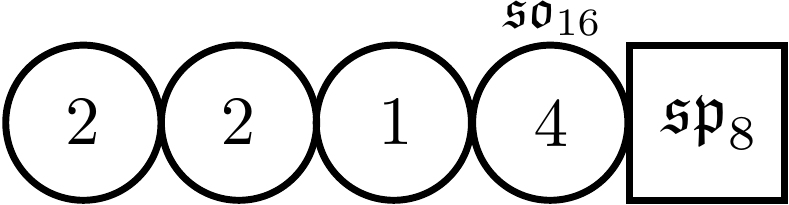} \end{array}
\end{align}
\begin{align}
\begin{array}{c}
\includegraphics[scale=0.7]{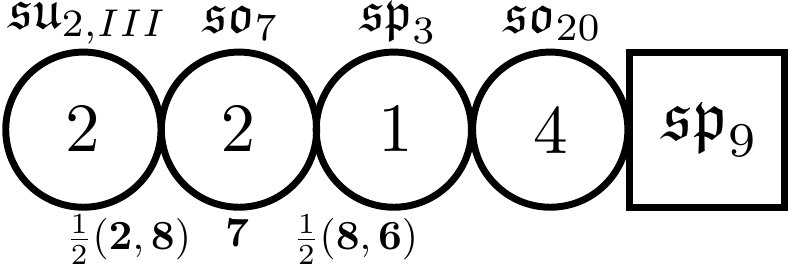} \end{array}
\end{align}
\begin{align}
\begin{array}{c}
\includegraphics[scale=0.7]{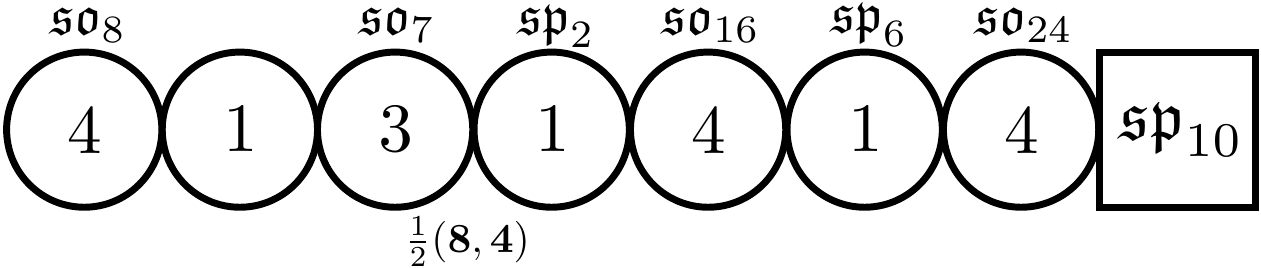} \end{array}
\end{align}
\begin{align}
\begin{array}{c}
\includegraphics[scale=0.7]{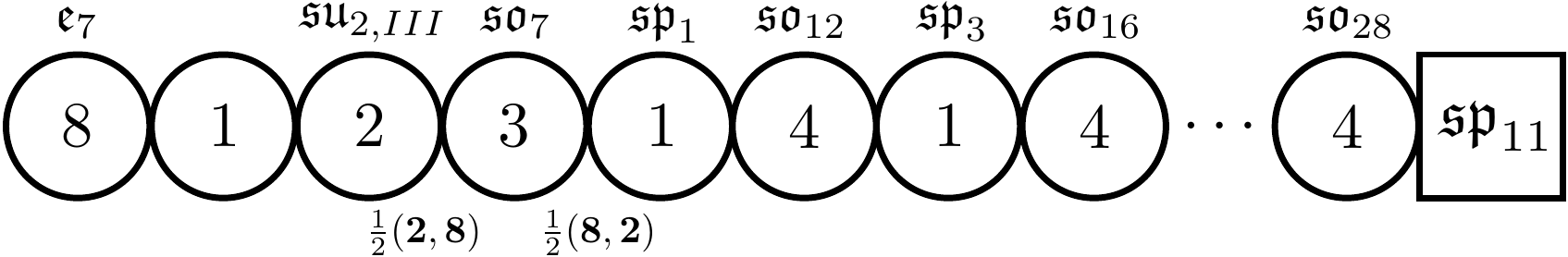} \end{array}
\end{align} 

We continue with the discrete holonomy instanton theory 9 with $[\text{SU}(2)_{III}\times \text{Sp}(3)]/\mathbb{Z}_2$ flavor group and continue enhancing the symplectic factor to an $\mathfrak{sp}_{3+n}$ by setting $a_2 = u^3$ and $a_4 = u v^{3+n}$. At the same time the $I_1$ component in the discriminant, which is of the form $I_1 = u^5+v^{3+2n}$, enhances as well. For $n>6$, one finds a non-crepantly resolvable singularity. Each chain is slightly different, given by the following tensor branches
\begin{align}
\begin{array}{c}
\includegraphics[scale=0.7]{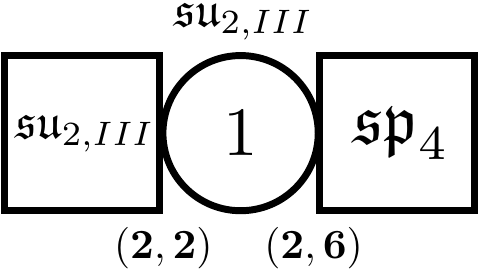}\end{array}
\end{align}
\begin{align}
\begin{array}{c}
\includegraphics[scale=0.7]{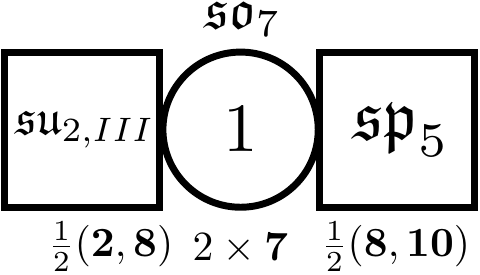}\end{array}
\end{align}
\begin{align}
\begin{array}{c}
\includegraphics[scale=0.7]{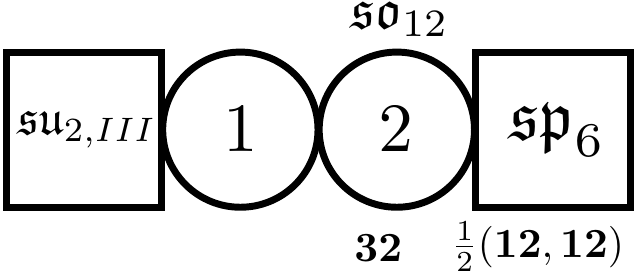}\end{array}
\end{align}
\begin{align}
\begin{array}{c}
\includegraphics[scale=0.7]{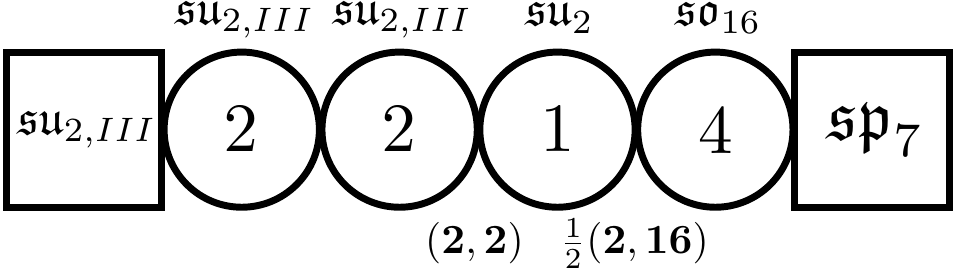}\end{array}
\end{align}
\begin{align}
\begin{array}{c}
\includegraphics[scale=0.7]{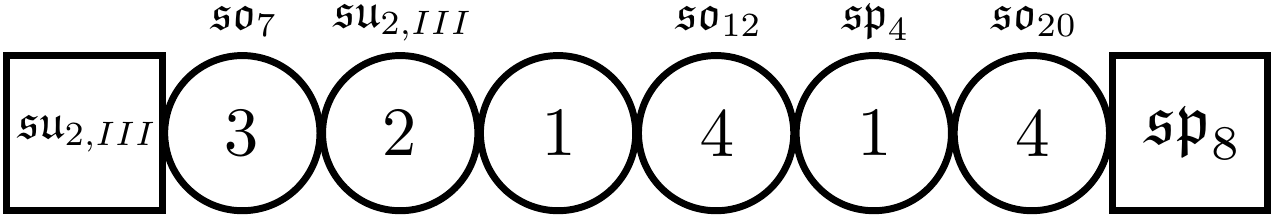}\end{array}
\end{align}
\begin{align}
\begin{array}{c}
\includegraphics[scale=0.5]{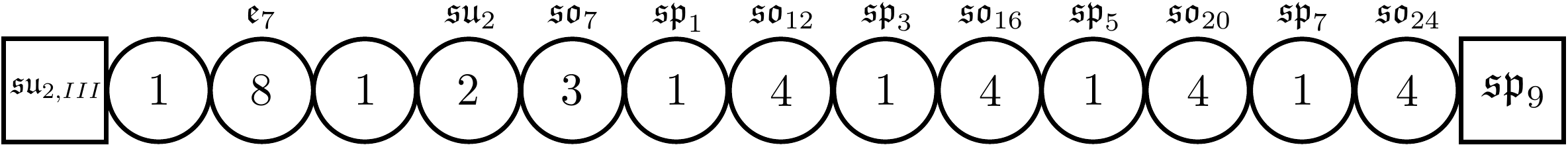}\end{array}
\end{align}

Next, we reconsider the discrete holonomy instanton theory 1 with $[$E$_7 \times $SU$(2)]/\mathbb{Z}_2$ flavor symmetry. For those theories we can enhance the $\mathfrak{su}_{2}$ flavor factor further to $\mathfrak{sp}_n$ by setting $a_2 = u ^3$ and $a_4= u^3 v^n$ with $n \leq 5$ to avoid non-minimal singularities,
\begin{align}
\begin{array}{c}
\includegraphics[scale=0.7]{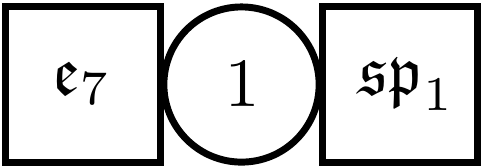}\end{array}
\end{align}
\begin{align}
\begin{array}{c}
\includegraphics[scale=0.7]{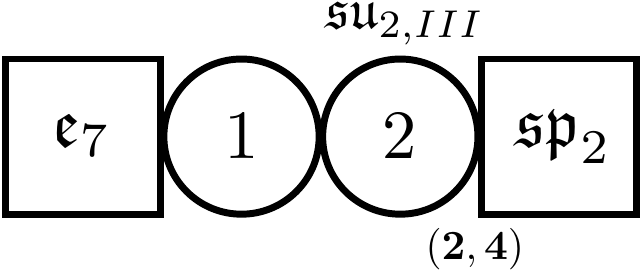}\end{array}
\end{align}
\begin{align}
\begin{array}{c}
\includegraphics[scale=0.7]{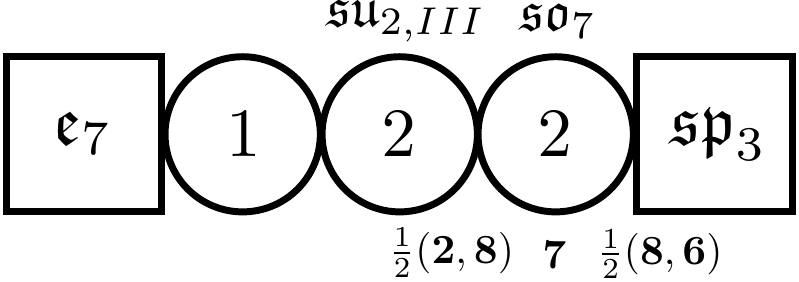}\end{array}
\end{align}
\begin{align}
\begin{array}{c}
\includegraphics[scale=0.7]{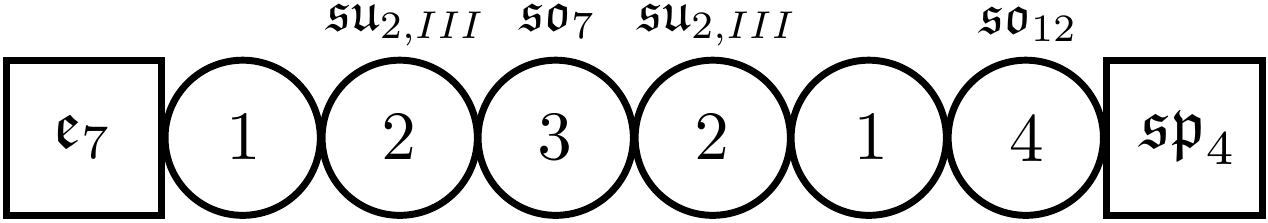}\end{array}
\end{align}
\begin{align}
\begin{array}{c}
\includegraphics[scale=0.5]{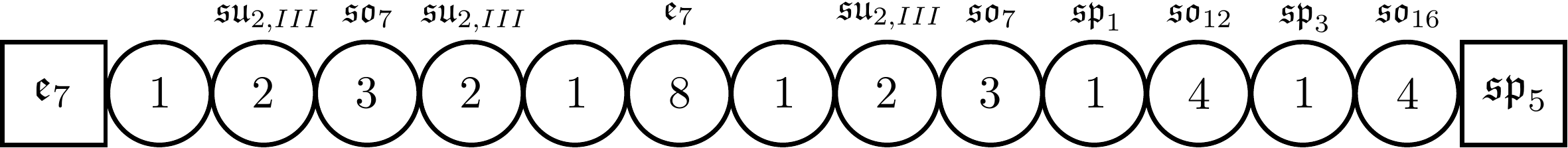}\end{array}
\end{align}

Let us next illustrate the importance of the exact form of the $I_1$ locus for the tensor branch. For that consider the family of models with tuning  $a_2 = u ^{3+2n}$ and $a_4= u^3 v^3$ with $n=0,1,2$. Those three theories all have an $\mathfrak{e}_7$ collision with an $\mathfrak{sp}_3$ flavor brane, but all lead to different tensor branches:
\begin{align}
\begin{array}{c}
\includegraphics[scale=0.7]{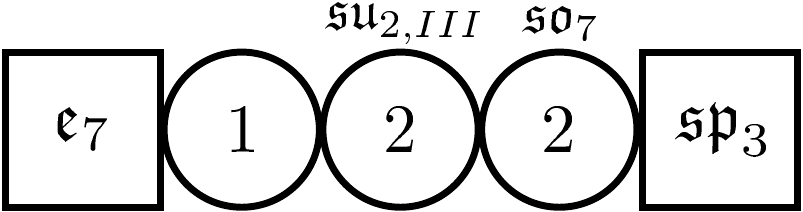} \end{array}
\end{align}
\begin{align}
\begin{array}{c}
\includegraphics[scale=0.7]{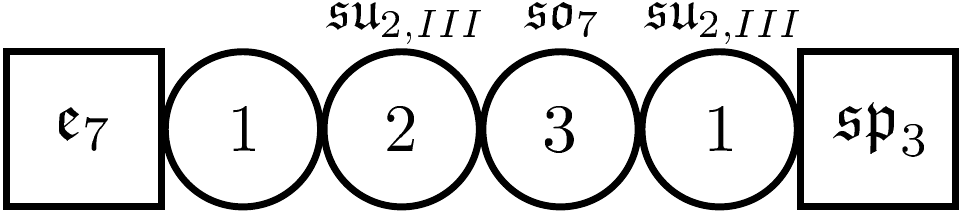} \end{array}
\end{align}
\begin{align}
\begin{array}{c}
\includegraphics[scale=0.7]{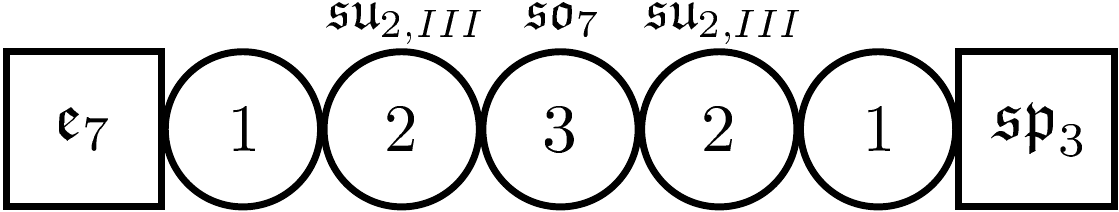} \end{array}
\end{align}

Finally, we consider the discrete holonomy instanton theory 10 with $[\text{Spin}(8) \times \text{Sp}(2)]/\mathbb{Z}_2$ flavor group, given by the factorization $a_2 = u^3$ and $a_4 = u^2 v^2$, and enhance the $\text{Sp}(2)$ side to $\text{Sp}(n)$ by setting $a_4 = u^2 v^{n}$. This chain is bounded by $n \leq 7$ in order to admit crepant resolutions. The flavor algebra further depends on whether $n$ is even or odd. For $n$ even, the flavor $\mathfrak{so}_7$ algebra enhances to an $\mathfrak{so}_8$ factor, leading to the chains
\begin{align}
\begin{array}{c}
\includegraphics[scale=0.7]{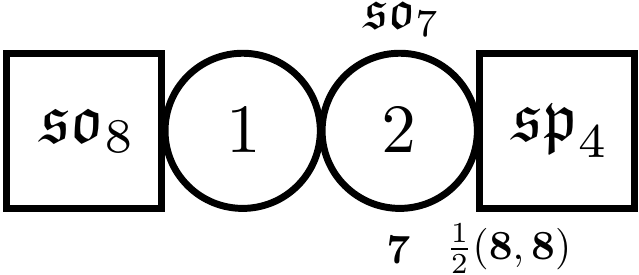}\end{array}
\end{align}
\begin{align}
\begin{array}{c}
\includegraphics[scale=0.7]{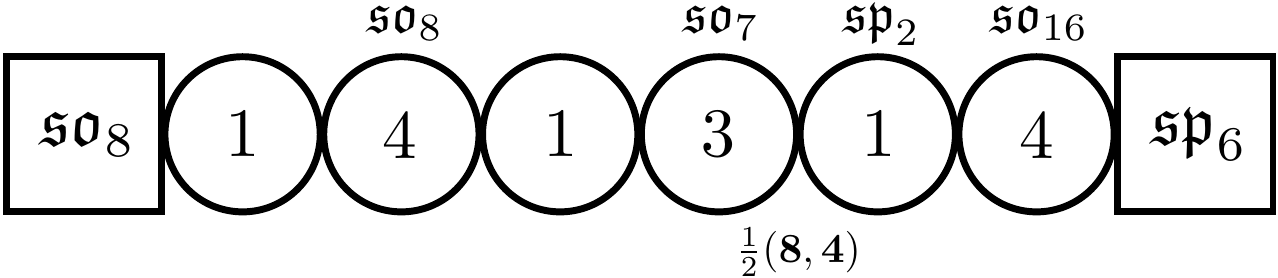}\end{array}
\end{align}

For $n$ odd, we get the three theories
\begin{align}
\begin{array}{c}
\includegraphics[scale=0.7]{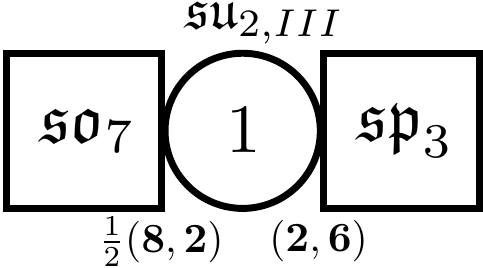}\end{array}
\end{align}
\begin{align}
\begin{array}{c}
\includegraphics[scale=0.7]{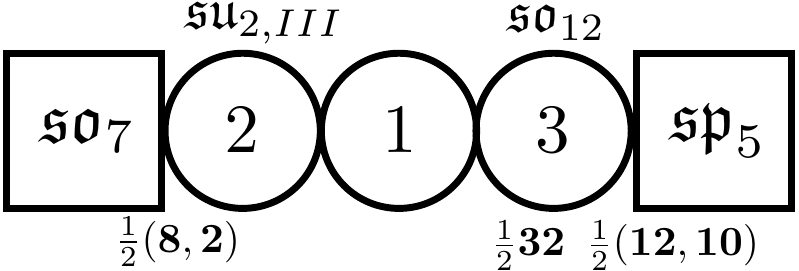}\end{array}
\end{align}
\begin{align}
\begin{array}{c}
\includegraphics[scale=0.5]{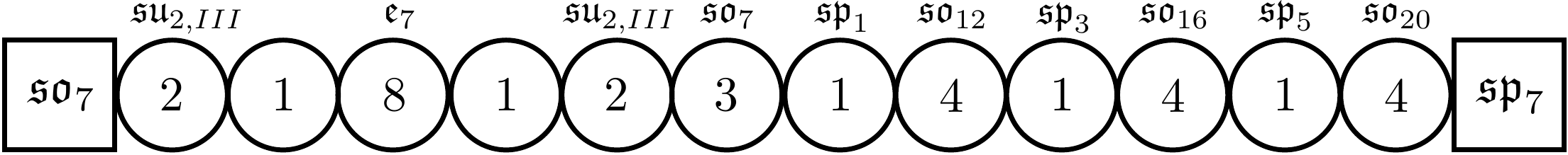}\end{array}
\end{align}

\subsection*{$\mathbb{Z}_3$ outlier theories}
In this section we study the factorization of an $\mathfrak{su}_{3n}$ flavor algebra into three different pieces. This is enforced by setting
\begin{align} 
 a_1 \rightarrow u\,,~a_3 \rightarrow (u+v)^k (u-v)^l (c_1 u+c_2 v)^m\,. 
\end{align} 
The model is an $[\text{SU}(3k)\times \text{SU}(3l) \times \text{SU}(3m)] /\mathbb{Z}_3$ theory with all groups intersecting at the origin. After blowing up in the base, we get an $\mathfrak{su}_{3(k+l+m-3)}$ gauge symmetry over the $(-1)$ curve and bi-fundamental matter at the intersections with the three flavor branes.
\begin{align}
\begin{array}{c}
\includegraphics[scale=0.7]{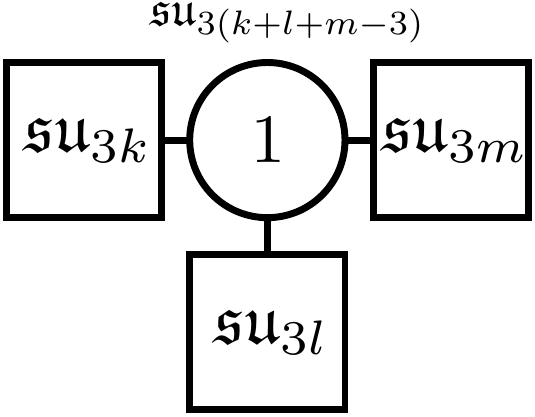} \end{array}
\end{align} 
In case that $k+l+m=2$, there is an $\mathfrak{su}_6$ gauge algebra found on the $(-1)$ curve that introduces an additional half-hyper in the triple anti-symmetric representation.
If $k+l+m>3$, we need to perform additional blow-ups until we end on a $(-1)$ curve with an $\mathfrak{su}_{3 (k+l+m \text{ mod } 3)}$ factor: 
\begin{align}
\label{eq:SUOutlierZ3}
\begin{array}{c}
\includegraphics[scale=0.7]{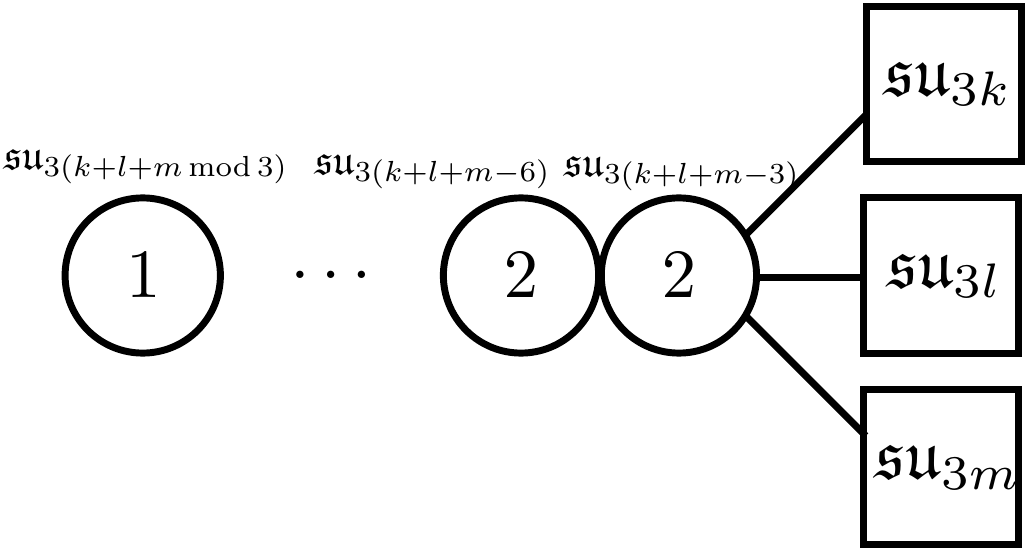} \end{array}
\end{align} 
As before, the above theory has a similar tensor branch structure as e.g.~\eqref{eq:Su3nZ3} and~\eqref{eq:SU_SU_SU_Z3}. Therefore, we might expect that theory~\eqref{eq:SUOutlierZ3} can be understood as a deformation of these theories.

\subsection*{$\mathbb{Z}_4$  outlier theories} 
In this section we consider theories of type $[\text{Spin}(10+4n) \times \text{SU}(4)]/\mathbb{Z}_4$ and enhance the $\mathfrak{su}_4$ to $\mathfrak{su}_8$ via the factorization $a_1 \rightarrow u^{1+n}\,,~  a_2 \rightarrow u v^2  $. This modifies the tensor branch to
\begin{align}
\begin{array}{c}
\includegraphics[scale=0.7]{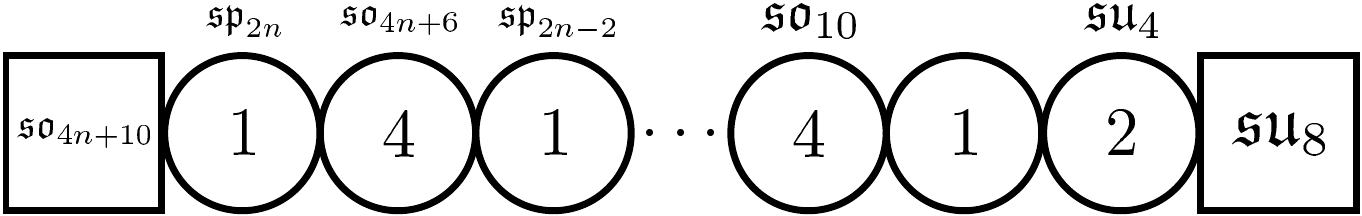} \end{array}
\end{align} 

Note that if $n>1$, one cannot enhance the $\mathfrak{su}_4$ higher than $\mathfrak{su}_8$, since one would encounters $(8,12,24)$ singularities in codimension two.
The structure of the tensor branch includes the superconformal matter theories of $\mathfrak{so}_{10+4m}\times \mathfrak{so}_{10+4n}$ type, as well as the discrete holonomy instanton theory of type $[\text{Spin}(10)\times \text{SU}(4)]/\mathbb{Z}_4$ at the end. Note that the above theory can also be enhanced to $[\text{Spin}(10) \times \text{SU}(12) ]/\mathbb{Z}_4$ (but not further) by setting $a_1 = u^1$ and $a_2 = u v^3$, with the tensor branch
\begin{align}
\begin{array}{c}
\includegraphics[scale=0.7]{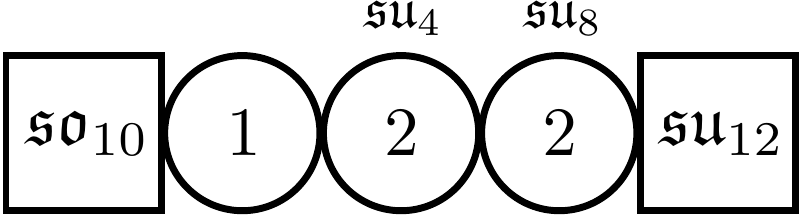} \end{array}
\end{align} 
Finally, we can turn the $\mathfrak{so}_{10}$ flavor algebra into $\mathfrak{sp}_2$ by setting $a_1 = u^2$ and $a_2=v^2$. The tensor branch then reads
 \begin{align}
 \begin{array}{c}
 \includegraphics[scale=0.7]{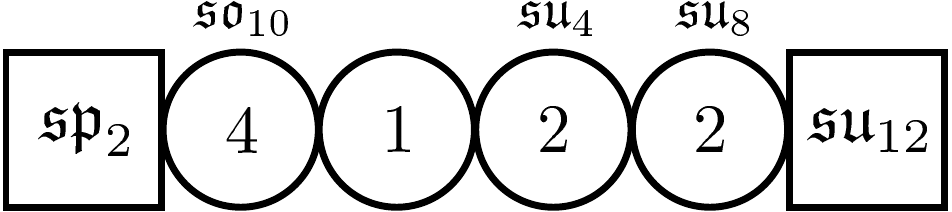} \end{array}
 \end{align} 
Any further enhancement on either the $\mathfrak{sp}_2$ or $\mathfrak{su}_{12}$ side would lead to non-crepantly resolvable singularities.

\subsection*{$\mathbb{Z}_2 \times \mathbb{Z}_2 $ outlier theories} 

 We start by enhancing $\mathfrak{so}_{12}$ to $\mathfrak{so}_{16}$ in the discrete holonomy instanton theory 2 by setting $b_1 = u v$ and $c_1=v^3$. This leads to the tensor branch
\begin{align}
\begin{array}{c}
\includegraphics[scale=0.7]{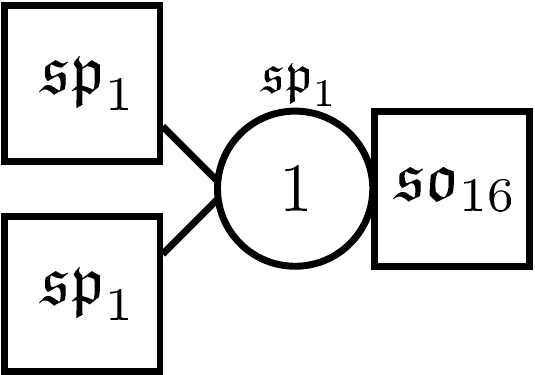} \end{array} 
\end{align}
When one of the $\mathfrak{sp}$ flavor factor is further enhanced, we obtain
\begin{align}
\begin{array}{c}
\includegraphics[scale=0.7]{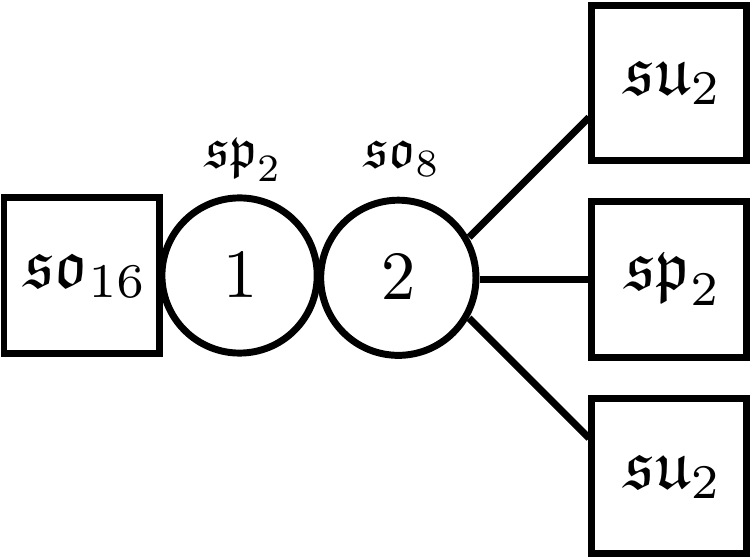} \end{array} 
\end{align}
as well as
\begin{align}
\begin{array}{c}
\includegraphics[scale=0.7]{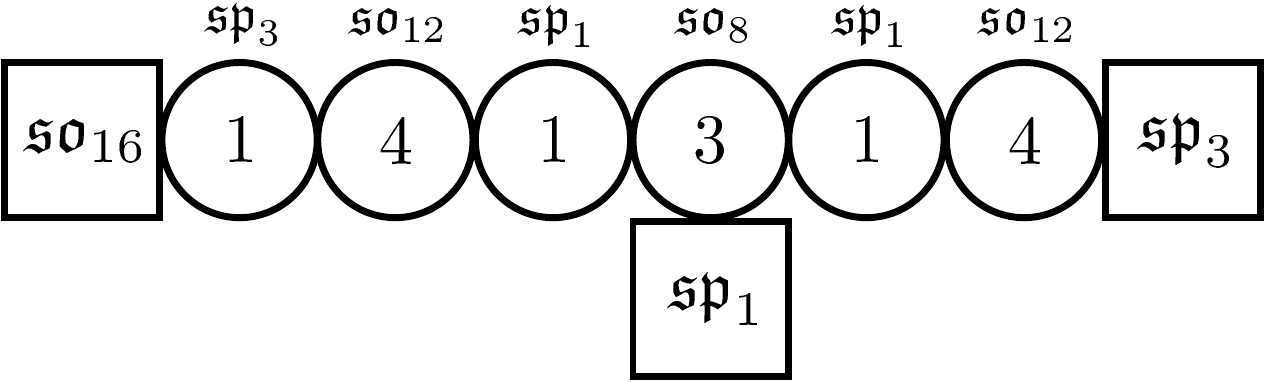} \end{array} 
\end{align}
For tunings of the type $b_2 =u^n$ and $c_2 = v^m$ there are a couple of models that have not yet been discussed for values $n+m<9$ and $n>2$. In the following, we fix $n=3$. For $m=3$, we find
\begin{align}
\begin{array}{c}
\includegraphics[scale=0.7]{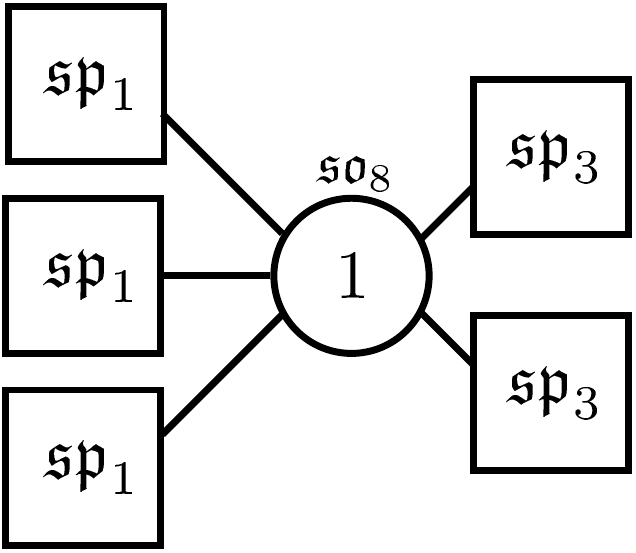} \end{array} 
\end{align}
The model has the special property that its $\mathfrak{sp}_1$ locus is of the form $z=u^3 - v^3$, i.e.\ it is reducible and can be reduced into three different pieces. Upon blow-up, all components intersect the $(-1)$ curve that hosts the $\mathfrak{so}_8$ at different points. This contributes the three spinor, vector and co-spinor representations required from anomaly cancellation. The flavor symmetry is the same as the one found in~\cite{Bertolini:2015bwa}.

For $m =4,5$ we obtain the following two theories
\begin{align}
\begin{array}{c}
\includegraphics[scale=0.7]{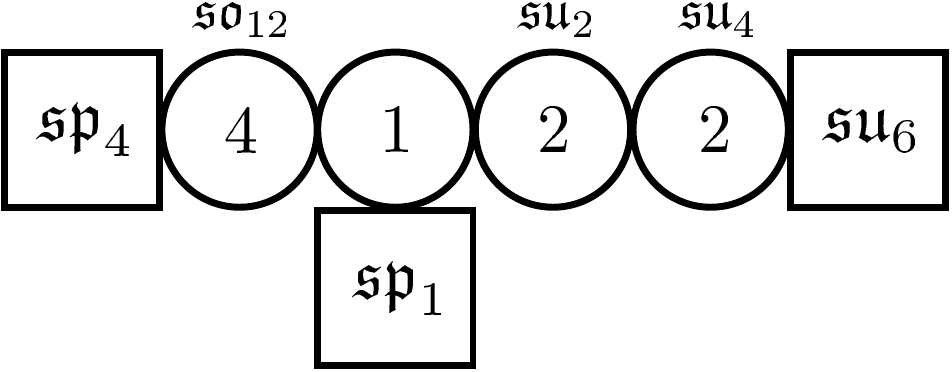} \end{array} 
\end{align}
and
\begin{align}
\begin{array}{c}
\includegraphics[scale=0.7]{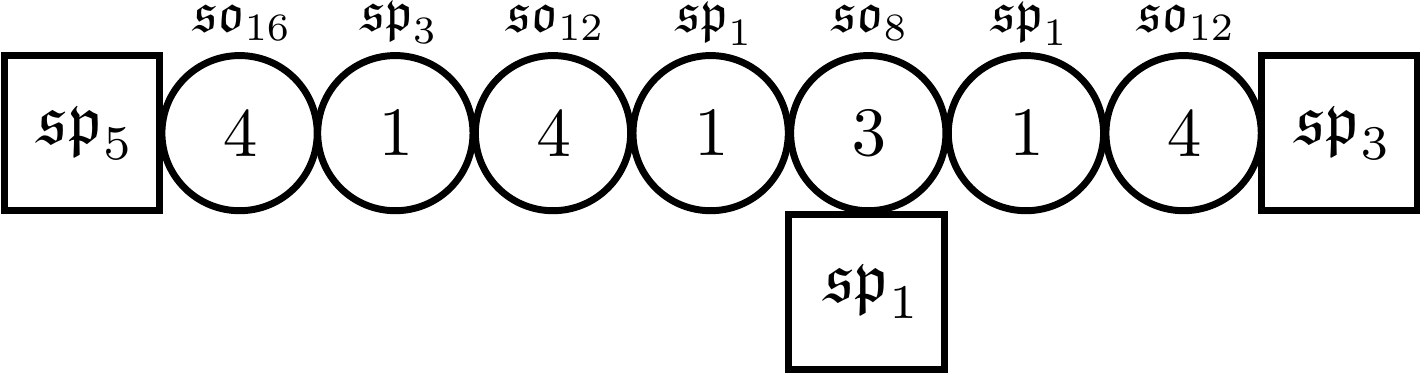} \end{array} 
\end{align}

\subsection*{$\mathbb{Z}_2 \times \mathbb{Z}_4$ outlier theories}
The outlier theories we consider here start from the $[\text{SU}(4)^2 \times \text{SU}(2)^2]/[\mathbb{Z}_2 \times \mathbb{Z}_4]$ theory. We enhance the flavor $\mathfrak{su}_{2n}$ algebra to $\mathfrak{su}_{2n} \times \mathfrak{su}_{2m}$ by setting
\begin{align}
a_1 \rightarrow (u+  v)^n (  u-v)^m \, , \quad b_1 \rightarrow v \, .
\end{align}
Hence, there are five flavor factors in total, all intersecting at the origin. The resulting tensor branch is
\begin{align}
\begin{array}{c}
\includegraphics[scale=0.7]{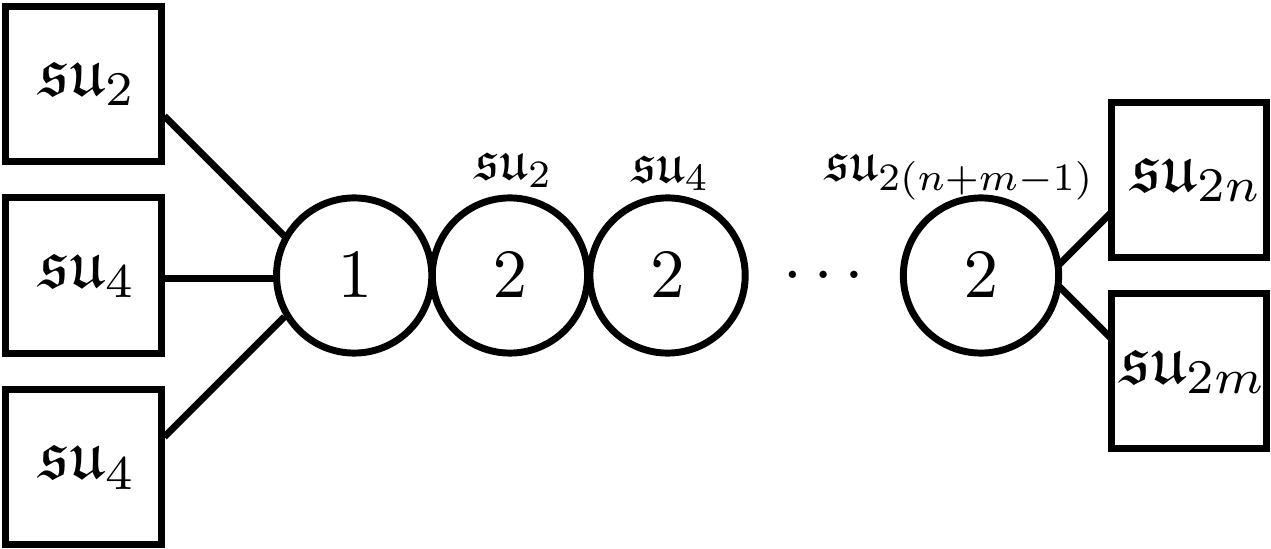} \end{array}  
\end{align}
A similar factorization can be done for the $\mathfrak{su}_4$ part, leading to a similar chain but with jumps by multiples of four in the rank of the gauge group on the tensor branch. Again, we note that the above tuning looks very similar to the theories constructed from an $[\text{SU}(4)^2 \times \text{SU}(2) \times \text{SU}(2k)]/[\mathbb{Z}_2 \times \mathbb{Z}_4]$ theory with $k=n+m$ as shown in \eqref{eq:SU2n_Z4Z2}. Hence, its is conceivable that these theories arise from a deformation of the theory with $\mathfrak{su}_{2(n+m)}$ flavor algebra.
 
\subsection*{$\mathbb{Z}_3 \times \mathbb{Z}_3$ outlier theories} 
The only simple outlier theories we can construct here come from further factorizing the components of the $\mathfrak{su_{3n}}$ loci as
\begin{align}
a_1 = u \, , \quad  b_1 = (u-   v)^m (u- \omega v)^n (u- \omega^2 v)^k (u+ v)^l \,,
\end{align}
which results in splitting the toric $\mathfrak{su}_{3(m+n+k+l)}$ locus into four individual blocks that all intersect at the origin. This is reflected in the tensor branch, where all four loci intersect the first resolution divisor, while the other three $\mathfrak{su}_3$ factors sit at the final $(-1)$ curve:
\begin{align}
\begin{array}{c}
\includegraphics[scale=0.7]{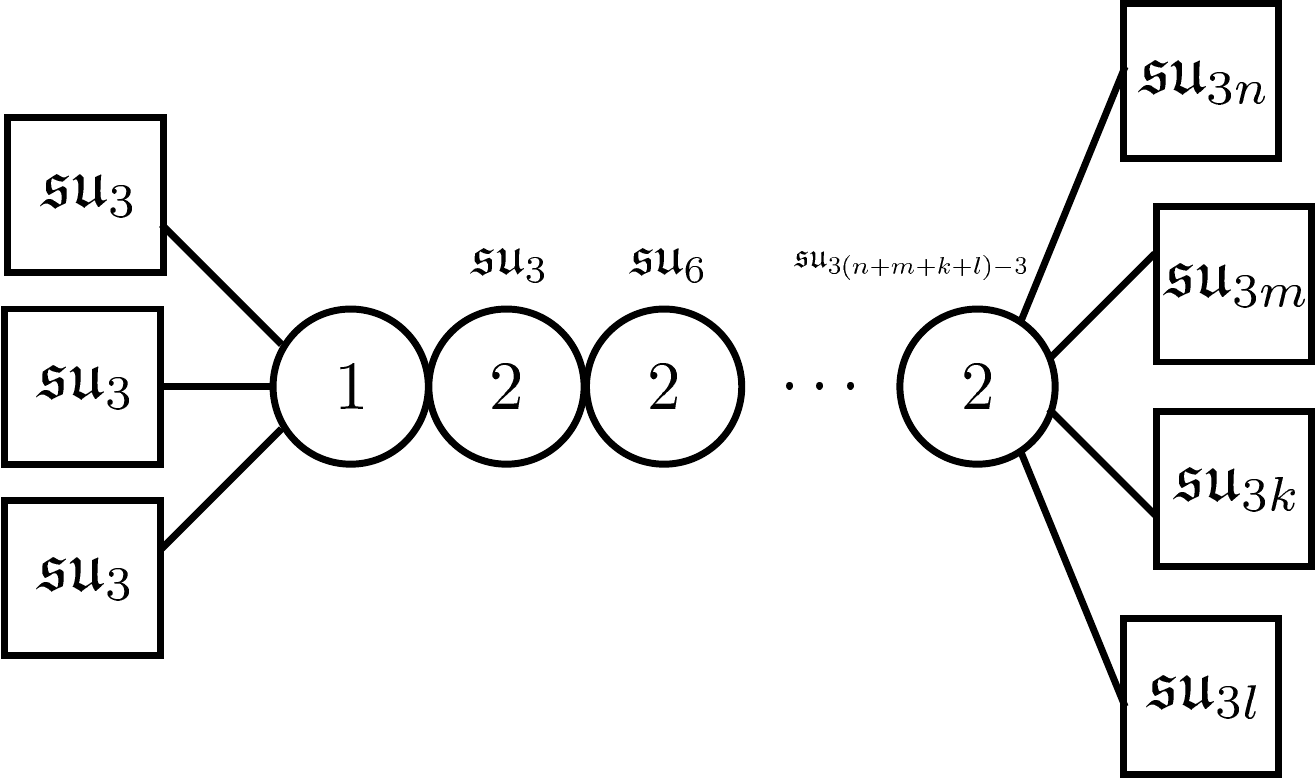} \end{array} 
\end{align}
The picture is very similar to the cases we had e.g.\ in the $\mathbb{Z}_2 \times \mathbb{Z}_4$ case in the section before. Hence also in this case we might expect the four flavor factors on the right to arise as a deformation of an $\mathfrak{su}_{3(n+m+l+l)}$ theory.

\section{Higher Order Torsion}
\label{sec:ExoticTorsion}
Throughout the paper we have worked with Weierstrass models whose torsion groups can appear in global models. However, it is also possible to construct higher order torsion models which are only consistent with non-compact 3-fold geometries. We construct one such quiver-like gauge theory with higher torsion that can flow to a non-trivial SCFT in the IR. For this we use a $\mathbb{Z}_7$ model over $\mathbb{F}_0 \equiv \mathbb{P}^1_s \times \mathbb{P}^1_t$ \cite{Hajouji:2019vxs}, given by
\begin{align}
 y^2& + (s_0 s_1 t_0 t_1 + (s_0 t_0 - s_1 t_1) (2 s_1 t_1-s_0t_0)) xy + s_0 s_1^3 t_0 t_1^3 (s_0 t_0 - 2 s_1 t_1) (s_1 t_1 - s_0 t_0) y \, \nonumber \\
    &= x^3 + s_1^2 t_1^2 (s_0 t_0 - 2 s_1 t_1) (s_1 t_1 - s_0 t_0)x^2  \, .
\end{align}
The model has two $(8,12,24)$ singularities over the loci $s_0 =t_1 =0$ and $s_1 = t_0 =0$. However, we ignore these points for now, decompactify the two $\mathbb{P}^1$ factors to $\mathbb{C}^2$, and set the coordinates $s_1$ and $t_1$ to one. The discriminant of the Weierstrass model then becomes
\begin{align}
 \Delta=\tfrac16 s_0^7 t_0^7 ( s_0 t_0-1)^7 (1 - 8 s_0 t_0 + 5 s_0^2 t_0^2 + s_0^3 t_0^3) \, .
\end{align}
Note that the $(8,12,24)$ singularities have been pushed off to infinity and the resulting model is crepantly resolvable. Moreover, since the local coordinates always appear in the combination $s_0 t_0$, we can add to this model a $\mathbb{Z}_n$ singularity at the origin via the action
\begin{align}
(s_0,~t_0)~\to~(e^{2\pi i/n}s_0,~e^{-2\pi i/n}t_0)\,.
\end{align}
Resolving the geometry with a chain of $n-1$ exceptional divisors $\{e_i =0\}$ leads to the discriminant
\begin{align}
\Delta=\tfrac16  e^7 s_0^7 t_0^7 \big(s_0 t_0 e-1\big)^7 \big(1- 8 s_0 t_0 e  + 5 s_0^2 t_0^2 e^2 + s_0^3 t_0^3 e^3\big) \, .
\end{align}
with $e=\prod_i^{n-1}e_i$. The tensor branch is given by 
\begin{align}
\begin{array}{c}
\includegraphics[scale=0.7]{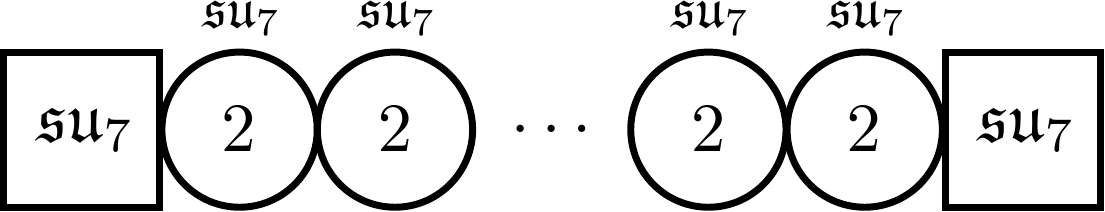}\end{array}
\end{align}
with bi-fundamental matter between adjacent symmetry group factors.

\bibliographystyle{JHEP}
\bibliography{biblio}

\providecommand{\href}[2]{#2}\begingroup\raggedright\begin{thebibliography}{10}

\bibitem{Nahm:1977tg}
W.~Nahm, \emph{{Supersymmetries and their Representations}},
  \href{https://doi.org/10.1016/0550-3213(78)90218-3}{\emph{Nucl. Phys. B}
  {\bfseries 135} (1978) 149}.

\bibitem{Witten:1995gx}
E.~Witten, \emph{{Small instantons in string theory}},
  \href{https://doi.org/10.1016/0550-3213(95)00625-7}{\emph{Nucl. Phys. B}
  {\bfseries 460} (1996) 541--559},
  [\href{https://arxiv.org/abs/hep-th/9511030}{{\ttfamily hep-th/9511030}}].

\bibitem{Witten:1995ex}
E.~Witten, \emph{{String theory dynamics in various dimensions}},
  \href{https://doi.org/10.1016/0550-3213(95)00158-O}{\emph{Nucl. Phys. B}
  {\bfseries 443} (1995) 85--126},
  [\href{https://arxiv.org/abs/hep-th/9503124}{{\ttfamily hep-th/9503124}}].

\bibitem{Strominger:1995ac}
A.~Strominger, \emph{{Open p-branes}},
  \href{https://doi.org/10.1016/0370-2693(96)00712-5}{\emph{Phys. Lett. B}
  {\bfseries 383} (1996) 44--47},
  [\href{https://arxiv.org/abs/hep-th/9512059}{{\ttfamily hep-th/9512059}}].

\bibitem{Seiberg:1996vs}
N.~Seiberg and E.~Witten, \emph{{Comments on string dynamics in
  six-dimensions}},
  \href{https://doi.org/10.1016/0550-3213(96)00189-7}{\emph{Nucl. Phys. B}
  {\bfseries 471} (1996) 121--134},
  [\href{https://arxiv.org/abs/hep-th/9603003}{{\ttfamily hep-th/9603003}}].

\bibitem{Ganor:1996mu}
O.~J. Ganor and A.~Hanany, \emph{{Small E(8) instantons and tensionless
  noncritical strings}},
  \href{https://doi.org/10.1016/0550-3213(96)00243-X}{\emph{Nucl. Phys. B}
  {\bfseries 474} (1996) 122--140},
  [\href{https://arxiv.org/abs/hep-th/9602120}{{\ttfamily hep-th/9602120}}].

\bibitem{Ohmori:2015pua}
K.~Ohmori, H.~Shimizu, Y.~Tachikawa and K.~Yonekura, \emph{{6d
  $\mathcal{N}=(1,0)$ theories on $T^2$ and class S theories: Part I}},
  \href{https://doi.org/10.1007/JHEP07(2015)014}{\emph{JHEP} {\bfseries 07}
  (2015) 014}, [\href{https://arxiv.org/abs/1503.06217}{{\ttfamily
  1503.06217}}].

\bibitem{Ohmori:2015pia}
K.~Ohmori, H.~Shimizu, Y.~Tachikawa and K.~Yonekura, \emph{{6d
  $\mathcal{N}=\left(1,\;0\right) $ theories on S$^{1}$ /T$^{2}$ and class S
  theories: part II}},
  \href{https://doi.org/10.1007/JHEP12(2015)131}{\emph{JHEP} {\bfseries 12}
  (2015) 131}, [\href{https://arxiv.org/abs/1508.00915}{{\ttfamily
  1508.00915}}].

\bibitem{DelZotto:2015rca}
M.~Del~Zotto, C.~Vafa and D.~Xie, \emph{{Geometric engineering, mirror symmetry
  and $ 6{\mathrm{d}}_{\left(1,0\right)}\to
  4{\mathrm{d}}_{\left(\mathcal{N}=2\right)} $}},
  \href{https://doi.org/10.1007/JHEP11(2015)123}{\emph{JHEP} {\bfseries 11}
  (2015) 123}, [\href{https://arxiv.org/abs/1504.08348}{{\ttfamily
  1504.08348}}].

\bibitem{Morrison:2016nrt}
D.~R. Morrison and C.~Vafa, \emph{{F-theory and $ \mathcal{N} $ = 1 SCFTs in
  four dimensions}}, \href{https://doi.org/10.1007/JHEP08(2016)070}{\emph{JHEP}
  {\bfseries 08} (2016) 070},
  [\href{https://arxiv.org/abs/1604.03560}{{\ttfamily 1604.03560}}].

\bibitem{Razamat:2016dpl}
S.~S. Razamat, C.~Vafa and G.~Zafrir, \emph{{4d $ \mathcal{N}=1 $ from 6d (1,
  0)}}, \href{https://doi.org/10.1007/JHEP04(2017)064}{\emph{JHEP} {\bfseries
  04} (2017) 064}, [\href{https://arxiv.org/abs/1610.09178}{{\ttfamily
  1610.09178}}].

\bibitem{Apruzzi:2016nfr}
F.~Apruzzi, F.~Hassler, J.~J. Heckman and I.~V. Melnikov, \emph{{From 6D SCFTs
  to Dynamic GLSMs}},
  \href{https://doi.org/10.1103/PhysRevD.96.066015}{\emph{Phys. Rev. D}
  {\bfseries 96} (2017) 066015},
  [\href{https://arxiv.org/abs/1610.00718}{{\ttfamily 1610.00718}}].

\bibitem{Bah:2017gph}
I.~Bah, A.~Hanany, K.~Maruyoshi, S.~S. Razamat, Y.~Tachikawa and G.~Zafrir,
  \emph{{4d $ \mathcal{N}=1 $ from 6d $ \mathcal{N}=\left(1,0\right) $ on a
  torus with fluxes}},
  \href{https://doi.org/10.1007/JHEP06(2017)022}{\emph{JHEP} {\bfseries 06}
  (2017) 022}, [\href{https://arxiv.org/abs/1702.04740}{{\ttfamily
  1702.04740}}].

\bibitem{DelZotto:2017pti}
M.~Del~Zotto, J.~J. Heckman and D.~R. Morrison, \emph{{6D SCFTs and Phases of
  5D Theories}}, \href{https://doi.org/10.1007/JHEP09(2017)147}{\emph{JHEP}
  {\bfseries 09} (2017) 147},
  [\href{https://arxiv.org/abs/1703.02981}{{\ttfamily 1703.02981}}].

\bibitem{Apruzzi:2018oge}
F.~Apruzzi, J.~J. Heckman, D.~R. Morrison and L.~Tizzano, \emph{{4D Gauge
  Theories with Conformal Matter}},
  \href{https://doi.org/10.1007/JHEP09(2018)088}{\emph{JHEP} {\bfseries 09}
  (2018) 088}, [\href{https://arxiv.org/abs/1803.00582}{{\ttfamily
  1803.00582}}].

\bibitem{Bhardwaj:2018yhy}
L.~Bhardwaj and P.~Jefferson, \emph{{Classifying 5d SCFTs via 6d SCFTs: Rank
  one}}, \href{https://doi.org/10.1007/JHEP07(2019)178}{\emph{JHEP} {\bfseries
  07} (2019) 178}, [\href{https://arxiv.org/abs/1809.01650}{{\ttfamily
  1809.01650}}].

\bibitem{Bhardwaj:2018vuu}
L.~Bhardwaj and P.~Jefferson, \emph{{Classifying 5d SCFTs via 6d SCFTs:
  Arbitrary rank}}, \href{https://doi.org/10.1007/JHEP10(2019)282}{\emph{JHEP}
  {\bfseries 10} (2019) 282},
  [\href{https://arxiv.org/abs/1811.10616}{{\ttfamily 1811.10616}}].

\bibitem{Bhardwaj:2019fzv}
L.~Bhardwaj, P.~Jefferson, H.-C. Kim, H.-C. Tarazi and C.~Vafa, \emph{{Twisted
  Circle Compactifications of 6d SCFTs}},
  \href{https://arxiv.org/abs/1909.11666}{{\ttfamily 1909.11666}}.

\bibitem{Vafa:1996xn}
C.~Vafa, \emph{{Evidence for F theory}},
  \href{https://doi.org/10.1016/0550-3213(96)00172-1}{\emph{Nucl. Phys.}
  {\bfseries B469} (1996) 403--418},
  [\href{https://arxiv.org/abs/hep-th/9602022}{{\ttfamily hep-th/9602022}}].

\bibitem{Morrison:1996na}
D.~R. Morrison and C.~Vafa, \emph{{Compactifications of F theory on Calabi-Yau
  threefolds. 1}},
  \href{https://doi.org/10.1016/0550-3213(96)00242-8}{\emph{Nucl. Phys. B}
  {\bfseries 473} (1996) 74--92},
  [\href{https://arxiv.org/abs/hep-th/9602114}{{\ttfamily hep-th/9602114}}].

\bibitem{Morrison:1996pp}
D.~R. Morrison and C.~Vafa, \emph{{Compactifications of F theory on Calabi-Yau
  threefolds. 2.}},
  \href{https://doi.org/10.1016/0550-3213(96)00369-0}{\emph{Nucl. Phys. B}
  {\bfseries 476} (1996) 437--469},
  [\href{https://arxiv.org/abs/hep-th/9603161}{{\ttfamily hep-th/9603161}}].

\bibitem{Heckman:2013pva}
J.~J. Heckman, D.~R. Morrison and C.~Vafa, \emph{{On the Classification of 6D
  SCFTs and Generalized ADE Orbifolds}},
  \href{https://doi.org/10.1007/JHEP06(2015)017,
  10.1007/JHEP05(2014)028}{\emph{JHEP} {\bfseries 05} (2014) 028},
  [\href{https://arxiv.org/abs/1312.5746}{{\ttfamily 1312.5746}}].

\bibitem{Heckman:2015bfa}
J.~J. Heckman, D.~R. Morrison, T.~Rudelius and C.~Vafa, \emph{{Atomic
  Classification of 6D SCFTs}},
  \href{https://doi.org/10.1002/prop.201500024}{\emph{Fortsch. Phys.}
  {\bfseries 63} (2015) 468--530},
  [\href{https://arxiv.org/abs/1502.05405}{{\ttfamily 1502.05405}}].

\bibitem{Bhardwaj:2015oru}
L.~Bhardwaj, M.~Del~Zotto, J.~J. Heckman, D.~R. Morrison, T.~Rudelius and
  C.~Vafa, \emph{{F-theory and the Classification of Little Strings}},
  \href{https://doi.org/10.1103/PhysRevD.93.086002}{\emph{Phys. Rev. D}
  {\bfseries 93} (2016) 086002},
  [\href{https://arxiv.org/abs/1511.05565}{{\ttfamily 1511.05565}}].

\bibitem{Bertolini:2015bwa}
M.~Bertolini, P.~R. Merkx and D.~R. Morrison, \emph{{On the global symmetries
  of 6D superconformal field theories}},
  \href{https://doi.org/10.1007/JHEP07(2016)005}{\emph{JHEP} {\bfseries 07}
  (2016) 005}, [\href{https://arxiv.org/abs/1510.08056}{{\ttfamily
  1510.08056}}].

\bibitem{Bhardwaj:2015xxa}
L.~Bhardwaj, \emph{{Classification of 6d $ \mathcal{N}=\left(1,0\right) $ gauge
  theories}}, \href{https://doi.org/10.1007/JHEP11(2015)002}{\emph{JHEP}
  {\bfseries 11} (2015) 002},
  [\href{https://arxiv.org/abs/1502.06594}{{\ttfamily 1502.06594}}].

\bibitem{Bhardwaj:2018jgp}
L.~Bhardwaj, D.~R. Morrison, Y.~Tachikawa and A.~Tomasiello, \emph{{The frozen
  phase of F-theory}},
  \href{https://doi.org/10.1007/JHEP08(2018)138}{\emph{JHEP} {\bfseries 08}
  (2018) 138}, [\href{https://arxiv.org/abs/1805.09070}{{\ttfamily
  1805.09070}}].

\bibitem{Heckman:2018jxk}
J.~J. Heckman and T.~Rudelius, \emph{{Top Down Approach to 6D SCFTs}},
  \href{https://doi.org/10.1088/1751-8121/aafc81}{\emph{J. Phys.} {\bfseries
  A52} (2019) 093001}, [\href{https://arxiv.org/abs/1805.06467}{{\ttfamily
  1805.06467}}].

\bibitem{Morrison:2012np}
D.~R. Morrison and W.~Taylor, \emph{{Classifying bases for 6D F-theory
  models}}, \href{https://doi.org/10.2478/s11534-012-0065-4}{\emph{Central Eur.
  J. Phys.} {\bfseries 10} (2012) 1072--1088},
  [\href{https://arxiv.org/abs/1201.1943}{{\ttfamily 1201.1943}}].

\bibitem{DelZotto:2014hpa}
M.~Del~Zotto, J.~J. Heckman, A.~Tomasiello and C.~Vafa, \emph{{6d Conformal
  Matter}}, \href{https://doi.org/10.1007/JHEP02(2015)054}{\emph{JHEP}
  {\bfseries 02} (2015) 054},
  [\href{https://arxiv.org/abs/1407.6359}{{\ttfamily 1407.6359}}].

\bibitem{Grassi:2011hq}
A.~Grassi and D.~R. Morrison, \emph{{Anomalies and the Euler characteristic of
  elliptic Calabi-Yau threefolds}},
  \href{https://doi.org/10.4310/CNTP.2012.v6.n1.a2}{\emph{Commun. Num. Theor.
  Phys.} {\bfseries 6} (2012) 51--127},
  [\href{https://arxiv.org/abs/1109.0042}{{\ttfamily 1109.0042}}].

\bibitem{Lee:2018ihr}
S.-J. Lee, D.~Regalado and T.~Weigand, \emph{{6d SCFTs and U(1) Flavour
  Symmetries}}, \href{https://doi.org/10.1007/JHEP11(2018)147}{\emph{JHEP}
  {\bfseries 11} (2018) 147},
  [\href{https://arxiv.org/abs/1803.07998}{{\ttfamily 1803.07998}}].

\bibitem{Apruzzi:2020eqi}
F.~Apruzzi, M.~Fazzi, J.~J. Heckman, T.~Rudelius and H.~Y. Zhang,
  \emph{{General prescription for global $U(1)$'s in 6D SCFTs}},
  \href{https://doi.org/10.1103/PhysRevD.101.086023}{\emph{Phys. Rev. D}
  {\bfseries 101} (2020) 086023},
  [\href{https://arxiv.org/abs/2001.10549}{{\ttfamily 2001.10549}}].

\bibitem{Aspinwall:1998xj}
P.~S. Aspinwall and D.~R. Morrison, \emph{{Nonsimply connected gauge groups and
  rational points on elliptic curves}},
  \href{https://doi.org/10.1088/1126-6708/1998/07/012}{\emph{JHEP} {\bfseries
  07} (1998) 012}, [\href{https://arxiv.org/abs/hep-th/9805206}{{\ttfamily
  hep-th/9805206}}].

\bibitem{Mayrhofer:2014opa}
C.~Mayrhofer, D.~R. Morrison, O.~Till and T.~Weigand, \emph{{Mordell-Weil
  Torsion and the Global Structure of Gauge Groups in F-theory}},
  \href{https://doi.org/10.1007/JHEP10(2014)016}{\emph{JHEP} {\bfseries 10}
  (2014) 016}, [\href{https://arxiv.org/abs/1405.3656}{{\ttfamily 1405.3656}}].

\bibitem{Cvetic:2017epq}
M.~Cvetic and L.~Lin, \emph{{The Global Gauge Group Structure of F-theory
  Compactification with U(1)s}},
  \href{https://doi.org/10.1007/JHEP01(2018)157}{\emph{JHEP} {\bfseries 01}
  (2018) 157}, [\href{https://arxiv.org/abs/1706.08521}{{\ttfamily
  1706.08521}}].

\bibitem{Hajouji:2019vxs}
N.~Hajouji and P.-K. Oehlmann, \emph{{Modular Curves and Mordell-Weil Torsion
  in F-theory}}, \href{https://doi.org/10.1007/JHEP04(2020)103}{\emph{JHEP}
  {\bfseries 04} (2020) 103},
  [\href{https://arxiv.org/abs/1910.04095}{{\ttfamily 1910.04095}}].

\bibitem{Monnier:2017oqd}
S.~Monnier, G.~W. Moore and D.~S. Park, \emph{{Quantization of anomaly
  coefficients in 6D $\mathcal{N}=(1,0)$ supergravity}},
  \href{https://doi.org/10.1007/JHEP02(2018)020}{\emph{JHEP} {\bfseries 02}
  (2018) 020}, [\href{https://arxiv.org/abs/1711.04777}{{\ttfamily
  1711.04777}}].

\bibitem{Heckman:2015ola}
J.~J. Heckman, D.~R. Morrison, T.~Rudelius and C.~Vafa, \emph{{Geometry of 6D
  RG Flows}}, \href{https://doi.org/10.1007/JHEP09(2015)052}{\emph{JHEP}
  {\bfseries 09} (2015) 052},
  [\href{https://arxiv.org/abs/1505.00009}{{\ttfamily 1505.00009}}].

\bibitem{Mekareeya:2016yal}
N.~Mekareeya, T.~Rudelius and A.~Tomasiello, \emph{{T-branes, Anomalies and
  Moduli Spaces in 6D SCFTs}},
  \href{https://doi.org/10.1007/JHEP10(2017)158}{\emph{JHEP} {\bfseries 10}
  (2017) 158}, [\href{https://arxiv.org/abs/1612.06399}{{\ttfamily
  1612.06399}}].

\bibitem{Heckman:2016ssk}
J.~J. Heckman, T.~Rudelius and A.~Tomasiello, \emph{{6D RG Flows and Nilpotent
  Hierarchies}}, \href{https://doi.org/10.1007/JHEP07(2016)082}{\emph{JHEP}
  {\bfseries 07} (2016) 082},
  [\href{https://arxiv.org/abs/1601.04078}{{\ttfamily 1601.04078}}].

\bibitem{Heckman:2018pqx}
J.~J. Heckman, T.~Rudelius and A.~Tomasiello, \emph{{Fission, Fusion, and 6D RG
  Flows}}, \href{https://doi.org/10.1007/JHEP02(2019)167}{\emph{JHEP}
  {\bfseries 02} (2019) 167},
  [\href{https://arxiv.org/abs/1807.10274}{{\ttfamily 1807.10274}}].

\bibitem{Anderson:2019kmx}
L.~B. Anderson, J.~Gray and P.-K. Oehlmann, \emph{{F-Theory on Quotients of
  Elliptic Calabi-Yau Threefolds}},
  \href{https://arxiv.org/abs/1906.11955}{{\ttfamily 1906.11955}}.

\bibitem{tHooft:1979rtg}
G.~'t~Hooft, \emph{{A Property of Electric and Magnetic Flux in Nonabelian
  Gauge Theories}},
  \href{https://doi.org/10.1016/0550-3213(79)90595-9}{\emph{Nucl. Phys. B}
  {\bfseries 153} (1979) 141--160}.

\bibitem{Aharony:2013hda}
O.~Aharony, N.~Seiberg and Y.~Tachikawa, \emph{{Reading between the lines of
  four-dimensional gauge theories}},
  \href{https://doi.org/10.1007/JHEP08(2013)115}{\emph{JHEP} {\bfseries 08}
  (2013) 115}, [\href{https://arxiv.org/abs/1305.0318}{{\ttfamily 1305.0318}}].

\bibitem{Tachikawa:2013hya}
Y.~Tachikawa, \emph{{On the 6d origin of discrete additional data of 4d gauge
  theories}}, \href{https://doi.org/10.1007/JHEP05(2014)020}{\emph{JHEP}
  {\bfseries 05} (2014) 020},
  [\href{https://arxiv.org/abs/1309.0697}{{\ttfamily 1309.0697}}].

\bibitem{Garcia-Etxebarria:2019cnb}
I.~García~Etxebarria, B.~Heidenreich and D.~Regalado, \emph{{IIB flux
  non-commutativity and the global structure of field theories}},
  \href{https://doi.org/10.1007/JHEP10(2019)169}{\emph{JHEP} {\bfseries 10}
  (2019) 169}, [\href{https://arxiv.org/abs/1908.08027}{{\ttfamily
  1908.08027}}].

\bibitem{Ohmori:2018ona}
K.~Ohmori, Y.~Tachikawa and G.~Zafrir, \emph{{Compactifications of 6d $N = (1,
  0)$ SCFTs with non-trivial Stiefel-Whitney classes}},
  \href{https://doi.org/10.1007/JHEP04(2019)006}{\emph{JHEP} {\bfseries 04}
  (2019) 006}, [\href{https://arxiv.org/abs/1812.04637}{{\ttfamily
  1812.04637}}].

\bibitem{Klevers:2014bqa}
D.~Klevers, D.~K. Mayorga~Pena, P.-K. Oehlmann, H.~Piragua and J.~Reuter,
  \emph{{F-Theory on all Toric Hypersurface Fibrations and its Higgs
  Branches}}, \href{https://doi.org/10.1007/JHEP01(2015)142}{\emph{JHEP}
  {\bfseries 01} (2015) 142},
  [\href{https://arxiv.org/abs/1408.4808}{{\ttfamily 1408.4808}}].

\bibitem{Oehlmann:2016wsb}
P.-K. Oehlmann, J.~Reuter and T.~Schimannek, \emph{{Mordell-Weil Torsion in the
  Mirror of Multi-Sections}},
  \href{https://doi.org/10.1007/JHEP12(2016)031}{\emph{JHEP} {\bfseries 12}
  (2016) 031}, [\href{https://arxiv.org/abs/1604.00011}{{\ttfamily
  1604.00011}}].

\bibitem{Weigand:2018rez}
T.~Weigand, \emph{{F-theory}}, {\emph{PoS} {\bfseries TASI2017} (2018) 016},
  [\href{https://arxiv.org/abs/1806.01854}{{\ttfamily 1806.01854}}].

\bibitem{Katz:2011qp}
S.~Katz, D.~R. Morrison, S.~Schafer-Nameki and J.~Sully, \emph{{Tate's
  algorithm and F-theory}},
  \href{https://doi.org/10.1007/JHEP08(2011)094}{\emph{JHEP} {\bfseries 08}
  (2011) 094}, [\href{https://arxiv.org/abs/1106.3854}{{\ttfamily 1106.3854}}].

\bibitem{Green:1984bx}
M.~B. Green, J.~H. Schwarz and P.~C. West, \emph{{Anomaly Free Chiral Theories
  in Six-Dimensions}},
  \href{https://doi.org/10.1016/0550-3213(85)90222-6}{\emph{Nucl. Phys. B}
  {\bfseries 254} (1985) 327--348}.

\bibitem{Sagnotti:1992qw}
A.~Sagnotti, \emph{{A Note on the Green-Schwarz mechanism in open string
  theories}}, \href{https://doi.org/10.1016/0370-2693(92)90682-T}{\emph{Phys.
  Lett. B} {\bfseries 294} (1992) 196--203},
  [\href{https://arxiv.org/abs/hep-th/9210127}{{\ttfamily hep-th/9210127}}].

\bibitem{Park:2011ji}
D.~S. Park, \emph{{Anomaly Equations and Intersection Theory}},
  \href{https://doi.org/10.1007/JHEP01(2012)093}{\emph{JHEP} {\bfseries 01}
  (2012) 093}, [\href{https://arxiv.org/abs/1111.2351}{{\ttfamily 1111.2351}}].

\bibitem{Marsano:2011hv}
J.~Marsano and S.~Schafer-Nameki, \emph{{Yukawas, G-flux, and Spectral Covers
  from Resolved Calabi-Yau's}},
  \href{https://doi.org/10.1007/JHEP11(2011)098}{\emph{JHEP} {\bfseries 11}
  (2011) 098}, [\href{https://arxiv.org/abs/1108.1794}{{\ttfamily 1108.1794}}].

\bibitem{Katz:1996xe}
S.~H. Katz and C.~Vafa, \emph{{Matter from geometry}},
  \href{https://doi.org/10.1016/S0550-3213(97)00280-0}{\emph{Nucl. Phys. B}
  {\bfseries 497} (1997) 146--154},
  [\href{https://arxiv.org/abs/hep-th/9606086}{{\ttfamily hep-th/9606086}}].

\bibitem{Aspinwall:2000kf}
P.~S. Aspinwall, S.~H. Katz and D.~R. Morrison, \emph{{Lie groups, Calabi-Yau
  threefolds, and F theory}},
  \href{https://doi.org/10.4310/ATMP.2000.v4.n1.a2}{\emph{Adv. Theor. Math.
  Phys.} {\bfseries 4} (2000) 95--126},
  [\href{https://arxiv.org/abs/hep-th/0002012}{{\ttfamily hep-th/0002012}}].

\bibitem{Cvetic:2018bni}
M.~Cveti\v~c and L.~Lin, \emph{{TASI Lectures on Abelian and Discrete
  Symmetries in F-theory}},
  \href{https://doi.org/10.22323/1.305.0020}{\emph{PoS} {\bfseries TASI2017}
  (2018) 020}, [\href{https://arxiv.org/abs/1809.00012}{{\ttfamily
  1809.00012}}].

\bibitem{Baume:2017hxm}
F.~Baume, M.~Cvetic, C.~Lawrie and L.~Lin, \emph{{When rational sections become
  cyclic --- Gauge enhancement in F-theory via Mordell-Weil torsion}},
  \href{https://doi.org/10.1007/JHEP03(2018)069}{\emph{JHEP} {\bfseries 03}
  (2018) 069}, [\href{https://arxiv.org/abs/1709.07453}{{\ttfamily
  1709.07453}}].

\bibitem{Gaiotto:2014kfa}
D.~Gaiotto, A.~Kapustin, N.~Seiberg and B.~Willett, \emph{{Generalized Global
  Symmetries}}, \href{https://doi.org/10.1007/JHEP02(2015)172}{\emph{JHEP}
  {\bfseries 02} (2015) 172},
  [\href{https://arxiv.org/abs/1412.5148}{{\ttfamily 1412.5148}}].

\bibitem{Mumford1961}
D.~Mumford, \emph{The topology of normal singularities of an algebraic surface
  and a criterion for simplicity},
  \href{https://doi.org/10.1007/BF02698717}{\emph{Publications
  Math{\'e}matiques de l'Institut des Hautes {\'E}tudes Scientifiques}
  {\bfseries 9} (Dec, 1961) 5--22}.

\bibitem{Tachikawa:2015wka}
Y.~Tachikawa, \emph{{Frozen singularities in M and F theory}},
  \href{https://doi.org/10.1007/JHEP06(2016)128}{\emph{JHEP} {\bfseries 06}
  (2016) 128}, [\href{https://arxiv.org/abs/1508.06679}{{\ttfamily
  1508.06679}}].

\bibitem{Heckman:2014qba}
J.~J. Heckman, \emph{{More on the Matter of 6D SCFTs}},
  \href{https://doi.org/10.1016/j.physletb.2015.05.046}{\emph{Phys. Lett.}
  {\bfseries B747} (2015) 73--75},
  [\href{https://arxiv.org/abs/1408.0006}{{\ttfamily 1408.0006}}].

\bibitem{Buchmuller:2017wpe}
W.~Buchmuller, M.~Dierigl, P.-K. Oehlmann and F.~Ruehle, \emph{{The Toric
  SO(10) F-Theory Landscape}},
  \href{https://doi.org/10.1007/JHEP12(2017)035}{\emph{JHEP} {\bfseries 12}
  (2017) 035}, [\href{https://arxiv.org/abs/1709.06609}{{\ttfamily
  1709.06609}}].

\bibitem{Dierigl:2018nlv}
M.~Dierigl, P.-K. Oehlmann and F.~Ruehle, \emph{{Global Tensor‐Matter
  Transitions in F‐Theory}},
  \href{https://doi.org/10.1002/prop.201800037}{\emph{Fortsch. Phys.}
  {\bfseries 66} (2018) 1800037},
  [\href{https://arxiv.org/abs/1804.07386}{{\ttfamily 1804.07386}}].

\bibitem{Paul-KonstantinOehlmann:2019jgr}
P.-K. Oehlmann and T.~Schimannek, \emph{{GV-Spectroscopy for F-theory on
  genus-one fibrations}},  \href{https://arxiv.org/abs/1912.09493}{{\ttfamily
  1912.09493}}.

\bibitem{Apruzzi:2018nre}
F.~Apruzzi, L.~Lin and C.~Mayrhofer, \emph{{Phases of 5d SCFTs from M-/F-theory
  on Non-Flat Fibrations}},
  \href{https://doi.org/10.1007/JHEP05(2019)187}{\emph{JHEP} {\bfseries 05}
  (2019) 187}, [\href{https://arxiv.org/abs/1811.12400}{{\ttfamily
  1811.12400}}].

\bibitem{Apruzzi:2019vpe}
F.~Apruzzi, C.~Lawrie, L.~Lin, S.~Schäfer-Nameki and Y.-N. Wang, \emph{{5d
  Superconformal Field Theories and Graphs}},
  \href{https://doi.org/10.1016/j.physletb.2019.135077}{\emph{Phys. Lett. B}
  {\bfseries 800} (2020) 135077},
  [\href{https://arxiv.org/abs/1906.11820}{{\ttfamily 1906.11820}}].

\bibitem{Apruzzi:2019opn}
F.~Apruzzi, C.~Lawrie, L.~Lin, S.~Schäfer-Nameki and Y.-N. Wang, \emph{{Fibers
  add Flavor, Part I: Classification of 5d SCFTs, Flavor Symmetries and BPS
  States}}, \href{https://doi.org/10.1007/JHEP11(2019)068}{\emph{JHEP}
  {\bfseries 11} (2019) 068},
  [\href{https://arxiv.org/abs/1907.05404}{{\ttfamily 1907.05404}}].

\bibitem{Apruzzi:2019enx}
F.~Apruzzi, C.~Lawrie, L.~Lin, S.~Schäfer-Nameki and Y.-N. Wang, \emph{{Fibers
  add Flavor, Part II: 5d SCFTs, Gauge Theories, and Dualities}},
  \href{https://doi.org/10.1007/JHEP03(2020)052}{\emph{JHEP} {\bfseries 03}
  (2020) 052}, [\href{https://arxiv.org/abs/1909.09128}{{\ttfamily
  1909.09128}}].

\bibitem{Braun:2018ovc}
A.~P. Braun, C.~R. Brodie, A.~Lukas and F.~Ruehle, \emph{{NS5-Branes and Line
  Bundles in Heterotic/F-Theory Duality}},
  \href{https://doi.org/10.1103/PhysRevD.98.126004}{\emph{Phys. Rev. D}
  {\bfseries 98} (2018) 126004},
  [\href{https://arxiv.org/abs/1803.06190}{{\ttfamily 1803.06190}}].

\bibitem{Ludeling:2014oba}
C.~Lüdeling and F.~Ruehle, \emph{{F-theory duals of singular heterotic K3
  models}}, \href{https://doi.org/10.1103/PhysRevD.91.026010}{\emph{Phys. Rev.
  D} {\bfseries 91} (2015) 026010},
  [\href{https://arxiv.org/abs/1405.2928}{{\ttfamily 1405.2928}}].

\bibitem{Aspinwall:1998he}
P.~S. Aspinwall and R.~Y. Donagi, \emph{{The Heterotic string, the tangent
  bundle, and derived categories}},
  \href{https://doi.org/10.4310/ATMP.1998.v2.n5.a4}{\emph{Adv. Theor. Math.
  Phys.} {\bfseries 2} (1998) 1041--1074},
  [\href{https://arxiv.org/abs/hep-th/9806094}{{\ttfamily hep-th/9806094}}].

\bibitem{deBoer:2001wca}
J.~de~Boer, R.~Dijkgraaf, K.~Hori, A.~Keurentjes, J.~Morgan, D.~R. Morrison
  et~al., \emph{{Triples, fluxes, and strings}},
  \href{https://doi.org/10.4310/ATMP.2000.v4.n5.a1}{\emph{Adv. Theor. Math.
  Phys.} {\bfseries 4} (2002) 995--1186},
  [\href{https://arxiv.org/abs/hep-th/0103170}{{\ttfamily hep-th/0103170}}].

\bibitem{Hassler:2019eso}
F.~Hassler, J.~J. Heckman, T.~B. Rochais, T.~Rudelius and H.~Y. Zhang,
  \emph{{T-Branes, String Junctions, and 6D SCFTs}},
  \href{https://doi.org/10.1103/PhysRevD.101.086018}{\emph{Phys. Rev.}
  {\bfseries D101} (2020) 086018},
  [\href{https://arxiv.org/abs/1907.11230}{{\ttfamily 1907.11230}}].

\bibitem{Anderson:2018heq}
L.~B. Anderson, A.~Grassi, J.~Gray and P.-K. Oehlmann, \emph{{F-theory on
  Quotient Threefolds with (2,0) Discrete Superconformal Matter}},
  \href{https://doi.org/10.1007/JHEP06(2018)098}{\emph{JHEP} {\bfseries 06}
  (2018) 098}, [\href{https://arxiv.org/abs/1801.08658}{{\ttfamily
  1801.08658}}].

\bibitem{Apruzzi:2017iqe}
F.~Apruzzi, J.~J. Heckman and T.~Rudelius, \emph{{Green-Schwarz Automorphisms
  and 6D SCFTs}}, \href{https://doi.org/10.1007/JHEP02(2018)157}{\emph{JHEP}
  {\bfseries 02} (2018) 157},
  [\href{https://arxiv.org/abs/1707.06242}{{\ttfamily 1707.06242}}].

\end{thebibliography}\endgroup

\end{document}